\documentclass[aps,prl,twocolumn,superscriptaddress,longbibliography]{revtex4-2}  

\usepackage{graphicx}
\usepackage{epstopdf}
\usepackage{color}
\usepackage{xcolor}
\usepackage{changes}

\usepackage{physics}
\usepackage{amsthm,amsmath,amssymb}
\usepackage{xr-hyper}
\usepackage[colorlinks=true, urlcolor=blue, linkcolor=blue, citecolor=blue, pdftex]{hyperref}
\usepackage{array}
\usepackage{multirow}
\usepackage{textcomp}
\newcommand{\SMsecref}[1]{SM Sec.~\ref{SM-#1}}

\usepackage{bm}

\begin{document}
\title{Ferromagnetic transition in a chiral spin chain}

\author{Yuan Xiao}
\affiliation{Physics Department, University of California, Santa Barbara, California 93106-4030, USA}

\author{Rimika Jaiswal}
\affiliation{Physics Department, University of California, Santa Barbara, California 93106-4030, USA}

\author{Oleg A. Starykh}
\affiliation{Department of Physics and Astronomy, University of Utah, Salt Lake City, UT 84112, USA}

\author{Leon Balents}
\email{balents@spinsandelectrons.com}
\affiliation{Kavli Institute for Theoretical Physics, University of California, Santa Barbara, California 93106-4030, USA}
\affiliation{French American Center for Theoretical Science, CNRS, KITP, Santa Barbara, California 93106-4030, USA}
\affiliation{Canadian Institute for Advanced Research, Toronto, Ontario, M5G 1M1 Canada}

\begin{abstract}
  We study the zero-temperature properties of a spin-$1/2$ Heisenberg chain with an additional three-site chiral term of strength $g$.  The model is known to be integrable and previous work using Bethe ansatz has identified a phase transition from the usual critical state for $g< g_c = 2/\pi$ to a chiral phase for $g>g_c$.  Here we combine analytical and computational approaches to demonstrate that the transition point comprises an unusual Lifshitz transition with dynamical critical exponent $z=3$, and the chiral phase is a partially spin-polarized phase which spontaneously breaks SU(2) symmetry.  We derive a strongly interacting chiral boson field theory for the critical point, and show that even its small-momentum, low-energy response is non-trivial.  We also determine the universal properties of the chiral phase, and specifically show that it has a simultaneous quadratically dispersing ferromagnetic ``magnon'' mode coexisting with the power-law singularities and linearly dispersing excitations typical of a Luttinger liquid.  Our conclusions are validated using numerics based on matrix product state methods and exact diagonalization.
\end{abstract}

\maketitle

\noindent \textit{Introduction:} The one-dimensional spin-1/2 Heisenberg nearest-neighbor antiferromagnetic chain is famous as a non-trivial soluble model of a gapless quantum phase of matter.  While the thermodynamics of the model are fully known from the Thermodynamic Bethe Ansatz (TBA)\cite{Takahashi1999}, a more complete understanding of the low-energy physics is obtained by a diverse set of methods from field theory and statistical mechanics.  Universal properties of the chain are fully described from the SU(2)$_1$ conformal field theory\cite{AffleckHaldane1987}, which has several useful representations (fermions, bosonization, current algebra, etc.) that allow calculation of various dynamical correlation functions and reproduce low-energy thermodynamics.  Many symmetry-preserving perturbations of the nearest-neighbor chain maintain these features as a robust {\em phase}, which shares the same theoretical description.

Recently, Sedrakyan \emph{et al.}\cite{Sed} studied in particular the perturbation of the Heisenberg chain by a three-site chiral term, i.e.
\begin{equation}
  \label{eq:1}
  H = \sum_n \left[\bm{S}_n\cdot\bm{S}_{n+1} - g\bm{S}_n\cdot\bm{S}_{n+1}\times \bm{S}_{n+2}\right],
\end{equation}
with strength $g$.  These authors showed that this model is Bethe-ansatz integrable for all $g$, and undergoes a transition to a ``chiral phase'' when $g \geq g_c = 2/\pi$.  More recently the same authors found\cite{sedrakyan2025magnetochiral} that the large $g$ limit of the state is ferromagnetic with a non-zero partial spin polarization.

In this paper, we argue that the transition occurs directly to a
ferromagnetic state, with continuous breaking of SU(2) symmetry.  This
is to our knowledge the only example of a continuous ferromagnetic
transition with full SU(2) symmetry in a 1d spin system, and the field
theory of the phase transition is consequently unique and interesting.  We note that a somewhat similar transition was studied by 
Kun Yang\cite{kunyang2004}, in a non-chiral situation and with easy-axis U(1)$\times \mathbb{Z}_2$ spin symmetry rather than SU(2), which applies here.

We show that in our case the critical theory is an integrable deformation of a
conformal field theory (CFT) with dynamical critical exponent $z=3$,
strikingly distinct from the usual $z=1$ CFT description which works
at low energies for $g<g_c$.   In particular, the critical theory has no free-field description, and can be consider as a strongly interacting chiral boson model.   The spontaneous
nature of the symmetry breaking furthermore implies the presence of
quadratically-dispersing excitations even in the ordered chiral phase,
which are visible in the dynamical spectra.  We describe the key
features of the spectral functions, exhibiting $z>1$ structures both
at the phase transition point and in the ferromagnetic phase.

\begin{figure}[hbtp]
  \centering \includegraphics[width=\columnwidth]{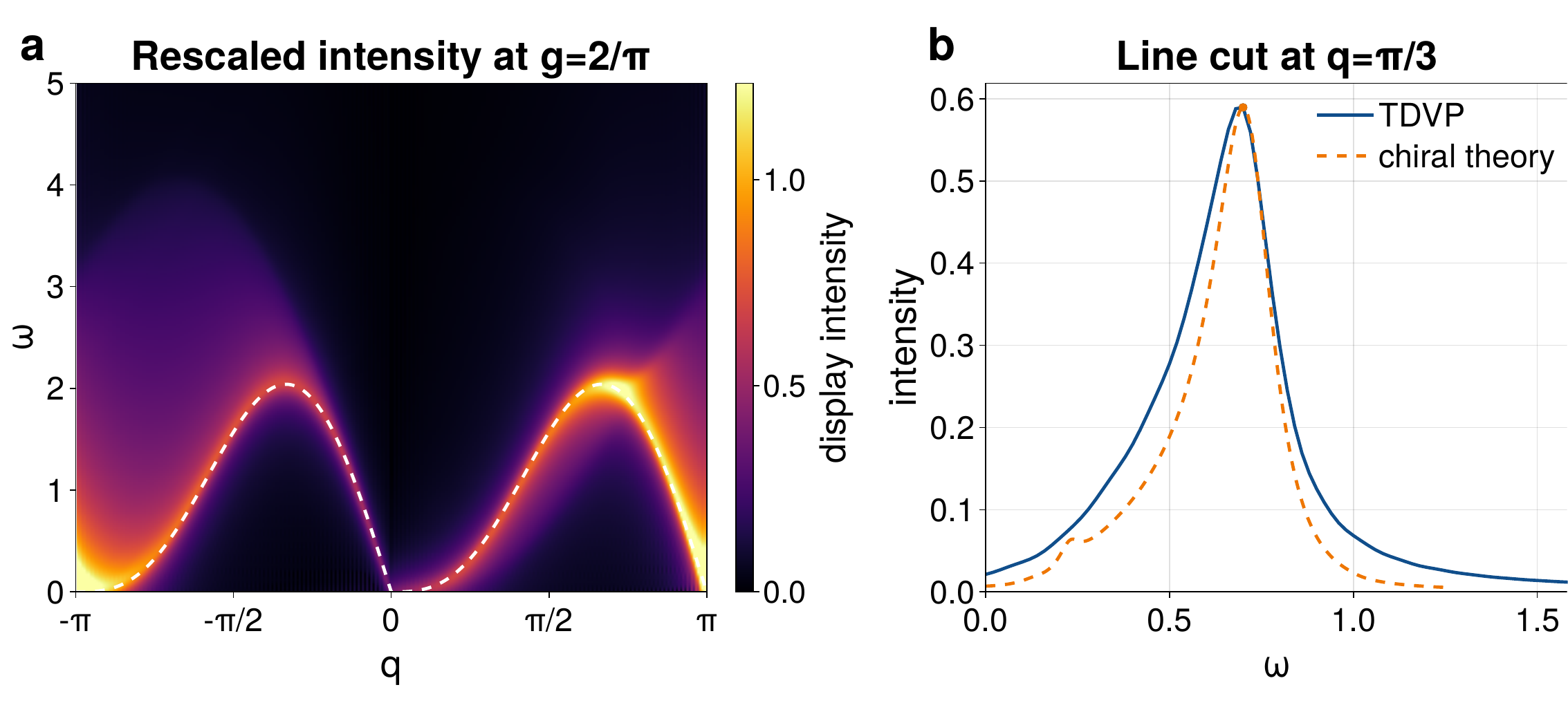}  
  \caption{Panel (a): Dynamical structure factor from DMRG+TDVP at the transition point. The color scale shows a rescaled display intensity derived from the positive part of the spectral density, $I(q,\omega)=\max\{S(q,\omega),0\}$. To improve visibility of weaker features, the intensity is first clipped at the 99.5th percentile of nonzero values and then square-root rescaled: $I_{\rm disp}(q,\omega)=[\min(I(q,\omega),I_{99.5})]^{1/2}$. The dashed white curve shows $\omega(q)=(\pi/2)|\sin q|(1-\operatorname{sgn}(q)\cos q)$. Panel (b): Comparison of lineshape for $q=\pi/3$ to the approximate low-energy chiral boson theory.  See text for details.}
  \label{fig:dynsf}
\end{figure}

\noindent \textit{Classical phase diagram:} An intuition for the phase diagram can be obtained already classically, i.e. by treating the spins as length $1/2$ unit vectors.  The Heisenberg term prefers antiparallel nearest-neighbors, while the chiral term can be lowered only by generating non-collinear triplets of three adjacent spins.  An optimal state can be found with the ansatz $\bm{S}_n = (\sqrt{\frac14-m^2}\cos q_{cl}n, \sqrt{\frac14-m^2} \sin q_{cl}n, m)$, which describes an ``umbrella'' state with wavevector $q_{cl}$ and uniform polarization $m$ along the z direction (which is arbitrary).  Inserting this into Eq.~\eqref{eq:1}, we find the energy per site $\mathcal{E}(m,q_{cl};g) = \frac14 \left(4m^2 + (1-4m^2)\cos q_{cl}\right) - \frac14 g m(1-4m^2) (2\sin q_{cl} - \sin 2q_{cl})$.  Optimizing with respect to $m$ and $q_{cl}$, one finds that for $g\leq 1$, $m=0$ and $q_{cl}=\pi$, which describes the antiferromagnetic solution.  For $g>1$, the magnetization continuously increases from zero as $|m| = \frac12\sqrt{(g-1)/(1+3g)}$ 
while $\cos q_{cl} =-(1+g)/(2g)$, maintaining the relation $g \sin q_{cl} = 4m/(1-12m^2)$.
Note that in the quantum system, the ``antiferromagnetic'' (i.e., SU(2)$_1$) state is more unstable than in the classical limit, since $g_c = 2/\pi <1$.

\noindent \textit{Bosonization:}  It is well-known that the low-energy physics of the Heisenberg chain can be described using a free boson field theory, and this provides a convenient way to analyze the approach to the transition point.  Using abelian bosonization \cite{extreme2012} (see \SMsecref{sec:bosonization} \cite{SM}),  the fixed point Hamiltonian density of the Heisenberg chain is  $\mathcal{H}(g=0) \sim v_0 [(\partial_x \varphi_L)^2 + (\partial_x \varphi_R)^2]$, where $\varphi_{R/L}$ are chiral boson fields with $v_0 = \pi/2$ the velocity for $g=0$.  Then the chiral term is just the energy current\cite{Zotos1997}, which bosonizes as $\bm{S}_n\cdot\bm{S}_{n+1}\times \bm{S}_{n+2} \sim v_0^2[(\partial_x \varphi_L)^2 - (\partial_x \varphi_R)^2]$.  Adding the two gives, to quadratic order in the bosons,  $\mathcal{H}_{\rm quad} (g) \sim v_L (\partial_x \varphi_L)^2 + v_R(\partial_x \varphi_R)^2$, where $v_L = v_0 + gv_0^2$ and $v_R=v_0 -g v_0^2$.  This result is correct up to operators that are irrelevant for $g<g_c$.  The critical value, determined by $v_R=0$, correctly gives $g_c=1/v_0 = 2/\pi$.  The basic physics revealed by bosonization is that the chiral term selectively slows the right-moving excitations until at $g=g_c$ their velocity vanishes. 

To make SU(2) symmetry manifest, we move to the current algebra/non-abelian formulation.  In particular, chiral SU(2)$_1$ currents can be introduced via $J_{L/R}^z = \frac{1}{\sqrt{2\pi}} \partial_x \varphi_{L/R}$, $J_{L/R}^+ = A \exp( \pm i 2\sqrt{2\pi} \varphi_{L,R})$, where $A$ is an amplitude that depends upon the cutoff scheme.  The currents obey the Kac-Moody SU(2)$_1$ algebra, and the quadratic Hamiltonian density in these variables becomes
\begin{align}
  \label{eq:25}
  \mathcal{H}_{\rm quad} = \frac{2\pi v_L}{3} : J_L^a J_L^a: + \frac{2\pi v_R}{3} : J_R^a J_R^a:.
\end{align}
Equivalently, we can introduce the chiral stress-energy tensors, $T_{L/R} = \frac13 :J_{L/R}^a J_{L/R}^a:$ and
\begin{align}
  \label{eq:26}
  \mathcal{H}_{\rm quad} = 2\pi v_L T_L + 2\pi v_R T_R.
\end{align}
As is well-known for non-chiral Heisenberg-like chains, there is also a ``backscattering'' contribution,
\begin{equation}
    \mathcal{H}_{\rm bs} = - g_{\rm bs} J_L^a J_R^a.
\end{equation}
In \SMsecref{sec:rg-bs} \cite{SM} we show that for $g_{\rm bs}>0$ (as expected) the backscattering interaction is marginally irrelevant, i.e. logarithmically suppressed at low energies, for any ratio $v_R/v_L$, and in particular even as $v_R \rightarrow 0$.  This argues that it remains irrelevant up to and including the critical point.  Henceforth we neglect $g_{\rm bs}$, though further study would be of interest.

Now we focus on the right-moving sector, which is becoming ``soft'' as $v_R \rightarrow 0$ at the transition point.  We expect that higher order corrections stabilize the system in this limit, and since it is now clear that the $T_{R/L}$ are SU(2) scalars,  the obvious non-trivial correction to the quadratic Hamiltonian is the ``$T^2$'' deformation
\begin{align}
  \label{eq:27}
  \mathcal{H}_R = 2\pi v_R T_R + \lambda : T_R^2:.
\end{align}
Here $\lambda$ is a coupling with dimensions of velocity$^3$ which remains finite and non-zero at the critical point, at which $v_R=0$.  The $\lambda$ term is present already in the pure Heisenberg chain and has been discussed in that context to describe corrections to scaling\cite{pereira2006dynamical,lukyanov1998low}.

Note that the $\lambda$ term represents an interaction in the boson theory.  It can be explicitly expressed in terms of the chiral boson field as
\begin{align}
    :T_R:^2 = \frac{1}{(2\pi)^2}\left[:(\partial_x \varphi_R)^4: + \frac{1}{2\pi} :(\partial_x^2 \varphi_R)^2:\right].\label{eq:28}
\end{align}
At the transition point, $v_R=0$, and the right-moving sector has no quadratic boson term, so it is extremely strongly interacting, i.e. $\mathcal{H}_R = \lambda :T_R^2:$.  What is immediately evident by scaling is that, since the right-moving Hamiltonian is the spatial integral of Eq.~\eqref{eq:28}, the energy scales like inverse length cubed, which defines the dynamical scaling exponent $z=3$.    

\emph{Critical structure factor:} This is confirmed by an examination of the dynamical spin spectral function
\begin{equation}
  \label{eq:2}
  S^{AB}(q,\omega) = \frac{\textrm{Re}}{L\pi}  \int_0^\infty \!\!\! dt \sum_{n, n'} \langle S^A_n(t) S_{n'}^B(0)\rangle e^{i q (n-n') + (i \omega  - \eta) t},
\end{equation}
with $\eta>0$ a damping parameter, and $S^A_n(t) = e^{iHt} S_n^A e^{-i H t}$.  Fig.~\ref{fig:dynsf} shows $S^{+-}(q,\omega) = 2 S^{zz}(q,\omega)$ at $g=2/\pi$ evaluated by  DMRG+TDVP for an open chain with $L=300$  (see \SMsecref{sec:comp-deta-furth} \cite{SM} for details of the method).  Here finite-size quantization is not visible and the spectrum appears continuous.  Although the spectral weight is continuous and there is no delta-function-like magnon mode, there is a dominant dispersing peak which evidences the chiral nature, being highly asymmetric around $q=0$ and $q=\pi$.  We find that the peak dispersion is well described by the functional form
\begin{align}
  \label{eq:70}
  \omega(q) = \omega_0 |\sin q| (1- \textrm{sign}(q) \cos q), \, \textrm{for} -\pi<q<\pi, 
\end{align}
which is inspired by the two-particle dispersion boundaries of free fermions (i.e. the Lindhard function).   This curve (with $\omega_0 = \pi/2$) matches the peak dispersion excellently.  Eq.~\eqref{eq:70} indeed satisfies $z=3$ scaling to the ``right'' of $q=0$ and $q=\pi$, but $z=1$ scaling to the ``left'', embodying the different behaviors of the two chiral sectors.  An additional intriguing feature above the peak near $q\approx 0.7\pi$ appears to echo a similar feature for free fermions. 

The structure factor for positive $q\ll \pi$ is a probe of the right-moving sector, and most directly manifests the strongly interacting nature of the chiral boson theory.  Here
\begin{align}
  \label{eq:29}
  \left\langle J_R^a(x,\tau) J_R^b(0,0)\right\rangle \sim \delta^{ab} \frac{1}{x^2} \mathfrak{C}(\tau/x^{1/3}),
\end{align}
where $\mathfrak{C}$ is a non-trivial scaling function, and we neglect any (i.e. logarithmic) corrections from irrelevant operators (see below).  Note that the power-law prefactor is fixed by the fact that at $\tau=0$, the correlator becomes a ground state expectation value, and the ground state is identical to that of the ordinary Heisenberg chain.  Eq.~\eqref{eq:29} governs the dynamical structure factor for small \emph{positive} momentum $q$ and low frequency.  It implies that
\begin{align}
  \label{eq:30}
  S(0<q\ll 1,\omega) \sim \frac{1}{q^2} \mathfrak{S}_0(\omega/q^3).
\end{align}
Notably, the function $\mathfrak{S}_0(\Omega)$ is non-trivial and determines a universal lineshape (up to overall scales), with a width of the same order as the frequency.  This implies that, unlike for the Heisenberg chain and generically for $g<g_c$, the chiral boson is itself not a good quasiparticle.

What is this lineshape $\mathfrak{S}_0(\Omega)$?  In \SMsecref{sec:continuum-many-boson} \cite{SM} we formulate this problem in the chiral strongly interacting boson field theory as a problem of ``splintering'' of an initial chiral boson created by the spin current into many constituent bosons carrying fractions of the momentum and energy.  We obtain an approximate numerical solution which is enabled by the strong positivity constraints on the momenta and energies of all the chiral bosons.  Details of the calculations are given in \SMsecref{sec:continuum-many-boson} \cite{SM}.    In Fig.~\ref{fig:dynsf}(b), the theoretical spectrum (emphatically \emph{not} an exact solution) is compared to the TDVP simulations at a judiciously chosen cut of  $S(q_0,\omega)$ with $q_0=\pi/3$, small enough to lie clearly within the right-moving chiral low-momentum branch, but not so small that the spectral weight is compressed to exceptionally low frequency that is not well calculated by TDVP.  The two lineshapes are similar, and notably feature a pronounced tail \emph{below} the peak.  

It is more subtle to apply scaling to the spin correlations near momentum $\pi$, which are governed by the staggered magnetization operator $\bm{N}$.  In the SU(2)$_1$ theory, this is a product of right- and left-moving (Kac-Moody) primary fields with conformal dimensions $(\tfrac14,\tfrac14)$.   Hence the spin correlations near $\pi$ involve both the $z=1$ (left) and $z=3$ (right) sectors.   Nevertheless, we expect that for $k=\pi +q$, with $0<q\ll 1$, the right-moving sector dominates the dynamical structure factor at low energy, because the energy of left-moving excitations is too high to be excited at these frequencies, and we postulate
\begin{align}
  \label{eq:31}
  S(\pi+q,\omega) \sim \frac{1}{q^3} \mathfrak{S}_\pi(\omega/q^3),
\end{align}
which applies in the scaling limit $\omega,q \ll 1$ but $\omega/q^3$ fixed, so that $\omega/q \approx 0$.  The scaling of the prefactor is again determined by the known ground state spin correlations.

\emph{Tower of states: } The same $T^2$ theory makes sharp predictions
for the structure of finite-size energy levels at criticality -- the
``tower of states'', which we check with exact diagonalization below.
While the states here do not form a \emph{conformal} tower because the
critical theory is not conformal, we can use results on integrable
deformations of conformal field theory\cite{bazhanov1996integrable} to
understand them.  In particular, the $T^2$ interaction can be entirely
expressed in terms of the (right-moving) generators $L_n$ of the Virasoro algebra (we drop the ``R'' subscript on the generators for compactness).
Taking the theory on a spatial circle, i.e. applying periodic boundary
conditions, one has the Fourier decomposition
\begin{align}
  \label{eq:32}
  L_n & = L\int_0^L \!\! dx\, e^{-2\pi i n x/L} T_R(x),
\end{align}
and these generators obey the Virasoro algebra $  [L_n, L_m] = (n-m)L_{n+m} + \frac{1}{12}n(n^2-1)\delta_{n+m, 0}$.  Note that $L_0$ can be identified with the momentum operator of the chiral theory, via $P_R = \frac{2\pi}{L} L_0$  (in the vacuum sector), and has integer eigenvalues $L_0=n=0,1,\cdots$.  Then the integral of the $T^2$ term is a higher conserved charge, and is expressed (see \SMsecref{sec:deriv-i_3-form} \cite{SM}) as 
\begin{align}
  \label{eq:33}
  H_R(v_R=0) & =   \lambda \int_0^L  \! dx\, :T_R^2(x): = \left(\frac{1}{L}\right)^3\lambda\, I_3, \nonumber \\
  I_3 & = 2\sum_{k=1}^{\infty} L_{-k}L_k + L_0^2 .
\end{align}
In our problem, the Virasoro generators are obtained from the (right-moving) Kac-Moody ones by $L_n = \frac13 \sum_m : J_m^a J_{n -m}^a:$, where the Kac-Moody commutation relations are $ \left[J_{m}^a,J_{n}^b\right] = i \epsilon^{abc} J_{m+n}^c + \frac12 m \,\delta_{m+n,0} \,\delta^{ab}$.

\begin{figure}[htbp]
  \centering
  \includegraphics[width=\columnwidth]{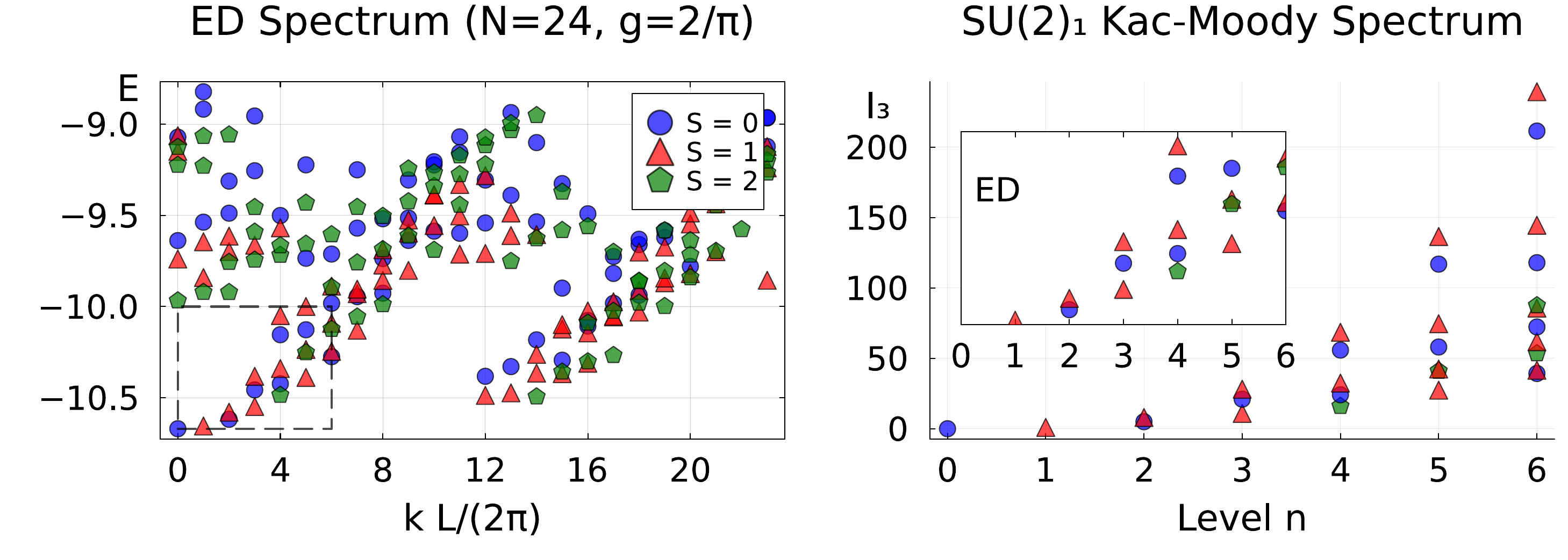}  
  \caption{Comparison of Exact Diagonalization (ED) spectrum for a 24-site chain (left) and the spectrum of the $:T^2:$ theory from Eq.~\eqref{eq:33} (right).  Inset in the right panel shows the corresponding range of momenta from the ED (dashed rectangle on the left).  The ordering of and number of low-energy levels for each spin at each momentum corresponds perfectly in the two plots.}
  \label{fig:ed-vs-kdv}
\end{figure}

Together, these relations allow for an algebraic calculation of the spectrum of $I_3$ for each value of the momentum $k_n = 2\pi n/L$ when $n$ is not too large.   The procedure, discussed in \SMsecref{sec:i_3-spectrum-circle} \cite{SM}, is to identify and construct the Virasoro and Kac-Moody towers of states (modules), and diagonalize $I_3$ within the states from a given Virasoro primary.  In Fig.~\ref{fig:ed-vs-kdv}, we compare the full theoretical spectrum (right panel) to exact diagonalization (left panel) on a 24-site system, and find that the  \emph{ordering of levels is exactly reproduced} for $n\leq 6$.    This is a striking validation of the field theory.  We note that one can further evaluate the finite-size dynamical structure factor by similar methods, as discussed in \SMsecref{sec:spectral-weights} \cite{SM}.

\begin{figure}[htbp]
  \centering
  \includegraphics[width=\columnwidth]{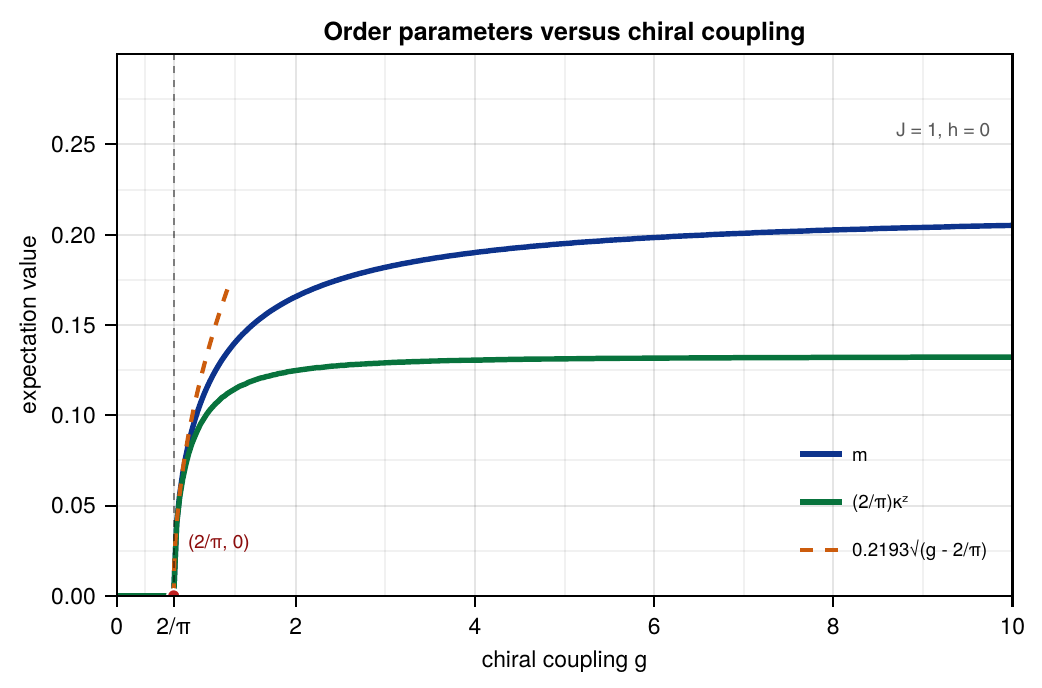}  
  \caption{Spontaneous magnetization $m$ and scaled vector spin chirality $\kappa^z$ versus $g$ at zero field calculated by Thermodynamic Bethe Ansatz (TBA).   The dashed curve shows the asymptotic behavior $m \sim m_0 \sqrt{g-g_c}$ with $g_c=2/\pi$ and $m_0=0.2193$ obtained from the TBA.  The agreement between the $m$ and $\frac{2}{\pi}\kappa^z$ near $g_c$ indicates the chiral nature of the critical point.}
  \label{fig:mag}
\end{figure}

\noindent \textit{Ordered phase:} For $\delta g = g-g_c>0$, the right-moving sector becomes unstable because $v_R<0$, favoring a spontaneous chiral spin current $\langle J_R^a\rangle \neq 0$.  Since $\langle J_L^a\rangle$ remains small, this immediately implies both a uniform magnetization $m^a \sim J_R^a+J_L^a$ and vector spin chirality $\chi^a \sim J_R^a-J_L^a$, consistent with the classical umbrella picture.  Moreover, because the critical instability is purely right-moving, the two order parameters are locked near the transition; in our normalization this predicts $(2/\pi)\kappa^z/m \to 1$ as $g\to g_c^+$.  The magnetization and chirality can be obtained exactly from the TBA, giving the curves in Fig.~\ref{fig:mag}, which confirm this prediction.  Scaling of the perturbation $-\delta g(\partial_x \varphi_R)^2$ gives $\xi \sim (\delta g)^{-1/2}$ and hence $M \sim (\delta g)^{\frac12}$, again in agreement with the TBA result.

The ordered phase has an unusual two-component low-energy dynamics.  Static long-wavelength fluctuations retain the power-law correlations of a Luttinger liquid, described by a linearly dispersing  boson, while spontaneous breaking of SU(2) also requires a ferromagnetic Goldstone mode with quadratic dispersion $\omega \simeq k^2/(2m^*)$ (see \SMsecref{sec:fm} and \SMsecref{sec:effective} \cite{SM}).   In a semiclassical umbrella description, these correspond respectively to rotations within the umbrella plane and precession of the umbrella axis.  This motivates an augmented bosonization theory, in which the in-plane phase and its conjugate become the standard bosonization fields, while the precession mode becomes an additional complex boson with non-relativistic dispersion; both enter the low-energy decomposition of the spin operators.   The semi-classical umbrella wavevector is promoted in this theory to a ``spiral'' wavevector $q_{cl} \rightarrow q_s$, where $q_s$ determines dominant correlations of $S^\pm$; it is renormalized from the classical value $q_{cl}$ but can be obtained exactly from TBA (see Fig.~\ref{SM-fig:wavevectors} in the SM). The semiclassical dynamics and augmented bosonization are summarized in the End Matter and detailed in \SMsecref{sec:classical} \cite{SM}.  Some parallels exist between other extensions of bosonization to ferromagnetism and beyond\cite{PhysRevLett.101.170403,PhysRevLett.99.240404,RevModPhys.84.1253,PhysRevB.76.140408}.


\begin{figure}[htbp]
  \centering
  \includegraphics[width=\columnwidth]{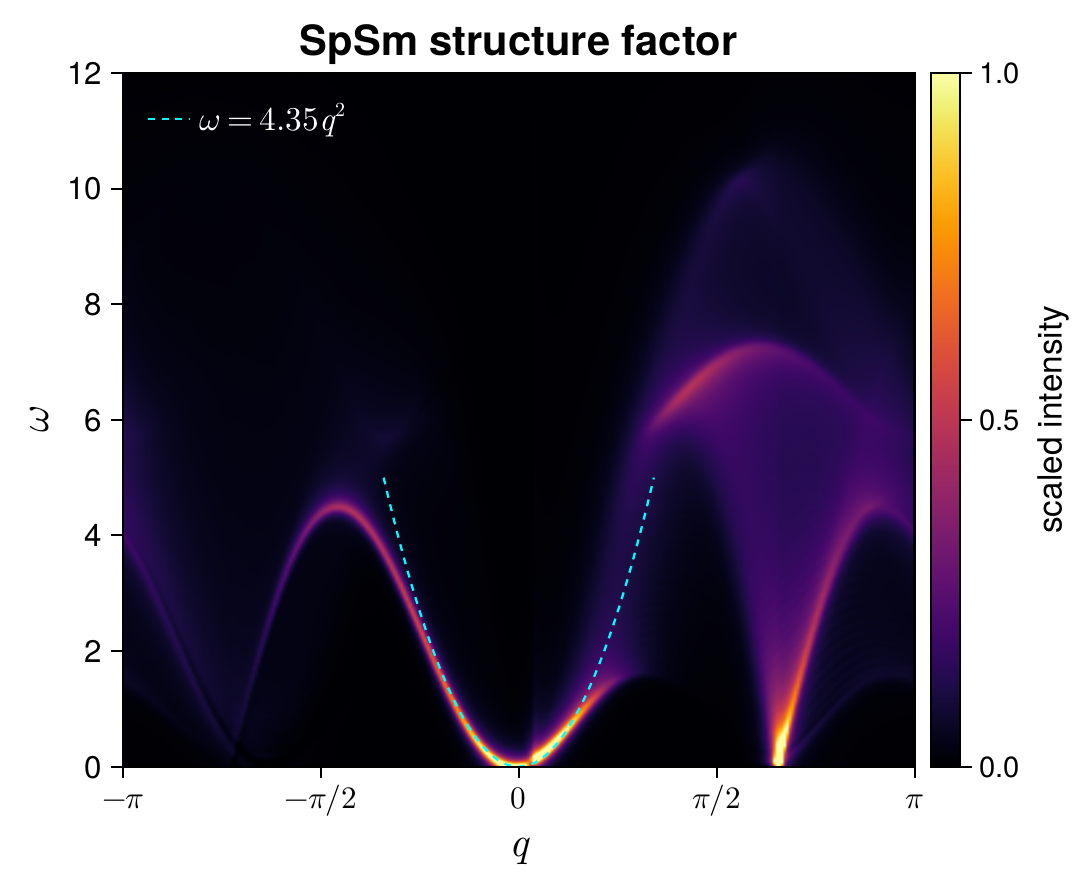}  
  \caption{Dynamical correlation function $S^{+-}(q,\omega)$ in the ordered phase with $g=4$, measured in the ground state with maximum magnetization along $z$ ($S^z_{\rm tot}=57$) calculated using DMRG+TDVP for a system of $N=300$ sites.   See \SMsecref{sec:comp-deta-furth} \cite{SM} for details of the method.   The color indicates the scaled intensity $I(q,\omega)=\min[\max(S^{+-}_\eta(q,\omega),0)/S_{\rm clip},1]$, where $S_{\rm clip}$ is 99.9\% of the maximum intensity  of $S^{+-}$; values above $S_{\rm clip}$ are saturated.  A prominent quadratically dispersing magnon is clearly visible near $q=0$: the dashed line is a parabola with curvature (inverse effective mass) determined from the exact TBA equations.}
  \label{fig:SpSm_g4}
\end{figure}

DMRG+TDVP calculations confirm this picture.  Ground-state Friedel oscillations\cite{hikihara2004}, shown in Fig.~\ref{SM-fig:friedel}, agree with the $z=1$ Luttinger-liquid sector (\SMsecref{sec:ground-state-results} \cite{SM}), while the ordered-phase dynamical structure factor, shown for $g=4$ in Fig.~\ref{fig:SpSm_g4}, displays the quadratic magnon in the $S^{+-}$ channel.  Its curvature agrees with the effective mass obtained independently from the TBA; further component-resolved spectra are presented in the End Matter, where a successful comparison with augmented bosonization is described.

The above results provide a comprehensive picture of the low-energy physics and quantum criticality of the chiral chain, highlighting an unconventional combination of ballistic and $z \geq 2$ dynamics.  The chirality and integrability of the quantum critical theory enabled substantial theoretical progress despite the lack of a free field or conformal field theory.  It would be interesting in future work to explore the effect of integrability-breaking perturbations, generalizations to higher dimensions, and possibilities to observe this physics in experiments or quantum simulators.

\textit{Acknowledgments.---}
LB thanks Ara Sedrakyan for hospitality at Landau Week 2025 in Yerevan, Armenia, where the idea for this project arose, and Tigran Sedrakyan for explaining his pioneering studies of the problem.  AI assistance (Gemini, Claude, Codex, and Mistral) was used in coding of the numerics, in aspects of the conformal field theory and TBA calculations, and in drafting parts of the summaries of these components of the work.  This research was supported in part by grant NSF PHY-2309135 to the KITP.  L.B. is supported by the NSF CMMT program under Grants No. DMR-2419871, the Simons Collaboration on Ultra-Quantum Matter, which is a grant from the Simons Foundation (Grant No. 651440), and the Gordon and Betty Moore Foundation through Grant GBMF8690 to the University of California, Santa Barbara.

\textit{Data availability.---}
Data and simulation codes are available from the authors upon reasonable request.

\textit{Author contributions.---}
Y.X. and R.J. carried out the initial numerical investigation of the problem.  O.S. and L.B. developed the analytical theory and wrote the manuscript, and L.B. performed the later numerical computations.  

\makeatletter
\let\combined@bibsection\bibsection
\renewcommand{\bibsection}{%
  \begingroup
  \renewcommand{\addcontentsline}[3]{}%
  \combined@bibsection
  \endgroup
}
\makeatother
\bibliography{chiral}

@article{sed,
	author = {Wei, Chenan and Mkhitaryan, Vagharsh V. and Sedrakyan, Tigran A.},
	date = {2024/06/19},
	doi = {10.1007/JHEP06(2024)125},
	id = {Wei2024},
	isbn = {1029-8479},
	journal = {Journal of High Energy Physics},
	number = {6},
	pages = {125},
	title = {Unveiling chiral states in the {XXZ} chain: finite-size scaling probing symmetry-enriched c = 1 conformal field theories},
	url = {https://doi.org/10.1007/JHEP06(2024)125},
	volume = {2024},
	year = {2024}}

@article{sedrakyan2025magnetochiral,
  title={Magnetochiral eigenstate of the {H}eisenberg chain with spontaneous symmetry breaking},
  author={Sedrakyan, Tigran A and Pang, Junjun and Wei, Chenan and Wang, Baigeng},
  journal={arXiv preprint arXiv:2512.09107},
  year={2025}
}

@article{extreme2012,
	author = {Starykh, Oleg A. and Katsura, Hosho and Balents, Leon},
	comment = {We report a thorough theoretical study of the low temperature phase diagram of Cs<sub>2</sub>CuCl<sub>4</sub>, a spatially anisotropic spin S=1/2 triangular lattice antiferromagnet, in a magnetic field. Our results, obtained in a quasi-one-dimensional limit in which the system is regarded as a set of weakly coupled Heisenberg chains, are in excellent agreement with experiment. The analysis reveals some surprising physics. First, we find that, when the magnetic field is oriented within the triangular layer, spins are actually most strongly correlated within planes perpendicular to the triangular layers. This is despite the fact that the inter-layer exchange coupling in Cs<sub>2</sub>CuCl<sub>4</sub> is about an order of magnitude smaller than the weakest (diagonal) exchange in the triangular planes themselves. Second, the phase diagram in such orientations is exquisitely sensitive to tiny interactions, heretofore neglected, of order a few percent or less of the largest exchange couplings. These interactions, which we describe in detail, induce entirely new phases, and a novel commensurate-incommensurate transition, the signatures of which are identified in NMR experiments. We discuss the differences between the behavior of Cs<sub>2</sub>CuCl<sub>4</sub> and an ideal two-dimensional triangular model, and in particular the occurrence of magnetization plateaux in the latter. These and other related results are presented here along with a thorough exposition of the theoretical methods, and a discussion of broader experimental consequences to Cs<sub>2</sub>CuCl<sub>4</sub> and other materials.},
	journal = {Phys. Rev. B},
	month = {Jul},
	number = {1},
	numpages = {40},
	pages = {014421},
	publisher = {American Physical Society},
	title = {Extreme sensitivity of a frustrated quantum magnet: {Cs$_{2}$CuCl$_{4}$}},
	url = {http://dx.doi.org/10.1103/PhysRevB.82.014421},
	volume = {82},
	year = {2010}}

@article{bazhanov1996integrable,
  title={Integrable structure of conformal field theory, quantum {K}d{V} theory and thermodynamic {B}ethe ansatz},
  author={Bazhanov, Vladimir V and Lukyanov, Sergei L and Zamolodchikov, Alexander B},
  journal={Communications in Mathematical Physics},
  volume={177},
  number={2},
  pages={381--398},
  year={1996},
  publisher={Springer}
}

@article{lukyanov1998low,
  title={Low energy effective {H}amiltonian for the {XXZ} spin chain},
  author={Lukyanov, Sergei},
  journal={Nuclear Physics B},
  volume={522},
  number={3},
  pages={533--549},
  year={1998},
  publisher={Elsevier}
}

@article{pereira2006dynamical,
  title = {Dynamical Spin Structure Factor for the Anisotropic Spin-$1/2$ {H}eisenberg Chain},
  author = {Pereira, R. G. and Sirker, J. and Caux, J.-S. and Hagemans, R. and Maillet, J. M. and White, S. R. and Affleck, I.},
  journal = {Phys. Rev. Lett.},
  volume = {96},
  issue = {25},
  pages = {257202},
  numpages = {4},
  year = {2006},
  month = {Jun},
  publisher = {American Physical Society},
  doi = {10.1103/PhysRevLett.96.257202},
  url = {https://link.aps.org/doi/10.1103/PhysRevLett.96.257202}
}

@article{hikihara2004,
  title = {Correlation amplitudes for the spin-$\frac{1}{2}$ {$\mathrm{XXZ}$} chain in a magnetic field},
  author = {Hikihara, T. and Furusaki, A.},
  journal = {Phys. Rev. B},
  volume = 69,
  issue = 6,
  pages = 064427,
  numpages = 11,
  year = 2004,
  month = {Feb},
  publisher = {American Physical Society},
  doi = {10.1103/PhysRevB.69.064427},
  url = {https://link.aps.org/doi/10.1103/PhysRevB.69.064427}
}

@book{francesco2012conformal,
  title={Conformal field theory},
  author={Francesco, Philippe and Mathieu, Pierre and S{\'e}n{\'e}chal, David},
  year=2012,
  publisher={Springer Science \& Business Media}
}

@article{kunyang2004,
  title = {Ferromagnetic Transition in One-Dimensional Itinerant Electron Systems},
  author = {Yang, Kun},
  journal = {Phys. Rev. Lett.},
  volume = {93},
  issue = {6},
  pages = {066401},
  numpages = {4},
  year = {2004},
  month = {Aug},
  publisher = {American Physical Society},
  doi = {10.1103/PhysRevLett.93.066401},
  url = {https://link.aps.org/doi/10.1103/PhysRevLett.93.066401}
}

@misc{SM,
  note = {See Supplementary Material for details.},
}

@article{Haldane1981,
  title = {Effective Harmonic-Fluid Approach to Low-Energy Properties of One-Dimensional Quantum Fluids},
  author = {Haldane, F. D. M.},
  journal = {Phys. Rev. Lett.},
  volume = {47},
  issue = {25},
  pages = {1840--1843},
  numpages = {0},
  year = {1981},
  month = {Dec},
  publisher = {American Physical Society},
  doi = {10.1103/PhysRevLett.47.1840},
  url = {https://link.aps.org/doi/10.1103/PhysRevLett.47.1840}
}

@book{Takahashi1999,
  title = {Thermodynamics of One-Dimensional Solvable Models},
  author = {Takahashi, Minoru},
  publisher = {Cambridge University Press},
  year = {1999},
  doi = {10.1017/CBO9780511524332}
}

@article{AffleckHaldane1987,
  title = {Critical Theory of Quantum Spin Chains},
  author = {Affleck, Ian and Haldane, F. D. M.},
  journal = {Phys. Rev. B},
  volume = {36},
  issue = {10},
  pages = {5291--5300},
  year = {1987},
  month = {Oct},
  publisher = {American Physical Society},
  doi = {10.1103/PhysRevB.36.5291},
  url = {https://link.aps.org/doi/10.1103/PhysRevB.36.5291}
}

@article{Zotos1997,
  title = {Transport and Conservation Laws},
  author = {Zotos, X. and Naef, F. and Prelov{\v{s}}ek, P.},
  journal = {Phys. Rev. B},
  volume = {55},
  issue = {17},
  pages = {11029--11032},
  year = {1997},
  month = {May},
  publisher = {American Physical Society},
  doi = {10.1103/PhysRevB.55.11029},
  url = {https://link.aps.org/doi/10.1103/PhysRevB.55.11029}
}

@article{PhysRevLett.101.170403,
  title = {Spectral Functions of Strongly Interacting Isospin-$\frac{1}{2}$ Bosons in One Dimension},
  author = {Matveev, K. A. and Furusaki, A.},
  journal = {Phys. Rev. Lett.},
  volume = {101},
  issue = {17},
  pages = {170403},
  numpages = {4},
  year = {2008},
  month = {Oct},
  publisher = {American Physical Society},
  doi = {10.1103/PhysRevLett.101.170403},
  url = {https://link.aps.org/doi/10.1103/PhysRevLett.101.170403}
}

@article{PhysRevB.76.140408,
  title = {Spin-charge separation in a strongly correlated spin-polarized chain},
  author = {Akhanjee, Shimul and Tserkovnyak, Yaroslav},
  journal = {Phys. Rev. B},
  volume = {76},
  issue = {14},
  pages = {140408(R)},
  numpages = {4},
  year = {2007},
  month = {Oct},
  publisher = {American Physical Society},
  doi = {10.1103/PhysRevB.76.140408},
  url = {https://link.aps.org/doi/10.1103/PhysRevB.76.140408}
}

@article{RevModPhys.84.1253,
  title = {One-dimensional quantum liquids: Beyond the {L}uttinger liquid paradigm},
  author = {Imambekov, Adilet and Schmidt, Thomas L. and Glazman, Leonid I.},
  journal = {Rev. Mod. Phys.},
  volume = {84},
  issue = {3},
  pages = {1253--1306},
  numpages = {0},
  year = {2012},
  month = {Sep},
  publisher = {American Physical Society},
  doi = {10.1103/RevModPhys.84.1253},
  url = {https://link.aps.org/doi/10.1103/RevModPhys.84.1253}
}

@article{PhysRevLett.99.240404,
  title = {Spin Dynamics in a One-Dimensional Ferromagnetic {B}ose Gas},
  author = {Zvonarev, M. B. and Cheianov, V. V. and Giamarchi, T.},
  journal = {Phys. Rev. Lett.},
  volume = {99},
  issue = {24},
  pages = {240404},
  numpages = {4},
  year = {2007},
  month = {Dec},
  publisher = {American Physical Society},
  doi = {10.1103/PhysRevLett.99.240404},
  url = {https://link.aps.org/doi/10.1103/PhysRevLett.99.240404}
}

\newpage
\onecolumngrid
\begin{center}
\large\bfseries End Matter
\end{center}
\twocolumngrid

\noindent \emph{Semi-classical analysis and augmented bosonization} --- Here we sketch the semi-classical analysis which qualitatively describes the transition and the ordered phase, and summarize the  augmented bosonization theory which derives from it.  Full detail is presented in \SMsecref{sec:classical} \cite{SM}.  We express the semi-classical spin operators as space-time fluctuations around a ``cone'' state with wavevector $q_s$ and uniform magnetization $m$, chosen to minimize the energy:
\begin{equation}
  \label{eq:3}
  \begin{aligned}
      \bm{S}(x,\tau)
      &= \tfrac12\sqrt{1-4(m+\delta m(x,\tau))^2} \big(\cos\theta(x,\tau) \bm{e}_1\\
      &\quad + \sin\theta(x,\tau) \bm{e}_2\big)
      + (m+\delta m(x,\tau)) \bm{e}_3,
  \end{aligned}
\end{equation}
where $\tau$ is imaginary time, $\theta(x,\tau) = q_sx + \delta\theta(x,\tau)$ and $\delta m$ describes deviation of $S^z$ from the equilibrium $m$ found above.   In the usual bosonization language, $\delta m(x,\tau) = \frac{1}{\pi}\partial_x \phi(x,\tau)$
with the local frame
\begin{eqnarray}
    \bm{e}_1 &\approx & (1, 0, -u)\nonumber\\
    \bm{e}_2 & \approx & (0, 1,-v)\nonumber\\
    \bm{e}_3 & \approx & (u, v, 1).
\end{eqnarray}
where $u(x,\tau)$ and $v(x,\tau)$ describe smooth variations of the orthonormal spin basis in space-time.

The Hamiltonian density for the fluctuations is given by a quadratic expansion around the semiclassical solution.  For the phase variables,
\begin{align}
\label{eq:36}
  H_{\rm eff}(\phi,\delta\theta) & = \frac{h_{mm}}{2\pi^2} (\partial_x \phi)^2 
    + \frac{h_{m\theta}}{\pi} \, \partial_x\phi \, \partial_x \delta\theta + \frac12 h_{\theta\theta} (\partial_x \delta \theta)^2,
\end{align}
while for the basis fluctuations,
\begin{equation}
  \label{eq:37}
  H_{\rm eff}(u,v) = \rho m^2 \big((\partial_x u)^2 +(\partial_x v)^2\big).
\end{equation}
The constants $h_{mm}$,$h_{m\theta}$ and $h_{\theta\theta}$ and $\rho$ are positive functions of $g$ whose semi-classical values are given in \SMsecref{sec:classical} \cite{SM}.  We expect that the \emph{form} of Eqs.~\eqref{eq:36}~and~\eqref{eq:37} remains valid at low energies beyond semi-classics, but with renormalized values of these parameters.  The quantum dynamics of these fields is then determined with the aid of their commutation relations,
\begin{align}
  [\delta\theta(x), \delta m(x')] & = i \,\delta(x-x') , \nonumber \\
  [u(x),v(x')] & = \frac{i}{m} \delta(x-x').
                 \label{eq:38}
\end{align}

The phase Hamiltonian in Eq.~\eqref{eq:36} resembles the usual bosonization one but with a cross term reflecting the chirality of the system.  It leads to unequal right and left-moving  velocities $v_{R/L} = \sqrt{h_{mm} h_{\theta\theta}} \, \mp h_{m\theta}$.  These appear in correlations of general vertex operators, $V_{r,p} = \exp [ i (r\delta\theta + 2p\phi)]$, expressed in momentum/frequency space
\begin{equation}
\label{eq:39}
\begin{aligned}
    \langle V_{r,p}V^\dagger_{r,p}\rangle_{k,\omega}
    &\propto \Theta(\omega+v_L k)\Theta(\omega-v_R k)\\
    &\quad\times(\omega+v_L k)^{\eta_R(r,p)-1}
    (\omega-v_R k)^{\eta_L(r,p)-1}.
\end{aligned}
\end{equation}
Here the exponents $ \eta_{R/L}(r,p) = (r\sqrt{K} \mp \tfrac{p}{2\sqrt{K}})^2$ with $K = \frac{1}{4\pi}\sqrt{\frac{h_{mm}}{h_{\theta\theta}}}$.

The fluctuations of the spin basis, Eq.~\eqref{eq:37}, describe a complex non-relativistic boson $\psi = u+iv$ with
\begin{equation}
  \label{eq:40}
  \langle \psi \psi^\dagger\rangle_{k,\omega} = \frac{2}{m} \frac{1}{\omega - \frac{k^2}{2m^*} + i 0^+},
\end{equation}
with effective mass $m^*$ such that $1/(2m^*) = 2 \rho m$.

Eqs.~\eqref{eq:39}~and~\eqref{eq:40} can then be used to evaluate spin correlations using the formulae
\begin{equation}
  \label{eq:45}
  \begin{aligned}
    S^z &= m + \frac{1}{\pi} \partial_x \phi + C_1 \cos\big(q_m x - 2\phi + \zeta\big)\\
    &\quad -\, {\rm Re}\Big\{D_0(u - i v)\, e^{i(q_s x + \delta\theta)}
    \\
    &\quad +\, \sum_{s=\pm} D_s (u - i v)\, e^{i[(q_s + s q_m)x + \delta\theta - 2s\phi]}\Big\} + \ldots ,\\
    S^\pm &= C_0 \, e^{\pm i (q_s x + \delta\theta)} + C_+ \, e^{\pm i [(q_s + q_m) x + \delta\theta - 2\phi]}\\
    &\quad +\, C_- \, e^{\pm i [(q_s - q_m) x + \delta\theta + 2\phi]}  + m \, (u \pm i v)\\
    &\quad + \, C_2 m (u \pm i v) \cos\big(q_m x - 2\phi + \zeta'\big) + \ldots
  \end{aligned}
\end{equation}
where $C_a,D_a,\zeta,\zeta'$ are constants.  These can be viewed as the decomposition of the lattice spin operators in terms of low energy fields, and have been augmented from the purely semi-classical forms as allowed by symmetry and discreteness of spin following Haldane\cite{Haldane1981}.  Here $q_m = \pi - 2\pi m$ is fully determined by the magnetization.  From the TBA in \SMsecref{sec:tba-magnetization-magnon} \cite{SM}, we obtain both $q_s$ and $q_m$ exactly for any $g$.  

\begin{figure}[htbp]
  \centering
  \begin{minipage}{0.48\columnwidth}
    (a)\par
    \centering
    \includegraphics[width=\columnwidth]{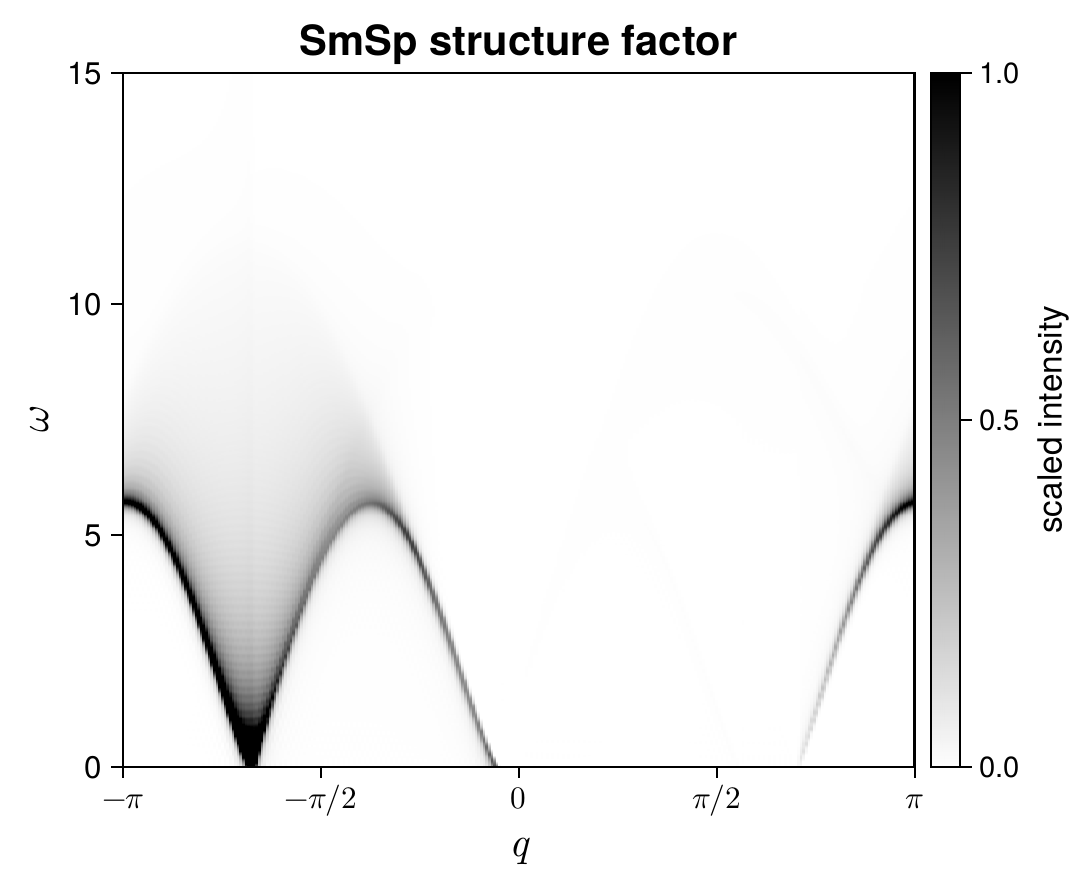}
  \end{minipage}
  \hfill
  \begin{minipage}{0.48\columnwidth}
    (b)\par
    \centering
    \includegraphics[width=\columnwidth]{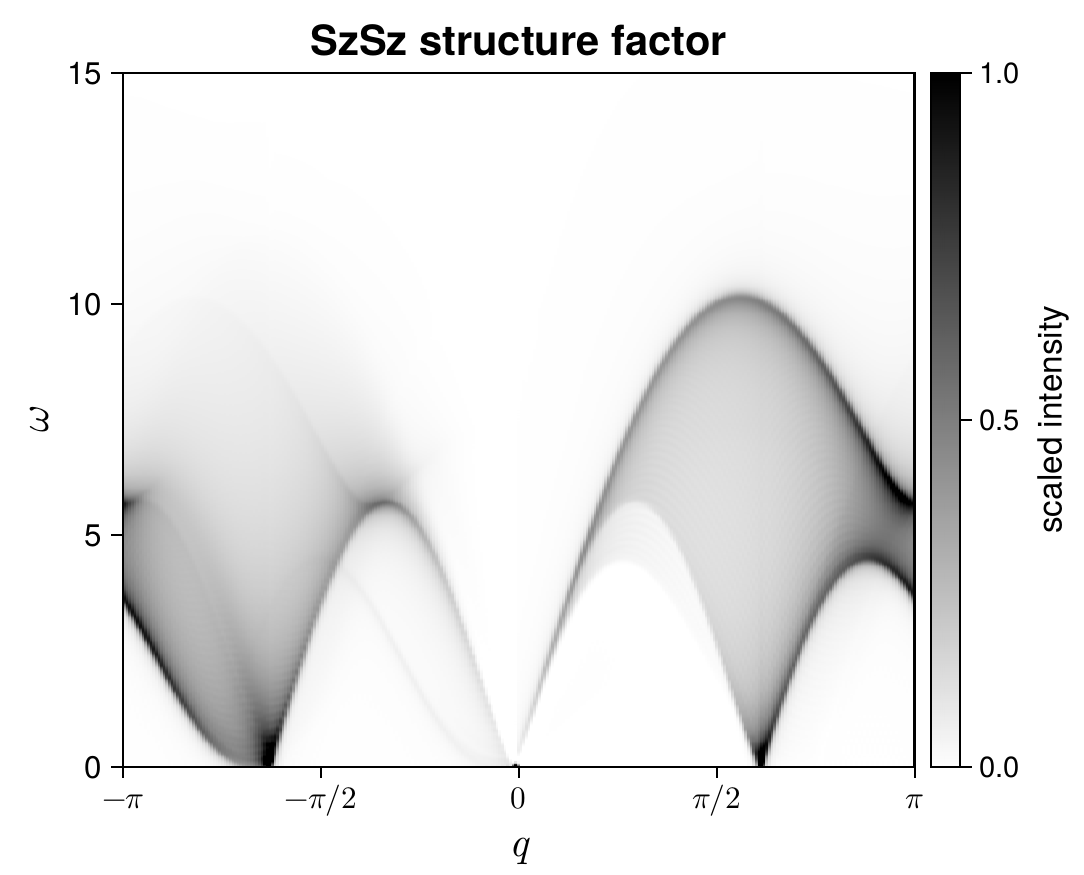}
  \end{minipage}
  \caption{Dynamical correlation functions in the ordered phase with
    $g=4$, measured in the ground state with maximum magnetization along
    $z$ ($S^z_{\rm tot}=57$) calculated using DMRG+TDVP for a system of
    $N=300$ sites: (a) $S^{-+}(q,\omega)$ and (b) $S^{zz}(q,\omega)$.
    The gray scale indicates the scaled intensity
    $I(q,\omega)=\min[\max(S^{AB}_\eta(q,\omega),0)/S_{\rm clip},1]$,
    with $AB=-+$ in panel (a) and $AB=zz$ in panel (b).  Here
    $S_{\rm clip}$ is 99.5\% of the maximum intensity of $S^{-+}$ in
    panel (a) and 99.9\% of the maximum intensity of $S^{zz}$ in panel
    (b); values above $S_{\rm clip}$ are saturated.}
  \label{fig:g4-other-components}
\end{figure}

\noindent \emph{Dynamical correlations in the ordered phase} ---
Dynamical spin correlations were calculated using DMRG+TDVP as described in \SMsecref{sec:comp-deta-furth} \cite{SM}, for various values of $g$ and system sizes.  The spectral function is reported in the main text for $g=2/\pi$~Fig.~\ref{fig:dynsf} (which has a single spectral function because SU(2) symmetry is preserved) and for $S^{+-}$ for $g=4.0$~Fig.~\ref{fig:SpSm_g4}.  Here we report the other components for $g=4$ and results for $g=1$, and discuss their features with respect to the augmented bosonization theory.   Fig.~\ref{fig:g4-other-components} shows the $S^{-+}$ and $S^{zz}$ components of the spectral function for $g=4$.  

\begin{figure*}[tbp]
  \centering
  \begin{minipage}{0.31\textwidth}
    (a)\par
    \centering
    \includegraphics[width=\linewidth]{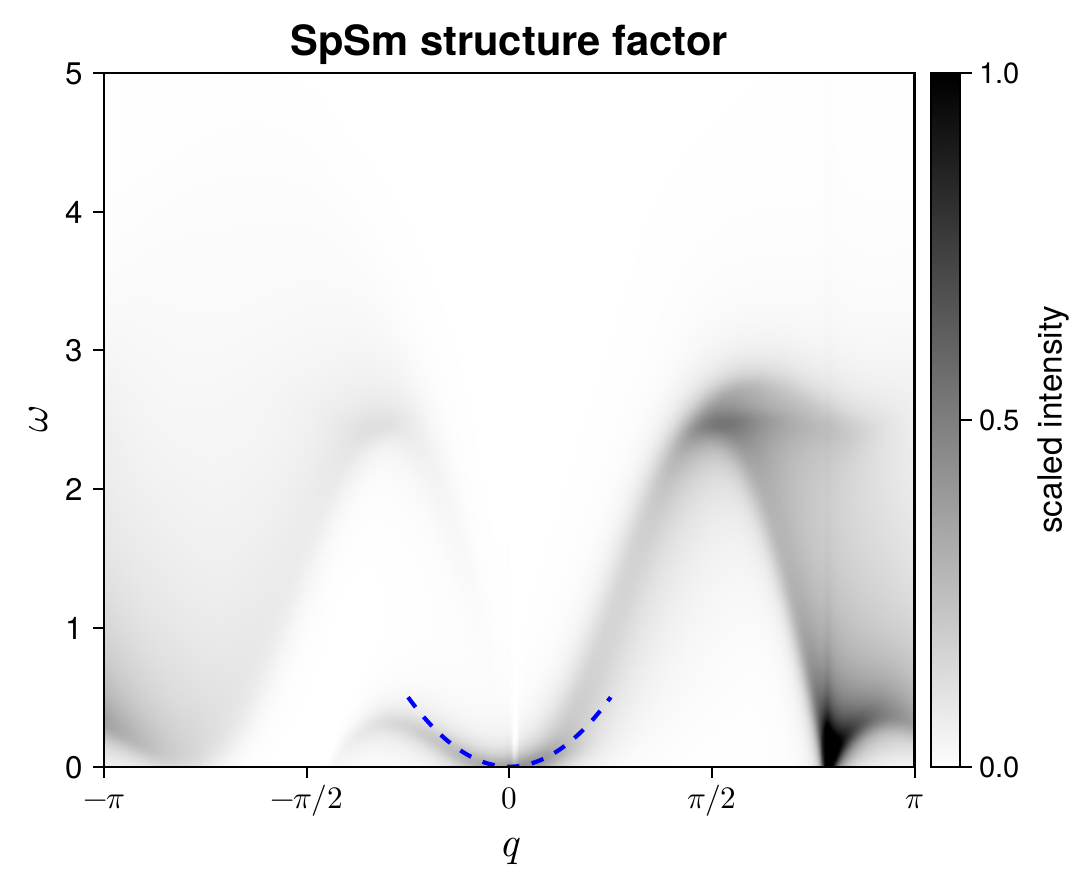}
  \end{minipage}
  \hfill
  \begin{minipage}{0.31\textwidth}
    (b)\par
    \centering
    \includegraphics[width=\linewidth]{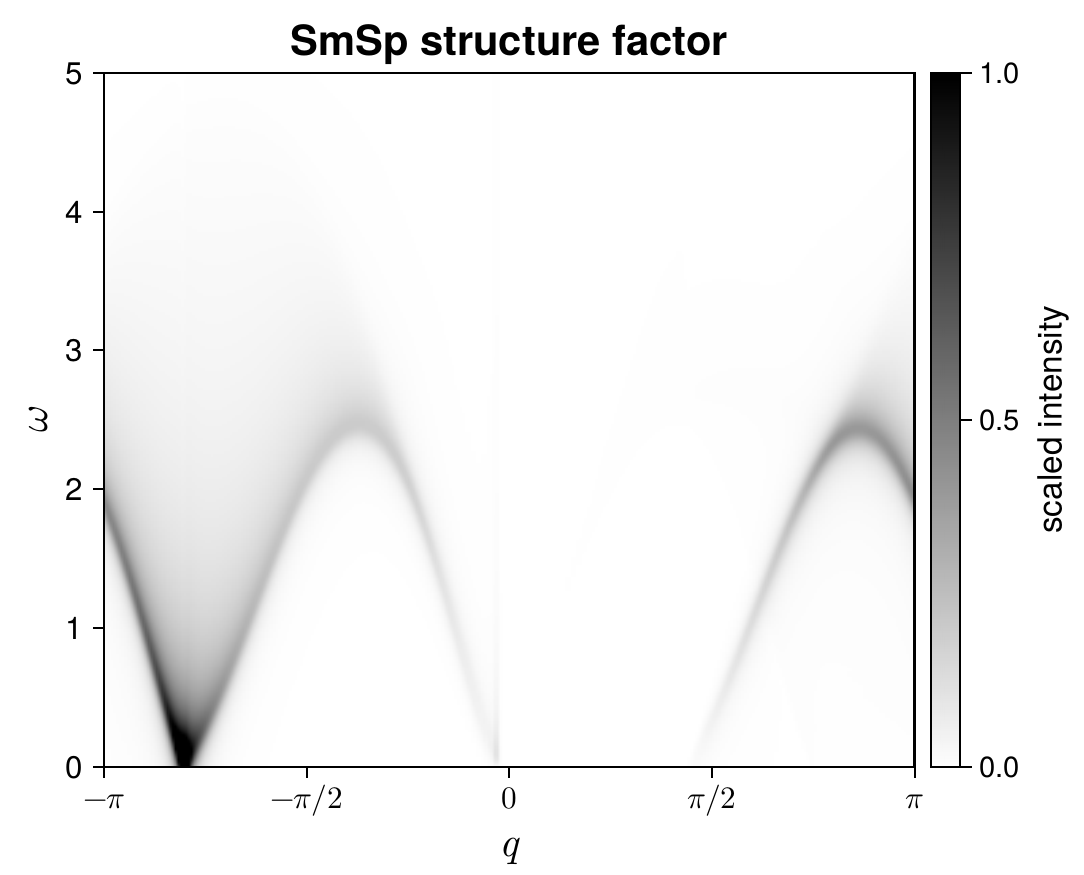}
  \end{minipage}
  \hfill
  \begin{minipage}{0.31\textwidth}
    (c)\par
    \centering
    \includegraphics[width=\linewidth]{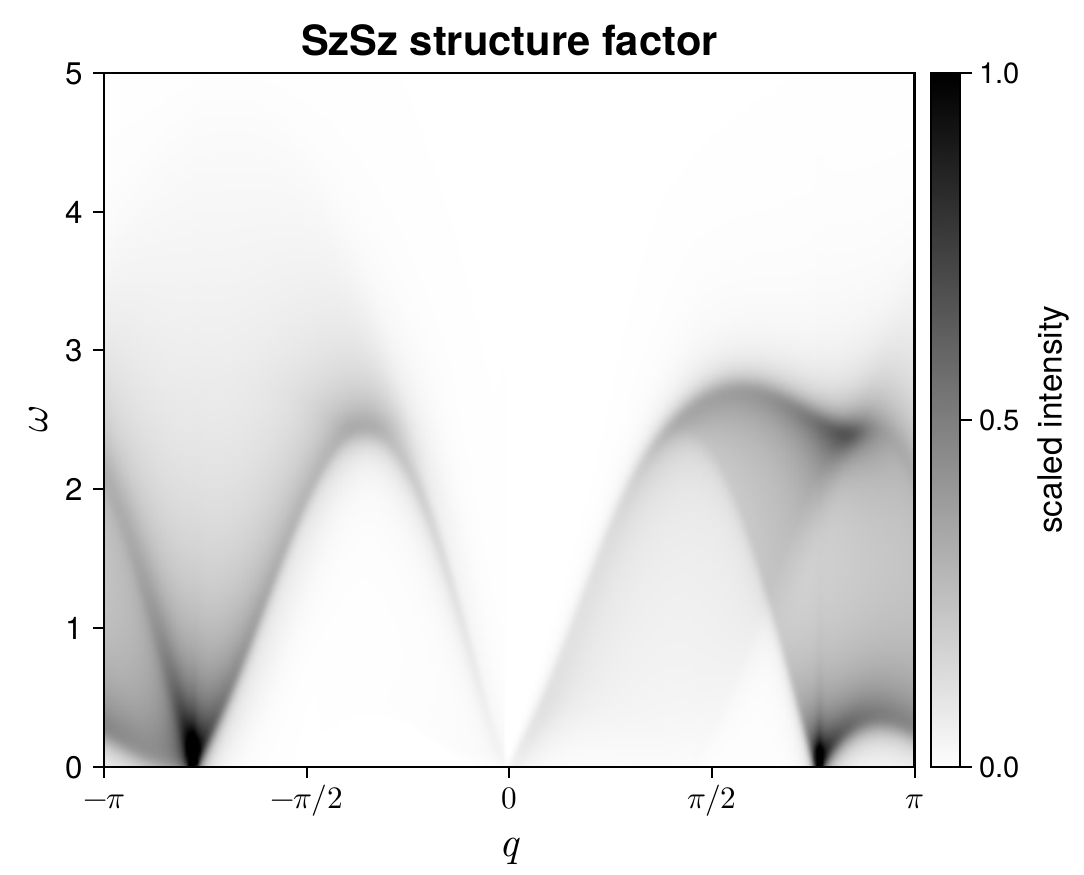}
  \end{minipage}
  \caption{Dynamical correlation functions in the ordered phase with
    $g=1$, measured in the ground state with maximum magnetization along
    $z$ ($S^z_{\rm tot}=34$) calculated using DMRG+TDVP for a system of
    $N=300$ sites: (a) $S^{+-}(q,\omega)$, (b) $S^{-+}(q,\omega)$ and (c) $S^{zz}(q,\omega)$.  In panel (a), the parabola shows the quadratic magnon with mass calculated from the TBA.  The gray scale indicates the scaled intensity
    $I(q,\omega)=\min[\max(S^{AB}_\eta(q,\omega),0)/S_{\rm clip},1]$,
    with $AB=+-$ in panel (a), $AB=-+$ in panel (b), and $AB=zz$ in panel (c).  Here
    $S_{\rm clip}$ is 99.9\% of the maximum intensity; values above $S_{\rm clip}$ are saturated.}
  \label{fig:g1-sf}
\end{figure*}

Fig.~\ref{fig:g1-sf} shows the spin structure factor at $g=1$. Several interesting features stand out in these plots. For concreteness, we focus on the low-energy structure of $S^{ab}$ in the ordered phase with $g=1$. 

We start with $S^{-+}$ by noting that the $\langle S^-(x,t) S^+(0,0)\rangle$ correlations involve only $C_0, C_\pm$ terms in Eq.~\eqref{eq:45}, because the magnon does not contribute  in this sector, $\langle \psi^\dagger(x,t) \psi(0,0)\rangle = 0$. Correspondingly, the main feature in Fig.~\ref{fig:g1-sf}b is located at momentum $-q_s$, the magnitude of which matches the TBA-derived result in Fig.~\ref{SM-fig:wavevectors} very well. Next, there is a left-moving branch that originates from momentum $-(q_s - q_m)$ just to the left of $q=0$. It corresponds to the $C_{-}$ term in \eqref{eq:45}. The correlations of the vertex operator $e^{i(\delta \theta + 2\phi)}$ are characterized by the exponents $\eta_R(1,1) < \eta_L(1,1)$. Correspondingly, Eq.~\eqref{eq:39} predicts a {\em left-moving} branch, in agreement with our data in Fig.~\ref{fig:g1-sf}b. Finally, the right-moving branch in the same figure originates from momentum $-(q_s + q_m)$: its magnitude exceeds $\pi$ and therefore it ``wraps" around the Brillouin zone to appear at positive $q$. Here, the spin excitations move right; they are produced by the vertex operator $e^{i(\delta \theta - 2\phi)}$, and, correspondingly, $\eta_R(1,-1) > \eta_L(1,-1)$ in \eqref{eq:39}. By fitting the low-energy data to straight lines, we find velocities of chiral spin excitations to be $v_L \approx 3, v_R \approx 2.3$. Observe that $S^{-+}$ at $g=4$, Fig.~\ref{fig:g4-other-components}a, has identical structure but with different values of $q_s$ and $v_{L/R}$. 

Turning to $S^{+-}$, we observe the appearance of the quadratically dispersing magnon mode Eq.~\eqref{eq:40} in Fig.~\ref{fig:g1-sf}a, as discussed in some detail in relation to Fig.~\ref{fig:SpSm_g4} in the main text. As noted there, the effective mass of the magnon, calculated by TBA, fits the numerical data well: $\omega = 4.35 q^2$ for $g=4$ in Fig.~\ref{fig:SpSm_g4}, and $\omega = 0.81 q^2$ for $g=1$ in Fig.~\ref{fig:g1-sf}a. The other prominent feature, at momentum $q_s$, is produced by the $C_{0}$ term in Eq.~\eqref{eq:45}.  A weaker feature at negative $q \approx -q_m$ may be attributed to a ``shadow'' of the magnon appearing due to the $C_2$ term.  Owing to the appearance of both $u-iv$ and the $V_{0,2}$ vertex operator in this term, the associated feature should be a convolution of two spectral functions, which explains the diffuse nature of this weight. 

The longitudinal structure factor $S^{zz}$, Fig.~\ref{fig:g1-sf}c, features chiral acoustic modes, centered around $q=0$ and produced by $\partial_x \phi$ in \eqref{eq:45}, as well as two intense continua at $\pm q_m$ produced by the $C_1$ term. Also present is a  ``half-parabola", centered at $-q_s$, which we associate with the composite operator 
$D_0 \psi^\dagger e^{i(q_s x + \delta\theta)} + {\rm h.c.}$ in $S^z$. It is most noticeable in $g=4$ data in Fig.~\ref{fig:g4-other-components}b. Note that it is not present at positive wave vector $q=+q_s$ because the corresponding magnon contribution vanishes, $\langle \psi^\dagger(x,t) \psi(0,0)\rangle = 0$. It appears that the ``half-parabola" asymmetry of the spectral weight has to do with the strong hybridization, on the right side of momentum $-q_s$, between the $D_0$ (centered around $-q_s$) and $C_1$ (originating from the wave vector $-q_m$, which is close to but different from $-q_s$, see Fig.~\ref{SM-fig:wavevectors}) contributions. Such hybridization is not possible to the left of $-q_s$, where the two corresponding contributions are well separated in energy and are seen as distinct quadratic and linear dispersing branches in the $S^{zz}(q,\omega)$ continuum.   A priori, at sufficiently low energy, the weight associated with the $D_0$ feature may be obtained by a convolution of the spectrum of the vertex operator, $V_{1,0}$, given in Eq.~\eqref{eq:39} and that of the magnon in Eq.~\eqref{eq:40}.

\clearpage
\onecolumngrid
\begin{center}
{\large\bfseries Supplemental Material for: Ferromagnetic transition in a chiral spin chain}
\end{center}
\tableofcontents
\twocolumngrid

\setcounter{section}{0}
\setcounter{subsection}{0}
\setcounter{subsubsection}{0}
\setcounter{equation}{0}
\setcounter{figure}{0}
\setcounter{secnumdepth}{3}
\renewcommand{\thesection}{\Alph{section}}
\renewcommand{\thesubsection}{\thesection.\arabic{subsection}}
\renewcommand{\thesubsubsection}{\thesubsection.\arabic{subsubsection}}
\renewcommand{\theequation}{S\arabic{equation}}
\renewcommand{\thefigure}{S\arabic{figure}}
\renewcommand{\theHsection}{SM.\arabic{section}}
\renewcommand{\theHsubsection}{SM.\arabic{section}.\arabic{subsection}}
\renewcommand{\theHsubsubsection}{SM.\arabic{section}.\arabic{subsection}.\arabic{subsubsection}}
\renewcommand{\theHequation}{SM.\arabic{equation}}
\renewcommand{\theHfigure}{SM.\arabic{figure}}
\makeatletter
\renewcommand{\p@subsection}{}
\renewcommand{\p@subsubsection}{}
\makeatother
\section{Bosonization}
\label{SM-sec:bosonization}

For concreteness, we follow the bosonization conventions of Ref.\cite{extreme2012}, taking the isotropic zero-field case, for which $2\pi R^2=1$ and $\beta=\sqrt{2\pi}$.  Then the Hamiltonian density is
\begin{align}
  \label{SM-eq:5}
  \mathcal{H} & = \frac{v}{2} \left[ (\partial_x \phi)^2 + (\partial_x \theta)^2\right].
\end{align}
with the commutation relations
\begin{align}
  \label{SM-eq:6}
  \left[\theta(x),\phi(x')\right] & = -i \Theta(x-x').
\end{align}
We can define
\begin{align}
  \label{SM-eq:7}
  \phi_{L/R} & = \frac12 (\phi \pm \theta),
\end{align}
so that
\begin{align}
  \label{SM-eq:8}
    \mathcal{H} & = v \left[ (\partial_x \phi_R)^2 + (\partial_x \phi_L)^2\right].
\end{align}
The commutation relations are then
\begin{align}
  \label{SM-eq:9}
  \left[\phi_L(x),\phi_L(x')\right] =  -\left[\phi_R(x),\phi_R(x')\right] = -\frac{i}{4} \textrm{sign}(x-x').
\end{align}
Inverting Eq.~\eqref{SM-eq:7} gives
\begin{align}
  \label{SM-eq:10}
  \phi = \phi_L + \phi_R, && \theta = \phi_L-\phi_R.
\end{align}
Now we see that the z-component of the uniform magnetization (Eq.~(9) of that paper) is
\begin{align}
  \label{SM-eq:11}
 M^z & = \frac{1}{\sqrt{2\pi}} \partial_x \phi = \frac{1}{\sqrt{2\pi}} \partial_x \phi_L + \frac{1}{\sqrt{2\pi}} \partial_x \phi_R .
\end{align}
Eq.~(11) of that paper gives
\begin{align}
  \label{SM-eq:12}
  M^+ & = \frac{i}{2}  A_2  \left( e^{i\sqrt{2\pi} (\theta+\phi)} - e^{i\sqrt{2\pi}(\theta-\phi)}\right).
\end{align}
Here $A_2$ is a cut-off dependent constant.  
Note that both terms here have the same exponential factor of $\theta$, which is required so that they both create solitons in $\phi$ with appropriate strength.  Let us make a small shift $\theta\rightarrow \theta - \sqrt{\pi/8}$ to make the formulas look nicer.  Then we get
\begin{align}
  \label{SM-eq:13}
  M^+ & \rightarrow \frac{1}{2}  A_2  \left( e^{i\sqrt{2\pi} (\theta+\phi)} + e^{i\sqrt{2\pi}(\theta-\phi)}\right) \nonumber \\
  & = \frac{A_2}{2} \left( e^{i2\sqrt{2\pi} \phi_L} + e^{-i2\sqrt{2\pi} \phi_R}\right).
\end{align}
In this way, we can write the vector magnetization
\begin{align}
  \label{SM-eq:14}
  \bm{M} & = \bm{J}_L + \bm{J}_R,
\end{align}
with
\begin{align}
  \label{SM-eq:15}
  \bm{J}_L & = \begin{pmatrix} A_2 \cos 2\sqrt{2\pi}\phi_L \\
                 A_2 \sin 2\sqrt{2\pi}\phi_L \\
                 \frac{\partial_x\phi_L}{\sqrt{2\pi}}
                \end{pmatrix}, &&   \bm{J}_R & = \begin{pmatrix} A_2 \cos 2\sqrt{2\pi}\phi_R \\
                 -A_2 \sin 2\sqrt{2\pi}\phi_R \\
                 \frac{\partial_x\phi_R}{\sqrt{2\pi}}.
                \end{pmatrix}.
\end{align}
Now one can consider the commutation relations of the currents.  This is easiest for the z components and the rest should follow from SU(2) symmetry.  We have
\begin{align}
  \label{SM-eq:16}
  \left[J_L^z(x),J_L^z(x')\right] & = \frac{1}{2\pi} \left[\partial_x \phi_L(x),\partial_{x'} \phi_L(x')\right] \nonumber \\
  & = -\frac{i}{8\pi} \partial_x \partial_{x'} \textrm{sign}(x-x') \nonumber \\
  & = \frac{i}{4\pi} \delta'(x-x').
\end{align}
The right-moving currents have the opposite sign commutation.  Now we can form the Fourier modes
\begin{align}
  \label{SM-eq:17}
  J_{L,n}^z = \int_0^L \! dx\, e^{-i 2\pi n x/L} J_L^z(x).
\end{align}
Integrate Eq.~\eqref{SM-eq:16} to get
\begin{align}
  \label{SM-eq:18}
  \left[J_{L,m}^z,J_{L,n}^z\right] & = \frac{i}{4\pi}\int_0^L \! dx dx' \, e^{-i 2\pi (m x+n x')/L }\delta'(x-x') \nonumber \\
  & = -\frac{m}{2L} \int_0^L \! dx  \, e^{-i 2\pi (m +n) x/L } = -\frac{m}{2} \delta_{m+n,0}.
\end{align}
If we define the same for the right-movers,
\begin{align}
  \label{SM-eq:19}
    J_{R,n}^z = \int_0^L \! dx\, e^{-i 2\pi n x/L} J_R^z(x),
\end{align}
then
\begin{align}
  \label{SM-eq:20}
   \left[J_{R,m}^z,J_{R,n}^z\right] & = \frac{m}{2} \delta_{m+n,0}.
\end{align}
This is the correct algebra for the Kac-Moody SU(2)$_1$ case, following at least one convention.  By SU(2) symmetry, we deduce then that 
\begin{align}
  \label{SM-eq:21}
  \left[J_{R,m}^a,J_{R,n}^b\right] = i \epsilon^{abc} J_{R,m+n}^c + \frac12 m \,\delta_{m+n,0} \, \delta^{ab}.
\end{align}
We can also rewrite the Hamiltonian.  We have
\begin{align}
  \label{SM-eq:22}
  H_R & = 2\pi v \int\! dx\, J_R^z J_R^z \nonumber \\
                & = \frac{2\pi v}{3} \int\! dx\, J_R^a J_R^a \nonumber \\
  & = \sum_{n=-\infty}^\infty \frac{2\pi v}{3 L} J_{R,-n}^a J_{R,n}^a,
\end{align}
which is (if we normal order it) the Sugarawa form.  Here, in going from the first to the second line, we used SU(2) symmetry.  From the  
\begin{align}
  \label{SM-eq:23}
  L_n = \frac13 \sum_m : J_m^a J_{n-m}^a:,
\end{align}
we have
\begin{align}
  \label{SM-eq:24}
  H_R & = \frac{2\pi v}{L} L_0.
\end{align}
Here, and in the main text, we omit the subscript $R$ in the right-moving operators $L_{n} \equiv L_{R, n}$ and $J^a_n \equiv J^a_{R, n}$ for brevity. 

\section{RG flow of the backscattering interaction}
\label{SM-sec:rg-bs}
This is a brief comment on the renormalization group flow of the backscattering interaction between the right and left currents. The spin Hamiltonian of the Luttinger phase ($g < g_c = 2/\pi$) is given by Eq.~\eqref{eq:25}. The right-left separation is violated by the backscattering term $H_{\rm bs} = - g_{\rm bs}\int dx  J^a_R J^a_L$. In the symmetric spin chain with $v_R = v_L$, this interaction is marginally irrelevant; it scales to zero with increasing system size (equivalently, with decreasing energy). It is important to ask if this behavior changes in the system with $v_R \neq v_L$.

The described RG flow result follows from the analysis of the corrections to the coupling constant $g_{\rm bs}$. The leading correction appears in second-order perturbation theory, when one expands the partition function in powers of $g_{\rm bs}$. This correction reads
\begin{equation}
\label{SM-eq:b1}
    g_{\rm bs}^2 \int dx d\tau \int dx' d\tau' J^a_R(x',\tau') J^a_L(x',\tau') J^b_R(x,\tau) J^b_L(x,\tau) 
\end{equation}
The leading, logarithmically divergent contribution arises from the nearby space-time points $(x,\tau) \approx (x',\tau')$, which allows one to evaluate it by fusing the pairs of right- and left-currents together according to the operator expansion rule
\begin{eqnarray}
    &&J^a_{R/L}(x',\tau') J^b_{R/L}(x,\tau) = \frac{\delta^{ab}}{8\pi^2 (v_{R/L} (\tau' - \tau) \mp i (x'-x))^2} \nonumber\\
    &&+ \frac{i \epsilon^{abc} J^c_{R/L}(x,\tau)}{2\pi (v_{R/L} (\tau' - \tau) \mp i (x'-x))}
\end{eqnarray}
Eq.\eqref{SM-eq:b1} becomes 
\begin{equation}
\label{SM-eq:b3}
\begin{aligned}
    &-\frac{g_{\rm bs}^2}{(2\pi)^2}\epsilon^{abc} \epsilon^{abd} \int dX dT J_R^c(X,T) J_L^d(X,T)\\
    &\quad\times \int d\Delta x d\Delta \tau \frac{1}{(v_R \Delta \tau - i \Delta x)(v_L \Delta \tau + i \Delta x)}
\end{aligned}
\end{equation}
where we introduced the center-of-mass $(X,T)$ and relative $(\Delta x, \Delta \tau)$ coordinates. In the standard case of equal velocities, $v_R = v_L = v_0$, the integral over the relative coordinates becomes, upon scaling the common velocity $v_0$ out, $\int d^2 \vec{r}/r^2 = 2\pi/v_0 \int_{r(\ell)}^{r(\ell+d\ell)} dr/r = 2\pi d\ell/v_0$. Here $r(\ell) = r_0 e^\ell$ describes the running cutoff scale that one progressively integrates out by small $d\ell$ steps.

This process is not possible in \eqref{SM-eq:b3}, and instead we choose the RG ``slice" to be rectangular by integrating over the full range of $\Delta x$ and then scaling in $\Delta\tau(\ell) = \Delta\tau_0 \, e^\ell$. The integral \eqref{SM-eq:b3} becomes
\begin{equation}
    \frac{2\pi}{v_R + v_L} \int_{\Delta\tau(\ell)}^{\Delta\tau(\ell+d\ell)} \frac{d\Delta\tau}{|\Delta\tau|} = \frac{2\pi d\ell}{v_R + v_L}
\end{equation}
As a result, the RG equation for the backscattering coupling retains its usual form
\begin{equation}
    \frac{dg_{\rm bs}}{d\ell} = - \frac{g_{\rm bs}^2}{\pi (v_R + v_L)} = -\frac{g_{\rm bs}^2}{2\pi v_0}
\end{equation}
where the velocity $v_0$ is replaced by the average one, $(v_R + v_L)/2$. Since $v_{R/L} = v_0 \mp g v_0^2$, the resulting RG equation is identical to that of the symmetric spin chain with $v_0 = v_R = v_L$.
Therefore, $g_{\rm bs}$ retains its marginal irrelevance property even when $v_R \neq v_L$.

\section{Derivation of the $I_3$ formula in the main text}
\label{SM-sec:deriv-i_3-form}

In this section, we derive Eq.~\eqref{eq:33} which expresses the chiral Hamiltonian at the critical point, $H_3 = H_R(v_R=0)$, which is proportional to Eq.~\eqref{eq:28}, in terms of the Virasoro generators.  The result is ``almost obvious'': if normal ordering is ignored, it is just the usual Fourier decomposition of the integral of a square, given that the Virasoro generators are the Fourier components of $T$.  Here we carefully track the normal ordering.

We begin from  Eq.~\eqref{eq:28}, and use the abelian representation, $J_R(x) = J_R^z(x) = \partial_x\phi_R/\sqrt{2\pi}$ to obtain
\begin{equation}
  \label{SM-eq:73}
  H_3 = \lambda \int_0^L \! dx\, \left[ : J_R^4: + \frac{1}{(2\pi)^2} : (\partial_x J_R )^2:\right].
\end{equation}
Then we invert Eq.~\eqref{SM-eq:19} to obtain
\begin{align}
  \label{SM-eq:41}
  J_R^z(x) & = \frac{1}{L}\sum_{n=-\infty}^\infty e^{i2\pi n x/L} J_n .
\end{align}
Then we obtain
\begin{align}
  \label{SM-eq:44}
  H_3 & = \frac{\lambda}{L^3} I_3,
\end{align}
where
\begin{equation}
  \label{SM-eq:74}
  I_3 = A_3 + \frac{1}{(2\pi)^2} B_3,
\end{equation}
with
\begin{align}
  \label{SM-eq:42}
  A_3  = &   \sum_{n_1\cdots n_4 \in \mathbb{Z}} : J_{n_1} J_{n_2} J_{n_3} J_{n_4}: \delta_{n_1+n_2+n_3+n_4,0}  \nonumber \\
  B_3 = &  (2\pi)^2 \sum_n n^2 :J_{-n} J_n:.
\end{align}
The second term is easily rewritten
\begin{align}
  \label{SM-eq:43}
   B_3 = &  (2\pi)^22 \sum_{n=1}^\infty  n^2 J_{-n} J_n.
\end{align}

To obtain Eq.~\eqref{eq:33} we need to express $I_3$ in terms of the Virasoro generators.  To do so, we use the abelian representation of the Virasoro generators:
\begin{align}
  \label{SM-eq:53}
  L_n & = \sum_p : J_{-p} J_{p+n}:
\end{align}
Since the $J_n$ commute unless their indices are equal and opposite, we can mostly drop the normal ordering:
\begin{align}
  \label{SM-eq:54}
  L_n & = \left\{ \begin{array}{cc} \sum_p J_{-p} J_{p+n} & n \neq 0 \\
                    J_0^2+2 \sum_{p=1}^\infty J_{-p} J_p  & n=0 \end{array}\right. .
\end{align}
It will be helpful to separate the terms in these sums to make explicit which $J_p$ operators are creation operators ($p<0$) and which are annihilation operators ($p>0$).  To do this we will write explicit formulas in which we add a dagger to those Kac-Moody generators which are creation operators.  That is, we can choose to write
\begin{align}
  \label{SM-eq:55}
  J_p & =  \left\{ \begin{array}{cc} J_p & p \geq 0 \\
                    J_{-p}^\dagger  & p<0 \end{array}\right. .
\end{align}
In this way we will only have non-negative indices.

In the following we always use the notation $n>0$.  Then we have
\begin{align}
  \label{SM-eq:56}
  L_n & = \sum_{p=-\infty}^{-n-1} J_{p+n} J_{-p} + \sum_{p=-n+1}^{-1} J_{-p} J_{p+n} + \sum_{p=1}^\infty J_{-p} J_{p+n} \nonumber \\
  & + 2 J_n J_0.
\end{align}
The final term is the contribution from $p=0$ and $p=-n$.  Now let us use Eq.~\eqref{SM-eq:55} to eliminate negative indices:
\begin{align}
  \label{SM-eq:57}
  L_n & = \sum_{p=-\infty}^{-n-1} J^\dagger_{-p-n} J_{-p} + \sum_{p=-n+1}^{-1} J_{-p} J_{p+n} + \sum_{p=1}^\infty J^\dagger_{p} J_{p+n} \nonumber \\
  & + 2 J_n J_0.
\end{align}
In the first term in Eq.~\eqref{SM-eq:57}, we can relabel $p\rightarrow -n -p$, after which it becomes equal to the third term.  In the second term we can relabel $p\rightarrow -p$.  We then get
\begin{align}
  \label{SM-eq:58}
  L_n & = 2\sum_{p=1}^\infty J_p^\dagger J_{p+n}^{\vphantom\dagger} + \sum_{p=1}^{n-1} J_p J_{n-p} + 2 J_n J_0.
\end{align}
Similar manipulations give (still $n>0$):
\begin{align}
  \label{SM-eq:59}
  L_{-n} & = 2 \sum_{q=1}^\infty J_{q+n}^\dagger J_q^{\vphantom\dagger} + \sum_{q=1}^{n-1} J_q^\dagger J_{n-q}^\dagger + 2 J_n^\dagger J_0^{\vphantom\dagger}.           
\end{align}

Now we are in a position to consider the $A_3$ term from Eq.~\eqref{SM-eq:42}.  It is equal to
\begin{align}
  \label{SM-eq:60}
  A_3 & = \sum_{n=-\infty}^\infty : L_{-n} L_n: \nonumber \\
  & = :L_0^2: + 2 \sum_{n=1}^\infty :L_{-n} L_n:.
\end{align}
This is true because the sum over $n$ plus internal sums over $p$ in each of the $L_{\pm n}$ factors, using \eqref{SM-eq:53}, covers all sets of indices summing to zero uniquely.

Consider the second term in \eqref{SM-eq:60}.  By examining Eqs.~\eqref{SM-eq:58}~and~\eqref{SM-eq:59}, we see that the only terms in the product $L_{-n} L_n$ which are not already normal-ordered are in the product of the first factors.  Hence we have
\begin{align}
  \label{SM-eq:61}
  :L_{-n} L_n: & = L_{-n} L_n \\
               & + 4\sum_{p,q=1}^\infty \left( : J_{q+n}^\dagger J_q^{\vphantom\dagger}  J_p^\dagger J_{p+n}^{\vphantom\dagger} : - J_{q+n}^\dagger J_q^{\vphantom\dagger}  J_p^\dagger J_{p+n}^{\vphantom\dagger}\right) \nonumber \\
               & = L_{-n} L_n - 4 \sum_{p,q=1}^\infty J_{q+n}^\dagger \left[J_q^{\vphantom\dagger} , J_p^\dagger\right] J_{p+n}^{\vphantom\dagger} \nonumber \\
  & = L_{-n} L_n - 2 \sum_{p=1}^\infty p J_{p+n}^\dagger J_{p+n}^{\vphantom\dagger},
\end{align}
where we used Eq.~\eqref{SM-eq:20}.  Now let us sum over $n$:
\begin{align}
  \label{SM-eq:62}
  \sum_{n=1}^\infty : L_{-n} L_n: & =   \sum_{n=1}^\infty  L_{-n} L_n - 2\sum_{n,p=1}^\infty p J_{p+n}^\dagger J_{p+n}^{\vphantom\dagger}. 
\end{align}
We can define a new index $m=p+n$.  $m$ ranges from $2$ to infinity.  For a given value of $m$, $p$ can range from $1$ to $m-1$.  Hence
\begin{align}
  \label{SM-eq:63}
   \sum_{n=1}^\infty : L_{-n} L_n: & =   \sum_{n=1}^\infty  L_{-n} L_n - 2\sum_{m=2}^\infty \sum_{p=1}^{m-1} p J_{m}^\dagger J_{m}^{\vphantom\dagger}.
\end{align}
The discrete sum $\sum_{p=1}^{m-1} p = m(m-1)/2$.  Hence
\begin{align}
  \label{SM-eq:64}
    \sum_{n=1}^\infty : L_{-n} L_n: & =   \sum_{n=1}^\infty  L_{-n} L_n - \sum_{m=2}^\infty (m^2-m) J_{m}^\dagger J_{m}^{\vphantom\dagger}.
\end{align}
Note that we can extend this sum down to $m=0$ because the $m^2-m$ factor vanishes for $m=0,1$.  Hence
\begin{align}
  \label{SM-eq:65}
  \sum_{n=1}^\infty : L_{-n} L_n: & =   \sum_{n=1}^\infty  L_{-n} L_n - \sum_{m=0}^\infty (m^2-m) J_{m}^\dagger J_{m}^{\vphantom\dagger} \nonumber \\
   & =   \sum_{n=1}^\infty  L_{-n} L_n - \sum_{m=0}^\infty (m^2-m) J_{-m} J_{m}.
\end{align}
Next consider the $:L_0^2:$ term in Eq.~\eqref{SM-eq:60}:
\begin{align}
  \label{SM-eq:66}
  :L_0^2: = :\left( J_0^2 + 2 \sum_{p=1}^\infty J_p^\dagger J_p^{\vphantom\dagger}\right)^2:
\end{align}
By similar reasoning as above, we can write
\begin{align}
  \label{SM-eq:67}
  :L_0^2: &= L_0^2 + 4\sum_{p,q=1}^\infty \left( :J_q^\dagger J_q^{\vphantom\dagger} J_p^\dagger J_p^{\vphantom\dagger}: - J_q^\dagger J_q^{\vphantom\dagger} J_p^\dagger J_p^{\vphantom\dagger}\right)\nonumber \\
          & = L_0^2 - 4 \sum_{p,q=1}^\infty :J_q^\dagger \left[J_q^{\vphantom\dagger}, J_p^\dagger\right] J_p^{\vphantom\dagger} \nonumber \\
  & = L_0^2 - 2 \sum_{p=1}^\infty p J_{-p} J_p.
\end{align}
Now we can combine Eq.~\eqref{SM-eq:65} and Eq.~\eqref{SM-eq:67} to put into Eq.~\eqref{SM-eq:60} and get
\begin{align}
  \label{SM-eq:68}
  A_3 & = L_0^2 - 2 \sum_{m=1}^\infty m J_{-m} J_m \nonumber \\
      & + 2 \sum_{n=1}^\infty  L_{-n} L_n - 2\sum_{m=0}^\infty (m^2-m) J_{-m} J_{m} \nonumber \\
  & = 2 \sum_{n=1}^\infty  L_{-n} L_n + L_0^2 - 2 \sum_{m=0}^\infty m^2 J_{-m} J_m.
\end{align}
Now if you compare to $B_3$ in Eq.~\eqref{SM-eq:43},  we see that the $B_3$ contribution precisely cancels the last term in Eq.~\eqref{SM-eq:68}, and so finally
\begin{align}
  \label{SM-eq:69}
  A_3 + \frac{1}{(2\pi)^2} B_3 & = 2 \sum_{n=1}^\infty  L_{-n} L_n + L_0^2 .
\end{align}
We have thus derived Eq.~\eqref{eq:33}.

Note that the result differs from the one given in Ref.\cite{bazhanov1996integrable} by the presence of an additional term linear in $L_0$ in that reference.  We suspect the difference is due to a different quantization convention in the latter work.

\newcommand{\su}{\mathfrak{su}}
\newcommand{\cH}{\mathcal H}
\newcommand{\cV}{\mathcal V}
\renewcommand{\dd}{\mathrm d}

\section{$I_3$ Spectrum on a circle}
\label{SM-sec:i_3-spectrum-circle}

In this appendix, we show how to construct the finite-size spectrum of the $I_3$ operator, which is \emph{quartic} in boson fields, for quantized (total) momenta $k_n = 2\pi n/L$ on the circle with periodic boundary conditions, for small integer $n$.  This computation is tractable because for the chiral theory at a given total momentum $k_n$, there are only a finite number of states.  The calculation is done algebraically using the Kac-Moody and Virasoro algebras, presented in the main text and repeated below for clarity.  The construction follows standard methods of conformal field theory as described in canonical texts such as Ref.\cite{francesco2012conformal}, incorporating elements from the comprehensive work of Ref.\cite{bazhanov1996integrable}.

As a reminder, we work with the affine SU(2) current algebra at level $k=1$.  The current modes
\begin{equation}
    J^a_m, \qquad a=1,2,3,\qquad m\in \mathbb Z,
\end{equation}
obey
\begin{equation}
    [J^a_m,J^b_n]
    =
    i\epsilon^{abc}J^c_{m+n}
    +
    \frac{1}{2}m\,\delta^{ab}\delta_{m+n,0}.
    \label{SM-eq:current-algebra}
\end{equation}

Here we consider the states that would exist in the lattice system near zero momentum.  This corresponds to the ``vacuum module'' in the conformal field theory language, consisting of states built from successive operation of Kac-Moody operators upon 
the vacuum state $|0\rangle$  which is annihilated by all nonnegative modes:
\begin{equation}
    J^a_m|0\rangle=0,\qquad m\ge 0.
    \label{SM-eq:vacuum}
\end{equation}
Descendant states are obtained by acting on the vacuum with negative current modes:
\begin{equation}
    J^{a_1}_{-n_1}J^{a_2}_{-n_2}\cdots J^{a_r}_{-n_r}|0\rangle,
    \qquad n_i>0.
\end{equation}

The level, or conformal grade, of such a descendant is
\begin{equation}
  n = \sum_{i=1}^r n_i.
  \label{SM-eq:71}
\end{equation}
The level $n$ corresponds precisely to the total momentum $k_n=2\pi n/L$.  One can see that since the $n_i>0$, there are only a finite set of $\{ n_i\}$ which can be used to satisfy Eq.~\eqref{SM-eq:71}.  

The full vacuum module decomposes as
\begin{equation}
    \cH = \bigoplus_{n\ge 0}\cH_n,
\end{equation}
where $\cH_n$ is the finite-dimensional level-$n$ subspace.

\subsection{Formal ordered current basis}

At fixed level $n$, one begins with a formal spanning set of ordered current monomials.

A formal basis state is represented by a sequence
\begin{equation}
    s = \bigl((m_1,a_1),(m_2,a_2),\ldots,(m_r,a_r)\bigr),
    \qquad m_i<0,
\end{equation}
corresponding to
\begin{equation}
    |s\rangle
    =
    J^{a_1}_{m_1}J^{a_2}_{m_2}\cdots J^{a_r}_{m_r}|0\rangle.
\end{equation}
The level of $s$ is
\begin{equation}
    |s| = -\sum_{i=1}^r m_i.
\end{equation}

An (overcomplete) basis can be formed by choosing a canonical ordering of the pairs $(m,a)$ and generating all nondecreasing sequences of such pairs with total level $n$.  This gives a spanning set for $\cH_n$.  It is not, in general, a linearly independent basis, because the current algebra imposes null relations and other dependencies.

We denote this formal spanning set by
\begin{equation}
    \mathcal B_n^{\rm form}
    =
    \{
        |s_i^{(n)}\rangle
    \}_{i=1}^{D_n}.
\end{equation}

\subsection{Action of the current modes}

The key algebraic operation is the action of $J^a_m$ on an ordered current monomial.  If the state is empty, then by the highest-weight condition
\begin{equation}
    J^a_m|0\rangle=0,\qquad m\ge 0,
\end{equation}
while for $m<0$ one obtains a one-current descendant.

For a nonempty ordered monomial, write
\begin{equation}
    |s\rangle
    =
    J^b_n |r\rangle,
\end{equation}
where $(n,b)$ is the first current in the ordered sequence and $|r\rangle$ is the remaining product.  If the pair $(m,a)$ is already in canonical position relative to $(n,b)$, one simply inserts the current.  Otherwise, one uses the current algebra to commute $J^a_m$ past $J^b_n$:
\begin{align}
    J^a_mJ^b_n|r\rangle
    &=
    J^b_nJ^a_m|r\rangle
    +
    [J^a_m,J^b_n]|r\rangle
    \nonumber \\
    &=
    J^b_nJ^a_m|r\rangle
    +
    i\epsilon^{abc}J^c_{m+n}|r\rangle
    +
    \frac{1}{2}m\,\delta^{ab}\delta_{m+n,0}|r\rangle.
    \label{SM-eq:recursive-current-action}
\end{align}
This recursion terminates because currents are moved toward canonical order and positive modes eventually annihilate the vacuum unless compensated by commutators.

Thus the action of any current mode on any formal descendant can be written as a finite linear combination
\begin{equation}
    J^a_m |s\rangle
    =
    \sum_t C^t_{(m,a),s}|t\rangle.
\end{equation}
In the implementation, these coefficients are kept exactly as complex rational numbers.

\subsection{Inner product and Gram matrix}

The Hermitian structure is fixed by
\begin{equation}
    \left(J^a_m\right)^\dagger = J^a_{-m},
    \qquad
    \langle 0|0\rangle = 1.
    \label{SM-eq:adjoint}
\end{equation}
Given two formal descendants
\begin{equation}
    |s\rangle
    =
    J^{a_1}_{m_1}J^{a_2}_{m_2}\cdots J^{a_r}_{m_r}|0\rangle,
\end{equation}
and $|t\rangle$, their inner product is
\begin{equation}
    \langle s|t\rangle
    =
    \langle 0|
    J^{a_r}_{-m_r}\cdots J^{a_2}_{-m_2}J^{a_1}_{-m_1}
    |t\rangle.
    \label{SM-eq:inner-product}
\end{equation}
Operationally, one acts successively on $|t\rangle$ with the modes
\begin{equation}
    J^{a_r}_{-m_r},\ldots,J^{a_1}_{-m_1},
\end{equation}
using the recursive current-algebra rule, and extracts the coefficient of the vacuum.

This gives the exact Gram matrix on the formal spanning set:
\begin{equation}
    G^{(n),{\rm form}}_{ij}
    =
    \langle s_i^{(n)}|s_j^{(n)}\rangle.
    \label{SM-eq:formal-gram}
\end{equation}
Because $\mathcal B_n^{\rm form}$ may contain null or dependent states, $G^{(n),{\rm form}}$ can be singular.

A linearly independent physical basis is obtained by selecting pivot columns of $G^{(n),{\rm form}}$ after row reduction.  Denote this basis by
\begin{equation}
    \mathcal B_n^{\rm phys}
    =
    \{
        |\phi_\alpha^{(n)}\rangle
    \}_{\alpha=1}^{d_n}.\label{SM-eq:72}
\end{equation}
The restricted physical Gram matrix is
\begin{equation}
    G^{(n)}_{\alpha\beta}
    =
    \langle \phi_\alpha^{(n)}|\phi_\beta^{(n)}\rangle.
    \label{SM-eq:physical-gram}
\end{equation}
This basis is not orthonormal, so spectral problems must be formulated as generalized eigenvalue problems.

\subsection{Virasoro algebra and $I_3$}

The $I_3$ operator is written entirely in terms of Virasoro generators, which we remind the reader are written in terms of the Kac-Moody ones as
\begin{equation}
    L_m
    =
    \frac{1}{3}
    \sum_{p\in\mathbb Z}
    :J^a_pJ^a_{m-p}:.
    \label{SM-eq:sugawara}
\end{equation}
The normal ordering places annihilation modes to the right of creation modes.  In the highest-weight representation this means, in effect, that positive modes are placed to the right, where they annihilate the vacuum.  For a fixed finite-level state, only finitely many terms in the Sugawara sum contribute.

This implies (again repeated from the main text) the Virasoro algebra
\begin{equation}
    [L_m,L_n]
    =
    (m-n)L_{m+n}
    +
    \frac{1}{12}m(m^2-1)\delta_{m+n,0}.
    \label{SM-eq:virasoro}
\end{equation}

The level operator is $L_0$.  In the vacuum module,
\begin{equation}
    L_0|\psi\rangle = n|\psi\rangle,
    \qquad |\psi\rangle\in \cH_n.
\end{equation}
Therefore, on $\cH_n$,
\begin{equation}
    L_0^2 = n^2.
    \label{SM-eq:L0-square}
\end{equation}

From the main text, the effective Hamiltonian in the right-moving sector at the critical point is $I_3$ in Eq.~\eqref{eq:33},
\begin{equation}
    I_3
    =
    2\sum_{k=1}^{\infty}L_{-k}L_k + L_0^2.
    \label{SM-eq:I3-definition}
\end{equation}
One can see that $L_k$ lowers the momentum/level by $k$:
\begin{equation}
    L_k:\cH_n\to \cH_{n-k},
\end{equation}
so that the term $L_k$ annihilates $\cH_n$ for $k>n$.  Thus the infinite sum truncates at fixed level:
\begin{equation}
    I_3\big|_{\cH_n}
    =
    n^2
    +
    2\sum_{k=1}^n L_{-k}L_k.
    \label{SM-eq:I3-level-n}
\end{equation}
This is the finite-dimensional operator whose spectrum is computed at level $n$.

\emph{N.B.} In the quantum KdV literature, the local integral usually denoted $I_3$ is often written with additional normal-ordering-dependent terms proportional to $L_0$ and to the identity.  Those terms shift the numerical eigenvalues and may change their linear dependence on $n$, but they do not alter the finite-level diagonalization procedure.  The convention used here is precisely Eq.~\eqref{SM-eq:I3-definition}.

\subsection{Matrix elements of $I_3$}

One can compute  the matrix of $I_3$ in the non-orthonormal basis of Eq.~\eqref{SM-eq:72}:
\begin{equation}
    M^{(n)}_{\alpha\beta}
    =
    \langle \phi_\alpha^{(n)}|
    I_3
    |\phi_\beta^{(n)}\rangle.
\end{equation}

The $L_0^2$ contribution is simply
\begin{equation}
    M^{(n,L_0^2)}_{\alpha\beta}
    =
    n^2 G^{(n)}_{\alpha\beta}.
\end{equation}

For the $L_{-k}L_k$ term, define the rectangular matrix $X_k^{(n)}$ by expanding the action of $L_k$ on the physical level-$n$ basis in the formal level-$(n-k)$ spanning set:
\begin{equation}
    L_k|\phi_\beta^{(n)}\rangle
    =
    \sum_i
    \left(X_k^{(n)}\right)_{i\beta}
    |s_i^{(n-k)}\rangle.
    \label{SM-eq:X-definition}
\end{equation}
Then
\begin{align}
    \langle \phi_\alpha^{(n)}|
    L_{-k}L_k
    |\phi_\beta^{(n)}\rangle
    &=
    \langle L_k\phi_\alpha^{(n)}|
    L_k\phi_\beta^{(n)}\rangle
    \nonumber \\
    &=
    \left[
    \left(X_k^{(n)}\right)^\dagger
    G^{(n-k),{\rm form}}
    X_k^{(n)}
    \right]_{\alpha\beta}.
\end{align}
Therefore
\begin{equation}
    M^{(n)}
    =
    n^2G^{(n)}
    +
    2\sum_{k=1}^n
    \left(X_k^{(n)}\right)^\dagger
    G^{(n-k),{\rm form}}
    X_k^{(n)}.
    \label{SM-eq:I3-matrix}
\end{equation}

Equation~\eqref{SM-eq:I3-matrix} is the central finite-level construction.

\subsection{Generalized eigenvalue problem}

Because the physical basis is not orthonormal, the eigenvalue problem is
\begin{equation}
    M^{(n)}v
    =
    \lambda\,G^{(n)}v.
    \label{SM-eq:generalized-eigenproblem}
\end{equation}
The eigenvector
\begin{equation}
    v=(v_1,\ldots,v_{d_n})
\end{equation}
represents the state
\begin{equation}
    |\Psi\rangle
    =
    \sum_{\alpha=1}^{d_n}
    v_\alpha |\phi_\alpha^{(n)}\rangle.
\end{equation}
The eigenvalue $\lambda$ is equivalently the stationary value of the Rayleigh quotient
\begin{equation}
    \lambda
    =
    \frac{
        \langle \Psi|I_3|\Psi\rangle
    }{
        \langle \Psi|\Psi\rangle
    }.
\end{equation}

The exact rational matrices are converted to floating point only after the exact construction of $M^{(n)}$ and $G^{(n)}$.  One then solves the Hermitian generalized eigenvalue problem Eq.~\eqref{SM-eq:generalized-eigenproblem}.

\subsection{Global SU(2) spin}

The global SU(2) generators are the zero modes
\begin{equation}
    S^a = J^a_0.
\end{equation}
The quadratic Casimir is
\begin{equation}
    S^2
    =
    \sum_{a=1}^3 J^a_0J^a_0.
\end{equation}

Since $I_3$ is built from SU(2)-invariant contractions of currents, it commutes with the global SU(2) action:
\begin{equation}
    [I_3,J^a_0]=0.
\end{equation}
Therefore eigenstates of $I_3$ can be organized into global SU(2) multiplets.

In the physical basis, define
\begin{equation}
    S^{2,(n)}_{\alpha\beta}
    =
    \langle \phi_\alpha^{(n)}|
    S^2
    |\phi_\beta^{(n)}\rangle.
\end{equation}
For a normalized generalized eigenvector $v$, with
\begin{equation}
    v^\dagger G^{(n)}v=1,
\end{equation}
one computes
\begin{equation}
    \langle S^2\rangle
    =
    v^\dagger S^{2,(n)}v.
\end{equation}
The spin is then obtained from
\begin{equation}
    S(S+1)=\langle S^2\rangle,
\end{equation}
or
\begin{equation}
    S
    =
    \frac{-1+\sqrt{1+4\langle S^2\rangle}}{2}.
\end{equation}

\subsection{Algorithmic summary}

While the procedure outlined here can be easily carried out by hand for $n=1,2$, we implement it numerically in julia code for $n \leq 8$.  Below is a summary of the computational procedure.

For each level $n$:

\begin{enumerate}
    \item Generate all ordered current monomials of total level $n$.
    \item Compute the exact formal Gram matrix using the SU(2)$_1$ current algebra.
    \item Select pivot columns of the Gram matrix to obtain a linearly independent physical basis.
    \item Construct the Sugawara Virasoro generators $L_k$.
    \item Build the finite-level matrix
    \begin{equation}
        M^{(n)}
        =
        n^2G^{(n)}
        +
        2\sum_{k=1}^n
        \left(X_k^{(n)}\right)^\dagger
        G^{(n-k),{\rm form}}
        X_k^{(n)}.
    \end{equation}
    \item Solve the generalized eigenvalue problem
    \begin{equation}
        M^{(n)}v=\lambda G^{(n)}v.
    \end{equation}
    \item Compute $S^2$ on each eigenvector and assign the global SU(2) spin $S$.
    \item Group eigenvalues by $(\lambda,S)$ and record degeneracies.
    \item Plot $\lambda$ versus level $n$.
\end{enumerate}

\section{Spectral weights of the spin current}
\label{SM-sec:spectral-weights}

The dynamical structure factor at quantized momentum $k_n = 2\pi n/L$ involves the spectral decomposition of the state $J^a_{-n}|0\rangle$ in terms of $I_3$ eigenstates.  Notably, since $I_3$ is entirely expressed in terms of the Virasoro generators $L_n$, only those states which can be generated by action of these $L_n$ upon the initial state can appear as intermediate states.  This is what is known as a Virasoro tower or Virasoro module, and the restriction to these states is a significant simplification.

By the usual spectral representation, we can express the finite-size structure factor for small positive momenta $q_n = 2\pi n/L$ with $n >0$ as
\begin{equation}
  \label{SM-eq:34}
  S(q_n,\omega) = \frac{1}{L}\sum _\alpha w_\alpha^{(n)} \delta(\omega - \lambda_{n,\alpha}),
\end{equation}
where $n,\alpha$ label the eigenstates of $I_3$ at level $n$, with eigenvalue $\lambda_{n,\alpha}$.  Here the spectral weight of each $I_3$ eigenstate $|v_\alpha^{(n)}\rangle$ is
\begin{equation}
    w_\alpha^{(n)}
    =
    \bigl|\langle v_\alpha^{(n)}|J^a_{-n}|0\rangle\bigr|^2.
    \label{SM-eq:weight-def}
\end{equation}
This appendix explains how these weights are computed using the Virasoro algebra, which is more efficient than the Kac-Moody approach of Sec.~\ref{SM-sec:i_3-spectrum-circle} and allows the spectrum to be extended to $n\leq 15$.

\subsection{The spin-1 Virasoro primary}

The state $J^a_{-n}|0\rangle$ lies in the spin-$1$ sector of $\cH_n$, since $J^a_{-n}$ transforms as an adjoint under the global SU(2) generators $S^b = J^b_0$.  We now identify this sector with a Virasoro module.

From the Sugawara construction, the Virasoro generators satisfy
\begin{equation}
    [L_m, J^a_n] = -n\, J^a_{m+n}.
    \label{SM-eq:L-J-commutator}
\end{equation}
Using this relation together with the vacuum condition $J^a_m|0\rangle = 0$ for $m\geq 0$ (including the zero mode), one verifies that $J^a_{-1}|0\rangle$ is a Virasoro primary of conformal dimension $h=1$:
\begin{equation}
    L_0\,J^a_{-1}|0\rangle = J^a_{-1}|0\rangle,
    \qquad
    L_m\,J^a_{-1}|0\rangle =  0
    \quad (m>0).
\end{equation}
We denote $|h\rangle \equiv J^a_{-1}|0\rangle$ (for any fixed component $a$; all three give degenerate results by SU(2) symmetry).

Applying Eq.~\eqref{SM-eq:L-J-commutator} iteratively with $m=-1$ gives
\begin{equation}
    L_{-1}\, J^a_{-n}|0\rangle = n\, J^a_{-(n+1)}|0\rangle,
\end{equation}
from which it follows by induction that
\begin{equation}
    J^a_{-n}|0\rangle
    =
    \frac{1}{(n-1)!}\,
    L_{-1}^{n-1}|h\rangle.
    \label{SM-eq:Jn-Virasoro}
\end{equation}
Thus the spin-$1$ sector of $\cH_n$ is spanned by the Virasoro descendants of $|h\rangle$ at level $N = n-1$, and the problem reduces to a Virasoro module calculation with $c=1$, $h=1$.

\subsection{Virasoro basis and Gram matrix}

The Virasoro basis at level $N$ above $|h\rangle$ is indexed by integer partitions of $N$.  For a partition $\mu = (\mu_1 \geq \mu_2 \geq \cdots \geq \mu_r \geq 1)$ with $\sum_i \mu_i = N$, the corresponding basis state is
\begin{equation}
    |\phi_\mu\rangle = L_{-\mu_1}L_{-\mu_2}\cdots L_{-\mu_r}|h\rangle.
\end{equation}
The Gram matrix entries
\begin{equation}
    G^{(N)}_{\mu\nu}
    =
    \langle h|L_{\mu_r}\cdots L_{\mu_1}L_{-\nu_1}\cdots L_{-\nu_s}|h\rangle
\end{equation}
are computed exactly using the Virasoro algebra Eq.~\eqref{SM-eq:virasoro} with $c=1$, $h=1$.  The Gram matrix may become singular at higher levels due to null vectors in the Virasoro module (the Kac determinant for $h=1$, $c=1$ vanishes first at level $N=3$).  As in Sec.~\ref{SM-sec:i_3-spectrum-circle}, a linearly independent physical basis is selected by Gaussian elimination (pivot columns of $G^{(N)}$).

\subsection{$I_3$ in the spin-1 sector}

On states at Virasoro level $N = n-1$ above $|h\rangle$, the eigenvalue of $L_0$ is $h+N = n$.  The operator $L_k$ lowers the Virasoro level by $k$, so $L_k$ annihilates any state at level $N < k$; in particular, the $k=n$ term in Eq.~\eqref{SM-eq:I3-level-n} vanishes on this subspace and the sum truncates at $k=N$:
\begin{equation}
    I_3\big|_{\text{spin-1}, \cH_n}
    =
    n^2
    +
    2\sum_{k=1}^{n-1} L_{-k}L_k.
    \label{SM-eq:I3-spin1}
\end{equation}
The matrix of $I_3$ in the Virasoro basis is then
\begin{equation}
    M^{(N)}_{\mu\nu}
    =
    n^2\, G^{(N)}_{\mu\nu}
    +
    2\sum_{k=1}^{N}
    \bigl(X_k^{(N)}\bigr)^\dagger\,
    G^{(N-k),{\rm form}}\,
    X_k^{(N)},
\end{equation}
where $(X_k^{(N)})_{i\beta}$ expands $L_k|\phi_\beta\rangle$ in the formal spanning set at level $N-k$, following the same construction as Eq.~\eqref{SM-eq:I3-matrix} in Sec.~\ref{SM-sec:i_3-spectrum-circle}.  The generalized eigenvalue problem $M^{(N)}v = \lambda G^{(N)}v$ yields the $I_3$ eigenvalues in the spin-$1$ sector.

\subsection{Overlap formula}

Let $|s\rangle = L_{-1}^{N}|h\rangle$ be the all-ones-partition basis state, corresponding to $J^a_{-n}|0\rangle$ via Eq.~\eqref{SM-eq:Jn-Virasoro}.  Denote by $g_s$ the row of $G^{(N)}$ at the all-ones partition, restricted to the physical basis.  For a G-orthonormal eigenvector $v_\alpha$ (satisfying $v_\alpha^T G^{(N)} v_\beta = \delta_{\alpha\beta}$), the matrix element $\langle v_\alpha|s\rangle = g_s \cdot v_\alpha$.

From Eq.~\eqref{SM-eq:Jn-Virasoro} and the current algebra Eq.~\eqref{SM-eq:current-algebra},
\begin{equation}
    \|J^a_{-n}|0\rangle\|^2
    =
    \langle 0|J^a_n J^a_{-n}|0\rangle
    =
    \tfrac{n}{2},
    \label{SM-eq:Jn-norm}
\end{equation}
while the Virasoro Gram matrix element at the all-ones partition gives
$\|s\rangle\|^2 = G^{(N)}_{ss}$.  Combining with Eq.~\eqref{SM-eq:Jn-Virasoro}:
\begin{equation}
    \bigl|\langle v_\alpha|J^a_{-n}|0\rangle\bigr|^2
    =
    \frac{\|J^a_{-n}|0\rangle\|^2}{\|s\rangle\|^2}\,
    (g_s\cdot v_\alpha)^2
    =
    \frac{n}{2\,G^{(N)}_{ss}}\,(g_s\cdot v_\alpha)^2.
    \label{SM-eq:overlap-formula}
\end{equation}
Completeness requires $\sum_\alpha w_\alpha^{(n)} = \|J^a_{-n}|0\rangle\|^2 = n/2$, which serves as a numerical check.

This result is a familiar one in another guise.  The sum over all $\alpha$ gives the total spectral weight, i.e. the frequency integral of the structure factor. This is therefore a sum rule. The proportionality to $n$, combined with the $1/L$ factor in Eq.~\eqref{SM-eq:34}, implies this weight is proportional to the momentum $q_n$, and hence vanishes in the zero-momentum limit.  The vanishing of the spectral weight at small momentum is a consequence of spin conservation: the spin at $q=0$ is the total spin which is conserved and equal to zero in the ground state (for $g \leq g_c$).  Hence the spectral weight must vanish on approaching zero momentum.  This behavior is well-known in the usual Heisenberg chain, in which this weight is concentrated in a narrow linearly-dispersing peak.  Here the peak disperses cubically and is \emph{not} narrow, but the sum rule remains true.  

\subsection{Algorithmic summary}

\begin{figure}[hbtp]
  \centering
    \includegraphics[width=\columnwidth]{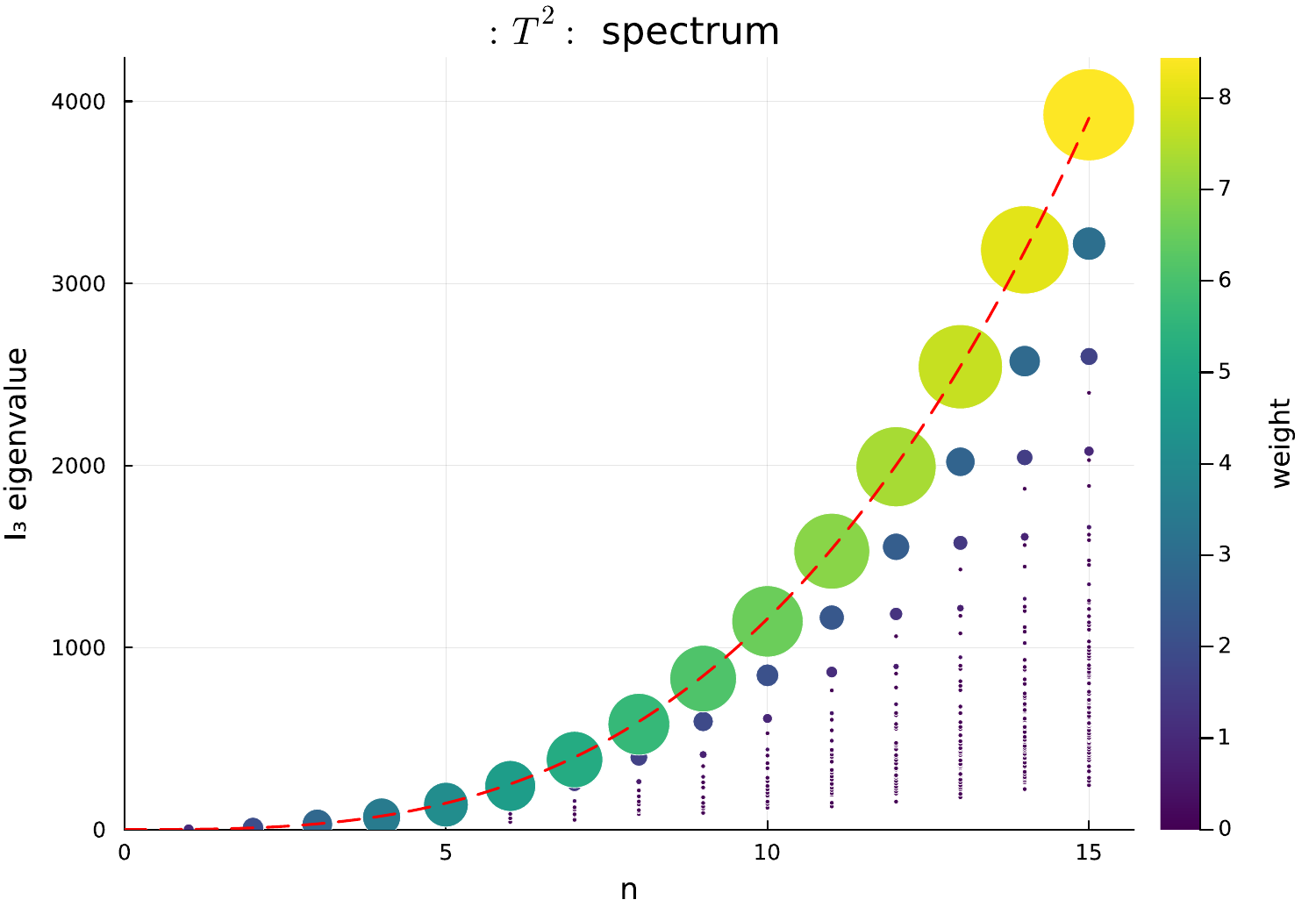}  
  \caption{Dynamical structure factor in the critical theory from $:T^2:$ theory for a finite-size system.  It consists of discrete delta-function peaks at various energies for each quantized momentum $k_n = 2\pi n/L$.  The weight of each state is indicated by the radius and color scale.  The dashed line is a fit of the maximum energies to $\epsilon = a n^3$, with $a=1.16$, which demonstrates $z=3$ scaling.  Unlike in the CFT case, the weight here is spread amongst many states at each level $n \gg 1$, indicating that the boson itself is \emph{not} quasiparticle-like. }
  \label{SM-fig:dkdv}
\end{figure}

For each level $n$ (with $N=n-1$):
\begin{enumerate}
    \item Generate all integer partitions of $N$.
    \item Identify the all-ones partition index $s$ (corresponding to $L_{-1}^N|h\rangle \propto J^a_{-n}|0\rangle$).
    \item Compute the exact Virasoro Gram matrix $G^{(N)}$ with $c=1$, $h=1$ using Eq.~\eqref{SM-eq:virasoro}.
    \item Build the $I_3$ matrix $M^{(N)}$ via Eq.~\eqref{SM-eq:I3-spin1}, using the same matrix element construction as in Sec.~\ref{SM-sec:i_3-spectrum-circle}.
    \item Solve the generalized eigenvalue problem $M^{(N)}v = \lambda G^{(N)} v$.
    \item Compute the spectral weight $w_\alpha = \frac{n}{2G^{(N)}_{ss}}(g_s\cdot v_\alpha)^2$ for each eigenstate.
\end{enumerate}
Fig.~\ref{SM-fig:dkdv} shows the resulting spectrum with dot size and color encoding the weight $w_\alpha$.

\section{Semi-classical analysis and bosonization of the ordered phase}
\label{SM-sec:classical}

We treat quantum spins as classical vectors of length $1/2$ and propose the variational ansatz $\bm{S}_n = (\sqrt{\frac14-m^2}\cos q_{cl}n, \sqrt{\frac14-m^2}\sin q_{cl}n, m)$. The classical energy per site reads $\mathcal{E}(m,q_{cl};g) = \frac14 \left(4m^2 + (1-4m^2)\cos q_{cl}\right) - \frac14 g m(1-4m^2) (2\sin q_{cl} - \sin 2q_{cl})$. We next optimize the energy by requiring $\partial\mathcal{E}(m,q_{cl};g)/\partial q_{cl} = 0 = \partial\mathcal{E}(m,q_{cl};g)/\partial m$. This leads to 
\begin{equation}
\label{SM-eq:mfenergy}
\begin{aligned}
    g \sin q_{cl} &= \frac{4m}{1-12m^2},\\
    2 g m (\cos q_{cl} - \cos 2q_{cl}) &= -\sin q_{cl}
\end{aligned}
\end{equation}
Somewhat unexpectedly, these equations admit an explicit algebraic solution. It proceeds by replacing $\sin q_{cl}$ in the second equation with the help of the first one. Then we set $D=1-12m^2$ and rewrite the first equation as $3g^2 D^2(1-\cos^2q_{cl}) = 4(1-D)$. But from the second equation $D = 2/(g^2 A)$, where $A = (2\cos q_{cl}+1)(\cos q_{cl}-1)$. Using this $D$, the first equation turns into a $m$-independent $3(1-\cos^2q_{cl}) = g^2 A^2 -2A$, which factorizes nicely into $(\cos q_{cl}-1)^2 \Big(g^2(2\cos q_{cl} +1)^2 -1\Big) =0$. 

For $g \leq 1$, the solution is trivial: $q_{cl} =\pi$ and $m=0$, corresponding to the classical antiferromagnetic state with energy $\mathcal{E} = -\frac14$. For $g > 1$, we find $\cos q_{cl} = -\frac{1+g}{2g}$ and $m = \pm \frac12\sqrt{\frac{g-1}{1+3g}}$, with energy $\mathcal{E} = -\frac{1+g^2}{8g}$. 
Note that the first equation in \eqref{SM-eq:mfenergy} ties the signs of $m$ and $\sin q_{cl}$, so that $m \sin q_{cl} > 0$. This corresponds to the minimization of the chiral three-spin term contribution. 
In the large-$g$ limit, one solution has $q_{cl}\to \frac{2\pi}{3}$ and $m\to \frac{1}{2\sqrt{3}}$, while the other $q_{cl}\to \frac{4\pi}{3}$ and $m\to -\frac{1}{2\sqrt{3}}$.

The next task is to derive the effective Hamiltonian of gapless low-energy spin modes by allowing for fluctuations of the spin spiral twist angle and magnetization, as well as the ``wobbling" of the spin axes triad ${\bm e}_{j=1,2,3}$. Namely, $\bm{S}(x,\tau) = \frac12 \bm{n}(x,\tau)$ ($x = na$),
\begin{align}
    \bm{n}(x,\tau) &= \sqrt{1-4(m+\delta m(x,\tau))^2} \big(\cos\theta(x,\tau) \bm{e}_1 + \nonumber\\
    &+ \sin\theta(x,\tau) \bm{e}_2\big) 
    + 2(m+\delta m(x,\tau)) \bm{e}_3,
\end{align}
where $\theta(x,\tau) = \frac{q_{cl}}{a}x + \delta\theta(x,\tau)$ and $\delta m$ describes deviation of $S^z$ from the equilibrium $m$ found above. In the usual bosonization language, $\delta m(x,\tau) = \frac{1}{\pi}\partial_x \phi(x,\tau)$; as shown below, this normalization of the field $\phi$ is fixed by the Berry-phase commutator together with the quantization of the total $S^z$. The spin axes are parameterized as
\begin{eqnarray}
    \bm{e}_1 &=& \frac{1}{\sqrt{1-v^2}} (\sqrt{1-u^2 - v^2}, 0, -u) \approx (1, 0, -u)\nonumber\\
    \bm{e}_2 &=& \frac{1}{\sqrt{1-v^2}} (- u v, 1-v^2, -v \sqrt{1-u^2-v^2}) \approx (0, 1,-v)\nonumber\\
    \bm{e}_3 &=& (u, v, \sqrt{1-u^2-v^2}) \approx (u, v, 1)
\end{eqnarray}
where $u(x,\tau)$ and $v(x,\tau)$ describe smooth variations of the orthonormal spin basis in space-time.

At the quadratic level, the $\delta m$ and $\delta\theta$ fluctuations decouple from the $u$ and $v$ ones. 

Performing the gradient expansion of \eqref{eq:1}, we find the effective Hamiltonian density
\begin{equation}
\label{SM-eq:eff-m-theta}
\begin{aligned}[t]
    H_{\rm eff}(\delta m, \delta\theta)
  &= {\cal E}(m,q_{cl}) \\
  &\hspace{-1.2em}{} + \frac12 h_{mm} (\delta m)^2
    + h_{m\theta} \, \delta m \, \partial_x \delta\theta
    + \frac12 h_{\theta\theta} (\partial_x \delta \theta)^2,
\end{aligned}
\end{equation}
where
\begin{equation}
  \label{SM-eq:1}
  \begin{aligned}
    h_{mm} &= \frac{9g^2-1}{2g}, \qquad
    h_{m\theta} = \frac{g+1}{2g},\\
    h_{\theta\theta} &= \frac{(g+1)(3g^2-1)}{4 g(3g+1)}
  \end{aligned}
\end{equation}
Expressions for the coefficients $h_{ij}$ of the quadratic form follow from using the equilibrium values of $m$ and $q_{cl}$ (we expand about the $m>0, \sin q_{cl} > 0$ solution). 

Similarly, writing the transverse stiffness as $\rho m^2$,
\begin{equation}
\label{SM-eq:Huv}
    H_{\rm eff}(u,v) = \rho m^2 \big((\partial_x u)^2 +(\partial_x v)^2\big)
\end{equation}
where classically $\rho m^2=\frac{3g^2-1}{8(3g+1)}$. Terms $\sim v\partial_x u$ vanish by virtue of the equilibrium condition \eqref{SM-eq:mfenergy}, second equation.

Fluctuations above the saddle point acquire dynamics from the Berry phase term in the action  with $s=1/2$
\begin{equation}
\label{SM-eq:berry1}
\begin{aligned}
    \frac{i}{2} \int dx d\tau\,
    \frac{n_x \partial_\tau n_y - n_y \partial_\tau n_x}{1+n_z}
    &= \int dx d\tau \, i m \, u \partial_\tau v\\
    &\quad + \frac{i}{2} (1 - 2m - 2\delta m) \partial_\tau \delta\theta
\end{aligned}
\end{equation}
In deriving these results, we have dropped fast-oscillating terms that average to zero under $x$-integration and, where needed, used integration by parts.

The term $-i\, \delta m \, \partial_\tau \delta\theta$ in \eqref{SM-eq:berry1} identifies $\delta m$ as the momentum canonically conjugate to $\delta\theta$,
\begin{equation}
  \label{SM-eq:comm-theta-m}
  [\delta\theta(x), \delta m(x')] = i \,\delta(x-x') .
\end{equation}
Similarly,
\begin{equation}
  \label{SM-eq:4}
  [u(x),v(x')] = \frac{i}{m} \delta(x-x'),
\end{equation}
forms a second conjugate pair.  
As a check on Eq.~\eqref{SM-eq:comm-theta-m}, using $S^+ \simeq \frac12 \sqrt{1-4m^2}\, e^{i\theta}$ and $S^z = m + \delta m$, \eqref{SM-eq:comm-theta-m} reproduces the microscopic relation $[S^+(x), S^z(x')]$ $= -S^+(x)\, \delta(x-x')$. With $\delta m = \frac{1}{\pi} \partial_x \phi$, it is equivalent to the standard dual-pair algebra $[\phi(x), \delta\theta(x')]$ $= -\frac{i \pi}{2} {\rm sgn}(x-x')$, so that $(\phi, \delta\theta)$ is a canonical Luttinger-liquid pair. Moreover, the change of the total spin, $\Delta S^z_{\rm tot} = \int dx \, \delta m$ $= \frac{1}{\pi}\big[\phi(\infty)-\phi(-\infty)\big]$, is quantized to integers, so that $\phi$ is a compact field with period $\pi$: the allowed spin-preserving vertex operators are $e^{2 i p \phi}$ with integer $p$, a fact we exploit below.

Writing $\delta m$ in terms of the conjugate phase, and dropping the constant term in the Hamiltonian gives the phase-only form
\begin{align}
  \label{SM-eq:35}
  H_{\rm eff}(\phi,\delta\theta) & = \frac{h_{mm}}{2\pi^2} (\partial_x \phi)^2 
    + \frac{h_{m\theta}}{\pi} \, \partial_x\phi \, \partial_x \delta\theta + \frac12 h_{\theta\theta} (\partial_x \delta \theta)^2.
\end{align}

Integrating out $\phi$ or $\delta m$ field produces the Euclidean action $S_\theta = \int dx d\tau L_\theta$, where
\begin{eqnarray}
    L_\theta &=& \frac{1}{2 h_{mm}} (\partial_\tau \delta\theta)^2 - \frac{i h_{m\theta}}{h_{mm}} \partial_\tau \delta\theta \,\partial_x \delta\theta \nonumber\\
    && + \frac12 \big(h_{\theta\theta} - \frac{h_{m\theta}^2}{h_{mm}}\big) (\partial_x \delta\theta)^2
\end{eqnarray}
At this point, it is convenient to continue this to real time $t$, $\partial_\tau \to - i \partial_t$, so that $L_\theta \to - {\cal L}_\theta$
\begin{eqnarray}
    {\cal L}_\theta &=& \frac{1}{2 h_{mm}} (\partial_t \delta\theta)^2 - \frac{ h_{m\theta}}{h_{mm}} \partial_t \delta\theta \,\partial_x \delta\theta \nonumber\\
    && - \frac12 \big(h_{\theta\theta} - \frac{h_{m\theta}^2}{h_{mm}}\big) (\partial_x \delta\theta)^2
\end{eqnarray}
The Euler-Lagrange equation of motion for the phase field $\delta\theta(x,t) \sim e^{i q x - i \omega t}$ results in the following chiral dispersion relation
\begin{equation}
\label{SM-eq-disp-theta}
    \omega_q = \sqrt{h_{mm} h_{\theta\theta}}\, |q| - h_{m \theta} \,q = 
    \left\{
\begin{aligned}
    v_R \, q, q>0  & \\
    v_L \, |q|, q < 0 &
\end{aligned}
\right.
\end{equation}
where $v_{R/L} = \sqrt{h_{mm} h_{\theta\theta}} \, \mp h_{m\theta}$. Given $h_{ij}$ values in \eqref{SM-eq:eff-m-theta}, we find that at the critical point $g=1$ the anisotropy of velocities is maximal, $v_L(g=1) =2, v_R(g=1) = 0$. For future convenience, we denote the {\em even} under $q \to -q$ part of the dispersion $\omega_q$ as $w_q$, so that \eqref{SM-eq-disp-theta} reads $\omega_q = w_q - h_{m\theta} q$.

To calculate correlation functions involving $\delta\theta$, we construct the corresponding Hamiltonian and quantize it. First, rescale $\delta\theta = \sqrt{h_{mm}}\, \tilde\theta$. Next, find the conjugate momentum $\Pi = \partial{\cal L}/\partial(\partial_t \tilde\theta) = \partial_t \tilde\theta - h_{m\theta} \partial_x \tilde\theta$, so that $H = \Pi \partial_t \tilde\theta - {\cal L}$. This gives
\begin{equation}
    H_\theta = \int dx \, \frac12 \Pi^2 + h_{m\theta} \Pi \partial_x \tilde\theta + \frac12 h_{\theta\theta} h_{mm} (\partial_x \tilde\theta)^2
\end{equation}
The quantization is achieved by the following mode decomposition
\begin{equation}
    \tilde{\theta}(x,t) = \frac{1}{\sqrt{L}} \sum_q \frac{1}{\sqrt{2 w_q}} \big(a_q e^{i q x - i \omega_q t} + {\rm h.c.}\big)
\end{equation}
where $[a_q, a^\dagger_k] = \delta_{q,k}$ and $w_q = (\omega_q + \omega_{-q})/2 = \sqrt{h_{mm} h_{\theta\theta}}\, |q|$ as described below \eqref{SM-eq-disp-theta}. The final Hamiltonian has the standard oscillator form $H_\theta = \sum_q \omega_q (a^\dagger_q a_q + 1/2)$ with dispersion \eqref{SM-eq-disp-theta}.

Note that eliminating $\delta m$ via its equation of motion following from \eqref{SM-eq:eff-m-theta}~and~\eqref{SM-eq:berry1} gives $\delta m = \frac{1}{h_{mm}}\big(\partial_t \delta\theta - h_{m\theta} \partial_x \delta\theta\big) = \Pi/\sqrt{h_{mm}}$, so that the canonical quantization above automatically implements the commutator \eqref{SM-eq:comm-theta-m}. Correspondingly, the dual field introduced earlier is $\phi(x) = \frac{\pi}{\sqrt{h_{mm}}} \int^x dx' \, \Pi(x')$.

The action for $u, v$ fields follows from \eqref{SM-eq:Huv},\eqref{SM-eq:berry1}
\begin{equation}
    S_{uv} = \int dx d\tau \, i m \, u \partial_\tau v
    + \rho m^2 \big( (\partial_x u)^2 + (\partial_x v)^2 \big).
    \label{SM-eq:Suv}
\end{equation} 

{\em Spin operators.} The components of $\bm{n}$ follow from the parameterization above:
\begin{equation}
\label{SM-eq:n-components}
\begin{aligned}
    n_x(x,\tau) &= \sqrt{1-4m^2} \cos\theta(x,\tau) + 2m \, u(x,\tau),\\
    n_y(x,\tau) &= \sqrt{1-4m^2} \sin\theta(x,\tau) + 2m \, v(x,\tau),\\
    n_z(x,\tau) &= 2m + 2\delta m - \sqrt{1-4m^2}\big(\cos\theta \, u + \sin\theta \, v\big)
\end{aligned}
\end{equation}
where $m = \frac12\sqrt{(g-1)/(3g+1)}$ is the equilibrium static magnetization.

Now we use Eq.~\eqref{SM-eq:n-components} as the basis to construct an augmented bosonization theory, in the spirit of Haldane\cite{Haldane1981}.  The terms in Eq.~\eqref{SM-eq:n-components} are only the smooth (gradient-expansion) parts of the spin operators: the compactness of $\phi$ generates additional oscillatory harmonics, which we now construct. Mapping the down spins to hard-core bosons, the boson density is $\rho_{\rm d} = \frac12 - S^z = \rho_0 - \delta m$ with $\rho_0 = \frac12-m$, and Haldane's counting field \cite{Haldane1981} $\vartheta(x) = \pi \int^x \rho_{\rm d} \, dx' = \pi \rho_0 x - \phi(x)$ is consistent with the normalization $\delta m = \frac{1}{\pi} \partial_x \phi$ adopted above. The density-wave harmonics $e^{2 i p \vartheta}$, which are exactly the vertex operators $e^{-2 i p \phi}$ allowed by the period-$\pi$ compactness of $\phi$, therefore each carry the oscillatory factor $e^{i p q_m x}$, with
\begin{equation}
\label{SM-eq:qm}
    q_m = 2\pi \rho_0 = \pi (1 - 2m) .
\end{equation}
At $g \leq 1$, where $m = 0$, $q_m$ reduces to the N\'eel wavevector $\pi$; at saturation, $m \to \frac12$, $q_m \to 0$ and the oscillation disappears, as it must. Building the harmonics from the up spins instead gives $\pi(1+2m) \equiv -q_m$ mod $2\pi$ --- the same set of harmonics, with the sign of the accompanying $2\phi$ reversed. Note also that the algebra above gives $[2\phi(x), \delta\theta(x')] = \mp i\pi$ at large separation, so the satellite operators $e^{i(\delta\theta \pm 2\phi)}$ carry integer conformal spin, as used below.

\begin{figure}[htbp]
  \centering
  \includegraphics[width=\columnwidth]{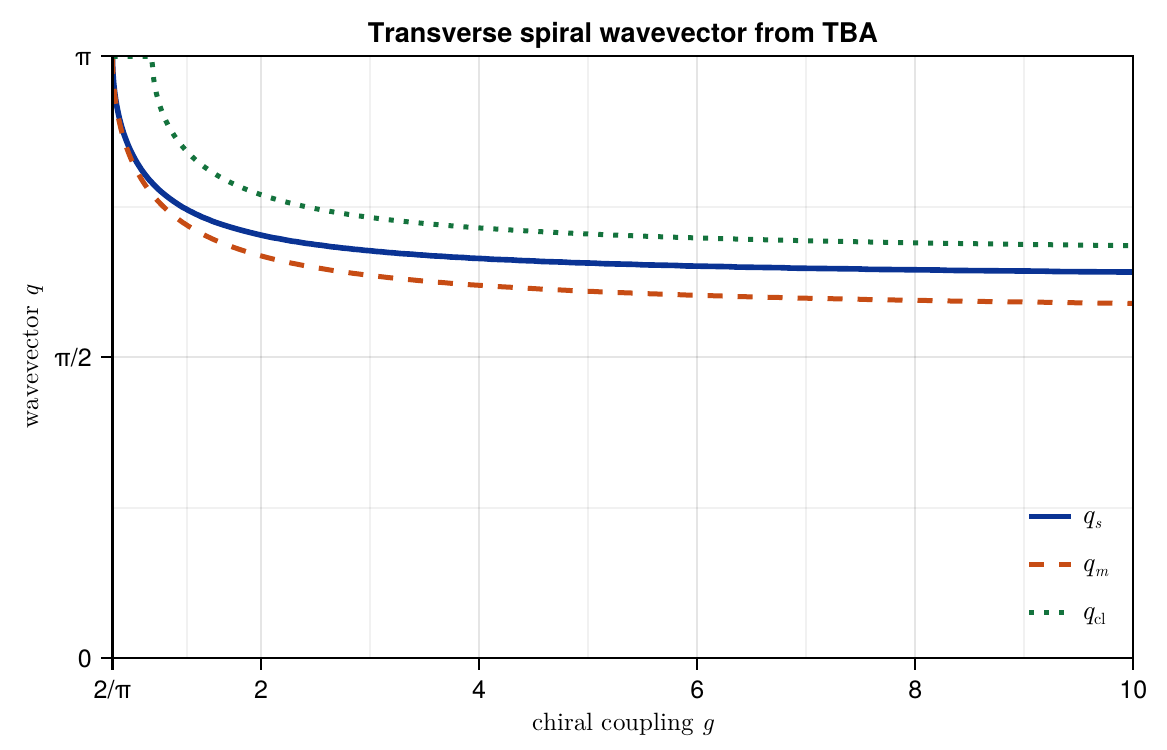}
  \caption{Wavevectors describing correlations in the ferromagnetic phase.  The exact wavevectors $q_s$ and $q_m$ are obtained from the TBA, and compared here with the classical spiral wavevector $q_{cl}$.  Note that $q_s$ and $q_m$ are quite close for all $g$.}
  \label{SM-fig:wavevectors}
\end{figure}

Now we allow the semi-classical umbrella wavevector $q_{cl}$ to be renormalized in the true low-energy theory, taking $q_{cl} \rightarrow q_s$.  Keeping the first harmonic and writing out $\theta = q_s x + \delta\theta$ explicitly (we set the lattice constant to unity), the spin operators become
\begin{equation}
\label{SM-eq:n-amended}
\begin{aligned}
    S^z &= m + \frac{1}{\pi} \partial_x \phi + C_1 \cos\big(q_m x - 2\phi + \zeta\big)\\
    &\quad -\, {\rm Re}\Big\{D_0(u - i v)\, e^{i(q_s x + \delta\theta)}\\
    &\qquad +\, D_+ (u - i v)\,
    e^{i[(q_s + q_m)x + \delta\theta - 2\phi]}\\
    &\qquad +\, D_- (u - i v)\,
    e^{i[(q_s - q_m)x + \delta\theta + 2\phi]}\Big\} + \ldots ,\\
    S^\pm &= C_0 \, e^{\pm i (q_s x + \delta\theta)}\\
    &\quad +\, C_+ \, e^{\pm i [(q_s + q_m) x + \delta\theta - 2\phi]}\\
    &\quad +\, C_- \, e^{\pm i [(q_s - q_m) x + \delta\theta + 2\phi]}\\
    &\quad +\, m \, (u \pm i v)
    + C_2 m (u \pm i v) \cos\big(q_m x - 2\phi + \zeta'\big) + \ldots
\end{aligned}
\end{equation}
with $C_0=D_0 = \frac12\sqrt{1-4m^2}$ at the semiclassical level. Each term now displays its wavevector explicitly: the transverse components oscillate at $q_s$ and at the satellites $q_s \pm q_m$, while the longitudinal one oscillates at $q_m$ and, through the $u$, $v$ cross terms, at $q_s$ and $q_s \pm q_m$. The amplitudes $C_i$, $D_i$, and the phase $\zeta$ are non-universal (cutoff dependent); since the chiral term breaks parity, $C_+ \neq C_-$ and $D_+ \neq D_-$ are allowed. Higher harmonics $e^{i p q_m x - 2 i p \phi}$, $|p| \geq 2$, follow the same pattern with larger scaling dimensions.  While we have motivated this expression using semi-classical arguments, we expect that it provides a correct low-energy description, provided the various constant coefficients in Eq.~\eqref{SM-eq:n-amended} and parameters in the action are allowed to deviate from their classical values.  Notably, the wavevector $q_s$ is generally different from the semiclassical value.  It can, however, be determined from the TBA, as shown in Fig.~\ref{SM-fig:wavevectors}.

{\em Spin correlations.} All correlations implied by \eqref{SM-eq:n-amended} are built from two Gaussian ingredients: the $u,v$ (magnon) propagator following from $S_{uv}$ \eqref{SM-eq:Suv}, and the vertex operators of the compact Luttinger field. It is useful to package the latter as
\begin{equation}
    V_{r,p}(x,t) = e^{i[r\delta\theta(x,t)+2p\phi(x,t)]},
    \qquad r,p\in{\mathbb Z},
\end{equation}
where $r$ is the spin raised by the operator and $p$ labels the Haldane harmonic. The case $r=1,p=0$ is the smooth transverse operator, while $r=0,p=\pm1$ is the first longitudinal density harmonic. Using $\delta\theta=\sqrt{h_{mm}}\,\tilde\theta$ and $\phi(x)=\frac{\pi}{\sqrt{h_{mm}}}\int^x dx'\,\Pi(x')$, the mode expansion above gives
\begin{equation}
\label{SM-eq:vertex-corr}
\begin{aligned}
    &\langle V_{r,p}(x,t)V^\dagger_{r,p}(0,0)\rangle\\
    &\quad \propto
    \Big(\frac{\alpha}{\alpha - i(x - v_R t)}\Big)^{\eta_R(r,p)}
    \Big(\frac{\alpha}{\alpha + i(x + v_L t)}\Big)^{\eta_L(r,p)} ,
\end{aligned}
\end{equation}
where $\alpha$ is the short-distance cutoff and
\begin{equation}
\label{SM-eq:eta-satellites}
    \eta_{R/L}(r,p) =
    \Big(r\sqrt{K} \mp \frac{p}{2\sqrt{K}}\Big)^2,\,
    K = \frac{1}{4\pi}\sqrt{\frac{h_{mm}}{h_{\theta\theta}}}.
\end{equation}
Thus $\eta_R+\eta_L=2r^2K+\frac{p^2}{2K}$ and $\eta_R-\eta_L=-2rp$. The integer difference is required by the mutual locality noted above. The dual density exponent is the special case $K_\phi=\eta_R(0,1)=\eta_L(0,1)=1/(4K)$, so that $K K_\phi=1/4$. The cross term $h_{m\theta}$, being proportional to the momentum density, tilts the dispersion ($v_R\neq v_L$) without changing the exponents themselves. If an operator carries an additional oscillating factor $e^{iQx}$, then for momentum $k$ measured from $Q$ the Fourier transform has the threshold form
\begin{equation}
\label{SM-eq:generic-chiral-fourier}
\begin{aligned}
    \langle V_{r,p}V^\dagger_{r,p}\rangle_{k,\omega}
    &\propto \Theta(\omega+v_L k)\Theta(\omega-v_R k)\\
    &\quad\times(\omega+v_L k)^{\eta_R(r,p)-1}
    (\omega-v_R k)^{\eta_L(r,p)-1}.
\end{aligned}
\end{equation}
Notably, the exponent $\eta_R$ is associated with the strength of the singularity of the \emph{left-moving} boundary  $\omega=-v_L k$ and vice-versa.  Thus if $\eta_R <\eta_L$, the operator has a dominant left-moving singularity in momentum space, and likewise if $\eta_L<\eta_R$, the dominant singularity is right-moving. Observe also that since for $r = \pm 1$ and $p = \pm 1$ the exponents are constrained by the relation $|\eta_R - \eta_L| = 2$, the presence of a singularity in one chiral channel (say, left-moving one for $\eta_R < 1$) implies the absence of a singularity for the opposite chirality (since then $\eta_L = 2 + \eta_R > 2$).  With these ingredients, we now go through the contributions to the spin correlations wavevector by wavevector.

{\em Transverse, $k \approx 0$.} The $m(u \pm i v)$ term of \eqref{SM-eq:n-amended} produces a ferromagnetic magnon mode, which follows from $S_{uv}$  \eqref{SM-eq:Suv}
\begin{equation}
  \label{SM-eq:spm-magnon}
    \langle S^+ S^-\rangle_{k,\omega_n}
    = \frac{2m^2}{i m \omega_n - 2\rho m^2 k^2}
    = \frac{2m}{i \omega_n - 2\rho m k^2}
\end{equation}
Here the coefficient of $i\omega_n$ follows by writing $\psi=u+i v$, for which \eqref{SM-eq:Suv} gives $\langle\psi\psi^\dagger\rangle_{k,\omega_n}=2/[i m \omega_n-2\rho m^2 k^2]$; the prefactor $m^2$ then comes from $S^+\simeq m\psi$. The expression is easily continued to the real frequency via $i \omega_n \to \omega + i0$.   The magnon pole has a residue of $2m$ and an inverse effective mass $1/(2m^*) = 2\rho m$.  The semi-classical expression for $\rho$ is qualitatively incorrect on approaching the critical point $g=1$, because semiclassically $\rho \sim 1/(16m^2)$ diverges in this limit.  However, the general form remains correct and Eq.~\eqref{SM-eq:spm-magnon} can still be applied if a corrected quantum value for $\rho$ is used.  

{\em Transverse, $k \approx q_s$.} The $C_0$ term is proportional to $e^{iq_s x}V_{1,0}$. Thus \eqref{SM-eq:vertex-corr} applies with $\eta_R=\eta_L=K$, and its equal-time prefactor is $(1-4m^2)/4$ at the semiclassical level. The corresponding frequency-momentum singularity follows from \eqref{SM-eq:generic-chiral-fourier}, with $k$ measured from $q_s$.

{\em Transverse, $k \approx q_s \pm q_m$.} The satellite terms $C_\pm$ are proportional to $e^{i(q_s \pm q_m)x}V_{1,\mp1}$. From \eqref{SM-eq:eta-satellites}, one satellite has $\eta_R-\eta_L=-2$ and the other has $R\leftrightarrow L$. Specifically, following the discussion after Eq.~\eqref{SM-eq:generic-chiral-fourier}, the coefficient $C_+$ is associated to a predominantly right-moving feature at $q=\pm(q_s+q_m)$, and the coefficient $C_-$ is associated to a predominantly left-moving feature at $q=\pm(q_s-q_m)$.  Because parity is broken, the two satellites carry independent non-universal weights, $|C_+|^2 \neq |C_-|^2$, unlike in a parity-symmetric chain.

{\em Longitudinal, $k \approx 0$.} The static part of $S^z$ gives the elastic Bragg peak of weight $m^2$ at $k = 0$, while the $\frac{1}{\pi} \partial_x \phi$ term produces the smooth density-fluctuation continuum, obtained from the gradient correlations of the dual field with the same mode expansion.

{\em Longitudinal, $k \approx \pm q_s$.} The ${\rm Re}\big[(u - i v) e^{i(q_s x + \delta\theta)}\big]$ term is a composite operator, so that the corresponding dynamic correlations are given by the convolution of the $V_{1,0}$ correlator in \eqref{SM-eq:vertex-corr} with the $u/v$ magnon propagator.

{\em Longitudinal, $k \approx \pm(q_s \pm q_m)$.} The $D_\pm$ terms are analogous composite operators dressed by the Haldane harmonics $e^{\mp 2 i \phi \pm i q_m x}$. Their correlations are therefore obtained by convolving the $u/v$ magnon propagator with the $V_{1,\mp1}$ correlators, with the same chiral exponents as the transverse satellites.

{\em Longitudinal, $k \approx \pm q_m$.} The $C_1$ harmonic is the $e^{iq_mx}V_{0,-1}$ density wave. It produces an algebraic singularity at $k=\pm q_m$ with equal-time decay $x^{-2K_\phi}=x^{-1/(2K)}$. The product of the transverse and longitudinal equal-time exponents, $(2K)(2K_\phi)=1$, is the standard Luttinger-liquid identity for spin chains at finite magnetization.

At the critical point $g \to 1^+$ these expressions evaluate to $K = 1/\pi$ and $2 K_\phi = \pi/2$, while deep in the chiral phase, $g \to \infty$, $K \to \frac{3}{2\pi\sqrt{2}}$.  We have not attempted to extract $K$ or $v_R$ and $v_L$ from the TBA in the ordered phase.

\section{Ferromagnetic ground state and its magnon excitations}
\label{SM-sec:fm}

As explained in the main text, for $\delta g = g - g_c > 0$ the ground state develops finite magnetization and vector spin chirality. We denote this state as $|S\rangle$. Here, we neglect the backscattering interaction that couples right- and left-moving sectors of the theory and focus exclusively on the right-moving sector. The Hamiltonian reads 
\begin{equation}
    H = - \frac{2\pi v}{L} L_0 + \frac{\lambda}{L^3} \big(L_0^2 + W\big), W = 2\sum_{n=1}^\infty L_{-n} L_n
\end{equation}
Here $-v = v_R$ is the negative velocity ($v > 0$), and, as before, $L_0 = \frac13 \sum_m : J^a_{-m} J^a_m:$ (see \eqref{eq:32}). 

The state $|S\rangle$ is the highest-weight state $L_0|S\rangle = S^2|S\rangle$ with the macroscopic spin $S$, $J^z_0 |S\rangle = S |S\rangle$. It satisfies $L_{n>0} |S\rangle = 0$, so that $W|S\rangle =0$, and is characterized by the finite energy density $H|S\rangle = L {\cal E}_0 |S\rangle$, where 
\begin{equation}
\label{SM-eq:E_0}
    {\cal E}_0 = - 2\pi v m^2 + \lambda m^4 ,
\end{equation}
and $m=S/L$ is the magnetization density. The minimum of \eqref{SM-eq:E_0} is achieved at $m = \sqrt{\pi v/\lambda}$.

Next, we construct an excited state, intended as a trial state for a magnon, 
\begin{equation}
    |p\rangle = J^{-}_{-p} |S\rangle, \, p > 0
\end{equation}
by lowering the spin $S$ by one unit with the current operator $J^{-}_{-p}$ which carries physical momentum $q = 2\pi p/L$.

Observe that $H |p\rangle = ([H, J^{-}_{-p}] + J^{-}_{-p} H)|S\rangle$, so that, subtracting the ground state energy $E_0 = \mathcal{E}_0 L$,
\begin{equation}
    (H - E_0)|p\rangle = [- \frac{2\pi v}{L} L_0 + \frac{\lambda}{L^3}L_0^2, J^{-}_{-p}] |S\rangle + \frac{\lambda}{L^3} [W, J^{-}_{-p}] |S\rangle,
    \label{SM-eq:e-magnon}
\end{equation}
where we split it into two contributions for future convenience. The first commutator on the right-hand side of the above equation is straightforwardly evaluated with the help of the basic commutator $[L_0, J^a_{-p}] = p J^a_{-p}$, valid for any $a$, and the defining property of the ferromagnetic state $L_0|S\rangle = S^2 |S\rangle$. We find the first contribution becomes
\begin{eqnarray}
    [- \frac{2\pi v}{L} L_0 & + & \frac{\lambda}{L^3}L_0^2, J^{-}_{-p}] |S\rangle  \nonumber\\
    &= & \frac{p}{L}\big(-2\pi v + 2 \lambda m^2\big) + \frac{\lambda p^2}{L^3}.
\end{eqnarray}
According to \eqref{SM-eq:E_0}, the first term in the second line is zero. The second term, when written in terms of the momentum $q = 2\pi p/L$, produces a vanishing contribution in the $L \to \infty$ limit, so that the first contribution entirely vanishes in the thermodynamic limit.

Therefore, \eqref{SM-eq:e-magnon} is determined by the second term. We reformulate it by noting that $J^{-}_{-p} W |S\rangle = 0$ and then projecting to the single magnon subspace, 
\begin{eqnarray}
  \Delta E_p &=& \langle p| H- E_0|p\rangle/\langle p|p\rangle \nonumber \\
  & = & \frac{\lambda}{L^3} \langle p| W  |p\rangle/\langle p|p\rangle \nonumber\\
    &=& \frac{2\lambda}{L^3} \sum_{n>0} \langle S| J^+_p L_{-n} L_n J^{-}_{-p} |S\rangle/\langle p|p\rangle,
\end{eqnarray}
where we used $W|S\rangle =0$ and $J^+_p = (J^{-}_{-p})^\dagger$. Using $L_n J^{-}_{-p} = J^{-}_{-p} L_n + [L_n, J^{-}_{-p}]$ and $[L_n, J^{-}_{-p}] = p J^{-}_{n-p}$, and its conjugate version for $J^+_p$, we obtain
\begin{equation}
\label{SM-eq:g7}
\begin{aligned}
    \langle p| W |p\rangle
    &= 2p^2 \sum_{n=1}^\infty \, \langle S| J^+_{p-n} J^{-}_{n-p}|S\rangle\\
    &= 2p^2 \sum_{k=1-p}^{2S-1} \, \langle S| J^+_{-k} J^{-}_{k}|S\rangle
\end{aligned}
\end{equation}
where we introduced a new index $k = n -p$. The upper limit of the last sum follows from $J^{-}_{k \geq 2S} |S\rangle =0$, which requires an explanation. The ferromagnetic state $|S\rangle$ is built from the vacuum $|0\rangle$ by acting on it with the spin raising operators with progressively increasing (negative) indices
\begin{equation}
    |S\rangle \propto J^+_{-(2S-1)} J^+_{-(2S-3)} \cdots J^+_{-3} J^+_{-1} |0\rangle .
    \label{SM-eq:g8}
\end{equation}
The product involves $S$ spin-raising operators $J^+$. This structure follows from the fact that in the affine Kac-Moody algebra at level $k=1$ states with ``denser" index packing are null. For example, state $|\chi\rangle = J^+_{-3} J^+_{-1} |0\rangle$ exists, and has finite norm $\langle \chi|\chi\rangle =1$. In contrast, 
$|\chi'\rangle = J^+_{-2} J^+_{-1} |0\rangle$ is a null state. That is, it has zero norm, $\langle \chi'|\chi'\rangle =0$, as follows from the application of the commutation relation $[J^+_m, J^{-}_n] = 2J^3_{m+n} + m \delta_{m+n,0}$ and the property $J^a_{n\geq 0} |0\rangle = 0$. This last property also means that, by acting on $|S\rangle$ with $J^{-}_{k}$ with sufficiently large positive index $k\geq 2S$, one can commute through all $J^+_{-n}$ terms and nullify the result. This explains the upper cutoff at $k_{\rm max} = 2S-1$ in \eqref{SM-eq:g7}. 

Another notable feature of \eqref{SM-eq:g8} is that its level, given by the sum of the indices $\sum_{k=1}^{S} (2k-1) = S^2$, is equal to the spin squared, $S^2$. This spin-level correspondence plays an important role below.

We next rewrite \eqref{SM-eq:g7} using $J^+_{-k} J^{-}_{k} = 2 J^3_0 - k + J^{-}_k J^+_{-k}$. The first two terms are easily evaluated, 
\begin{equation}
\label{SM-eq:g9}
\begin{aligned}
    \sum_{k=1-p}^{2S-1} \langle S| 2J^3_0 -k |S\rangle
    &= \sum_{k=1-p}^{2S-1} (2S -k)\\
    &= \frac12 (2S+p)(2S+p-1).
\end{aligned}
\end{equation}

Next, consider the sum involving $\langle S| J^{-}_k J^+_{-k} |S\rangle$ and split the sum over $k$ into contributions 
 over negative $k \in \{1-p,0\}$ and positive $k \in \{1,2S-1\}$. The contribution with $k \leq 0$ involves $J^+_{-k} |S\rangle = J^+_{|k|} |S\rangle =0$ because: (1) $J^+$ with the positive $|k|$ index commutes with all $J^+_{-n}$ making up the state $|S\rangle$ (it is important here that these operators have the same upper index ``+"), and ends up annihilating it, $J^+_{|k|}|0\rangle = 0$ and (2) for $k=0$, $J_0^+|S\rangle =0$ because $|S\rangle$ is a highest weight state.  

 The sum involving $\langle S| J^{-}_k J^+_{-k} |S\rangle$ with positive $k \in \{1,2S-1\}$ is less obvious but also ends up giving zero. This is so because it involves the state produced by the operator string $J^+_{-k} |S\rangle \propto J^+_{-k} J^+_{-(2S-1)} J^+_{-(2S-3)} \cdots J^+_{-3} J^+_{-1} |0\rangle$. As discussed below \eqref{SM-eq:g8}, for this combination not to be a null state, its level, which is given by $k + S^2$, must be greater than or equal to its spin squared, $(S+1)^2$. That is, the minimal $k$ required for this must satisfy $k = (S+1)^2 - S^2 = 2S+1$. But the sum ends at $k_{\rm max} = 2S-1$, and, therefore, every state appearing in the sum is null,
 \begin{equation}
   \sum_{k=1}^{2S-1} \langle S| J^{-}_k J^+_{-k} |S\rangle  = 0.
 \end{equation}
 Therefore, \eqref{SM-eq:g7} reduces to $2p^2$ times \eqref{SM-eq:g9}: $p^2 (2S+p)(2S+p-1)$. Dividing by the norm of the state $|p\rangle$, $\langle p|p\rangle = 2S+p$, we finally arrive at 
 \begin{equation}
     \Delta E_p = \frac{\lambda}{L^3} p^2 (2S + p-1).
 \end{equation}
 Expressing the index $p$ via the physical momentum $q$, $p = L q/(2\pi)$, and using $S = m L$, this turns, in the $L\rightarrow \infty$ limit, into the final result for the variational magnon dispersion
 \begin{equation}
     \Delta E_p = \frac{\lambda m}{2\pi^2} q^2 + \frac{\lambda }{8\pi^3} q^3.
     \label{SM-eq:g12}
 \end{equation}
 Note that it is fully controlled by the interaction $\lambda$.

We conclude this section with a few remarks.  First, the above calculation is for the \emph{expectation value} of the Hamiltonian in the trial magnon state $|p\rangle$.  This does not mean that $|p\rangle$ is an exact eigenstate (as can sometimes be constructed in simple ferromagnets).  In fact, one can show that it is \emph{not} an eigenstate, and indeed the present argument does not fix the spectral weight of the magnon.  Our expectation is that for sufficiently small $q$ a single magnon excitation disperses quadratically with a mass which is upper bounded by the estimate from $\Delta E_p$, and becomes quasiparticle-like for small $q$.  Second, the above calculation is only for positive magnon momentum $q$, but a similar calculation holds as well for negative $q$, confirming that the quadratic magnon exists and is approximately symmetric at small momentum (the cubic term provides an asymmetry).

\section{Effective Hamiltonian}
\label{SM-sec:effective}

The ferromagnetic magnon dispersion admits a simple physical interpretation in terms of the effective Hamiltonian, which captures the smooth deformation of the quantum ground state $|S\rangle$ with finite spin polarization. Namely, we postulate a semiclassical ansatz for the local magnetization vector $m^a(x) = \langle S| J^a(x)|S\rangle$ and the effective classical low-energy Hamiltonian
\begin{equation}
    H_{\rm eff} = \int dx \, \frac12 \rho (\partial_x \vec{m})^2
\end{equation}
where $\rho$ is the stiffness. The dynamics of the $\vec{m}$ field follows from the Poisson bracket 
\begin{eqnarray}
    \{m^+(x), m^-(y)\} &=& - 2i m^z(x) \delta(x-y) - \frac{1}{2\pi} \partial_x \delta(x-y), \nonumber\\
    \{m^+(x), m^z(y)\} &=& i m^+(x) \delta(x-y)
\end{eqnarray}
which the $\vec{m}$ field inherits from the Kac-Moody commutators of the right-moving currents
\begin{equation}
    [J^a(x), J^b(y)] = i \epsilon^{abc} J^c(x) \delta(x-y) - \frac{i}{4\pi} \partial_x \delta(x-y) \delta^{ab} 
\end{equation}
as well as the commutator-Poisson bracket correspondence: $\{A, B\} \to -i [A,B]$. 

The equation of motion $\partial_t m^+(x) = \{m^+(x), H_{\rm eff}\}$ follows right away
\begin{equation}
    \partial_t m^+ =  i \rho\big(m^z \partial_x^2 m^+ - m^+\partial_x^2 m^z\big) + \frac{\rho}{4\pi} \partial_x^3 m^+
\end{equation}
For small fluctuations about the ordered state with $m^z(x) \approx m$, the derivative of $m^z$ can be set to zero. The plane-wave ansatz $m^+(x,t) \propto e^{i q x - i \omega t}$ leads to the dispersion relation of the chiral magnon:
\begin{equation}
    \omega = \rho m q^2 + \frac{\rho}{4\pi} q^3 \, .
    \label{SM-eq:h5}
\end{equation}
Notably, the ratio of the coefficients of the $m q^2$ and $q^3$ terms is $4\pi$, matching the result of the algebraic consideration. The present analysis shows that the cubic term in the dispersion originates from the ``quantum anomaly" $\partial_x \delta(x-y)$ in the Kac-Moody commutator.

Comparing \eqref{SM-eq:h5} with \eqref{SM-eq:g12}, we obtain $\rho = \lambda/(2\pi^2)$.

\section{Continuum many-boson formulation}
\label{SM-sec:continuum-many-boson}

This appendix gives a complementary many-boson form of the same chiral problem studied in Secs.~\ref{SM-sec:deriv-i_3-form}~and~\ref{SM-sec:i_3-spectrum-circle}.  The starting point is not a new continuum field theory, but the Kac-Moody oscillator representation already introduced above.  We first rewrite the finite-size Hamiltonian directly in terms of boson creation and annihilation operators at integer momenta.  The continuum integral equations then follow by taking the large-$L$ limit at fixed physical momentum.  This form is useful for numerical calculations of the scaling functions in Eqs.~\eqref{eq:30}~and~\eqref{eq:31}, because it keeps the total momentum fixed and converts the problem to finite-dimensional approximations on momentum simplices.

\subsection{Finite $L$, discrete momenta}

We work in the right-moving sector and set $v_R=0$.  Recall from Eq.~\eqref{SM-eq:20} that the abelian current modes obey
\begin{equation}
    [J_m,J_n]=\frac{m}{2}\delta_{m+n,0}.
    \label{SM-eq:abelian-current-algebra-final-app}
\end{equation}
Thus, for $n>0$, it is natural to define canonical boson operators by
\begin{equation}
    J_n=\sqrt{\frac{n}{2}}\,a_n,
    \qquad
    J_{-n}=\sqrt{\frac{n}{2}}\,a_n^\dagger,
    \qquad
    [a_n,a_m^\dagger]=\delta_{nm}.
    \label{SM-eq:Jn-to-an}
\end{equation}
The vacuum condition $J_n|0\rangle=0$ for $n>0$ becomes the usual oscillator vacuum condition $a_n|0\rangle=0$.

The current form of the critical Hamiltonian was derived in Eqs.~\eqref{SM-eq:73}~and~\eqref{SM-eq:74}.  Equivalently,
\begin{align}
    H_3&=\frac{\lambda}{L^3}I_3,
    \nonumber\\
    I_3&=
    \sum_{\{n_i\}\in\mathbb Z}
    :J_{n_1}J_{n_2}J_{n_3}J_{n_4}:
    \delta_{n_1+n_2+n_3+n_4,0}
    \nonumber\\
    &\quad
    +2\sum_{n=1}^\infty n^2J_{-n}J_n .
    \label{SM-eq:I3-current-discrete}
\end{align}
The second term immediately gives the one-boson cubic dispersion
\begin{equation}
    2\sum_{n>0}n^2J_{-n}J_n
    =
    \sum_{n>0}n^3 a_n^\dagger a_n^{\vphantom\dagger} .
    \label{SM-eq:discrete-cubic-dispersion}
\end{equation}
The normal-ordered quartic term conserves the total integer momentum
\begin{equation}
    N=\sum_{n>0} n\,a_n^\dagger a_n^{\vphantom\dagger},
    \label{SM-eq:integer-total-momentum}
\end{equation}
but it can change the number of bosons by two.  Written in oscillator variables, it contains only the number-preserving $2\leftrightarrow2$ process and the number-changing $1\leftrightarrow3$ process:
\begin{align}
    I_3
    &=
    \sum_{n>0}n^3a_n^\dagger a_n^{\vphantom\dagger}
    +
    \frac{3}{2}
    \sum_{\substack{p,q,k,l>0\\ p+q=k+l}}
    \sqrt{pqkl}\,
    a_p^\dagger a_q^\dagger
    a_k^{\vphantom\dagger}a_l^{\vphantom\dagger}
    \nonumber\\
    &\quad
    +
    \sum_{\substack{p,q,r,k>0\\ p+q+r=k}}
    \sqrt{pqrk}\,
    \left(
        a_p^\dagger a_q^\dagger a_r^\dagger
        a_k^{\vphantom\dagger}
        +
        a_k^\dagger
        a_p^{\vphantom\dagger}
        a_q^{\vphantom\dagger}
        a_r^{\vphantom\dagger}
    \right).
    \label{SM-eq:I3-discrete-oscillator}
\end{align}
The factors $3/2$ and $1$ come directly from the six ways to choose two creation and two annihilation currents, and the four ways to choose one annihilation current among three creation currents, together with the factors $\sqrt{n/2}$ in Eq.~\eqref{SM-eq:Jn-to-an}.  Because the boson number changes by $0$ or $\pm2$, the parity of the boson number is conserved.  The spin-current spectral function starts from the one-boson state, so only the odd sectors are needed.

At fixed total integer momentum $N$, a general odd-sector state can therefore be written as
\begin{equation}
    |\Psi_N\rangle
    =
    \sum_{\substack{r=1,3,5,\ldots}}
    \frac{1}{r!}
    \sum_{\substack{n_i>0\\ \sum_i n_i=N}}
    \psi_r(n_1,\ldots,n_r)
    a_{n_1}^\dagger\cdots a_{n_r}^\dagger |0\rangle ,
    \label{SM-eq:discrete-many-boson-state}
\end{equation}
with completely symmetric wavefunctions $\psi_r$.  With the $1/r!$ convention in Eq.~\eqref{SM-eq:discrete-many-boson-state}, the norm is
\begin{equation}
    \langle \Psi_N|\Psi_N\rangle
    =
    \sum_{\substack{r=1,3,5,\ldots}}
    \frac{1}{r!}
    \sum_{\substack{n_i>0\\ \sum_i n_i=N}}
    |\psi_r(n_1,\ldots,n_r)|^2 .
    \label{SM-eq:discrete-wavefunction-norm}
\end{equation}
This is already a finite-dimensional problem for each $N$, and is the integer-momentum precursor of the continuum formulation below.

\subsection{Continuum limit}

Now take the large-$L$ limit with fixed physical momentum
\begin{equation}
    q=\frac{2\pi N}{L},
    \qquad
    p_i=\frac{2\pi n_i}{L}.
\end{equation}
We choose the continuum normalization $[a_p,a_{p'}^\dagger]=2\pi\delta(p-p')$, with the harmless powers of $2\pi$ absorbed into the definition of the continuum coupling $\lambda$.  The oscillator Hamiltonian inherited from Eq.~\eqref{SM-eq:I3-discrete-oscillator} is then
\begin{equation}
    H_3=H_0+H_{22}+H_{13}.
    \label{SM-eq:continuum-oscillator-H}
\end{equation}
The three terms are, with all integration variables in $H_{22}$ and $H_{13}$ positive,
\begin{align*}
    H_0&=
    \lambda
    \int_0^\infty\frac{dp}{2\pi}\,
    p^3a_p^\dagger a_p^{\vphantom\dagger}
    \\
    H_{22}&=
    \frac{3\lambda}{2}
    \int
    \frac{dp_1dp_2dk_1dk_2}{(2\pi)^3}
    \delta(p_1+p_2-k_1-k_2)
    \\
    &\quad\times
    \sqrt{p_1p_2k_1k_2}\,
    a_{p_1}^\dagger a_{p_2}^\dagger
    a_{k_1}^{\vphantom\dagger}a_{k_2}^{\vphantom\dagger}
    \\
    H_{13}&=
    \lambda
    \int
    \frac{dp_1dp_2dp_3dk}{(2\pi)^3}
    \delta(p_1+p_2+p_3-k)
    \\
    &\quad\times
    \sqrt{p_1p_2p_3k}
    \left(
    a_{p_1}^\dagger a_{p_2}^\dagger
    a_{p_3}^\dagger a_k^{\vphantom\dagger}
    +{\rm h.c.}
    \right).
\end{align*}
This Hamiltonian is homogeneous of degree three in the momenta.  Hence, on the fixed-total-momentum subspace, its eigenvalues have the form $E=\lambda q^3\epsilon$ in the normalization used below.

At fixed total momentum $q$, a general state in the spin-current sector can be expanded as
\begin{align}
    |\Psi_q\rangle
    &=
    \sum_{\substack{r=1,3,5,\ldots}}
    \frac{1}{r!}
    \int_{p_i>0}\!\!\prod_{i=1}^r \frac{dp_i}{2\pi}\,
    \delta\!\left(q-\sum_{i=1}^r p_i\right)
    \nonumber\\
    &\quad\times
    \psi_r(p_1,\ldots,p_r)
    a_{p_1}^\dagger\cdots a_{p_r}^\dagger |0\rangle .
    \label{SM-eq:many-boson-state}
\end{align}
Only odd $r$ appear when the state is generated from a single current operator, because the quartic interaction changes the number of bosons by $0$ or $\pm 2$.  The wavefunctions $\psi_r$ are completely symmetric.  The continuum normalization corresponding to Eq.~\eqref{SM-eq:discrete-wavefunction-norm} is
\begin{align}
    \langle \Psi_q|\Psi_{q'}\rangle
    &=
    2\pi\delta(q-q')\,\mathcal{N}_q,
    \nonumber\\
    \mathcal{N}_q
    &=
    \sum_{\substack{r=1,3,5,\ldots}}
    \frac{1}{r!}
    \int_{p_i>0}\!\!\prod_{i=1}^r\frac{dp_i}{2\pi}\,
    \delta\!\left(q-\sum_i p_i\right)
    \nonumber\\
    &\quad\times
    |\psi_r(p_1,\ldots,p_r)|^2 .
    \label{SM-eq:continuum-wavefunction-norm}
\end{align}

We next scale out the total momentum by writing $p_i=q x_i$, where
\begin{equation}
    x_i>0,\qquad \sum_{i=1}^r x_i=1.
    \label{SM-eq:simplex-def}
\end{equation}
Thus each $r$-boson sector is represented on the simplex $\Delta_r$.  The square-root factors generated by the boson oscillator normalization are removed by writing
\begin{equation}
    \psi_r(qx_1,\ldots,qx_r)
    =
    q^{-(r-1)/2}
    \sqrt{x_1x_2\cdots x_r}\,
    P_r(x_1,\ldots,x_r),
    \label{SM-eq:P-convention}
\end{equation}
where $P_r$ is again symmetric.  In this convention the inner product in the $r$-boson sector is
\begin{equation}
    \langle P_r,Q_r\rangle_r
    =
    \int_{\Delta_r} d^{r-1}x\,
    \left(\prod_{i=1}^r x_i\right)
    P_r(x)^* Q_r(x),
    \label{SM-eq:P-inner-product}
\end{equation}
up to an overall $r$-dependent convention for symmetrized states.  With the same convention used in every sector, the Hamiltonian blocks below are mutually adjoint with respect to Eq.~\eqref{SM-eq:P-inner-product}.

\subsection{First-quantized Hamiltonian}

\subsubsection{Block structure}

We now convert Eq.~\eqref{SM-eq:continuum-oscillator-H} into dimensionless integral equations for the functions $P_r$.  Dividing out the overall factor $\lambda q^3$, the eigenvalue equation has the block form
\begin{align}
    \epsilon P_r
    &=
    H_{r,r-2}P_{r-2}
    +
    H_{rr}P_r
    +
    H_{r,r+2}P_{r+2},
    \nonumber\\
    &\hspace{2.5cm} r=1,3,5,\ldots ,
    \label{SM-eq:block-eigen-equation}
\end{align}
where absent sectors are omitted.  Here $H_{rr}$ preserves the number of bosons, while $H_{r+2,r}$ and $H_{r,r+2}$ are the splitting and fusion blocks which change the boson number by two.

\subsubsection{Same-sector block}

The same-sector block $H_{rr}$ receives contributions from $H_0$ and $H_{22}$ in Eq.~\eqref{SM-eq:continuum-oscillator-H}.  The quadratic term $H_0$ gives the multiplication operator $\sum_i x_i^3$.  For the $2\leftrightarrow2$ part $H_{22}$, pick a pair of momenta $x_i,x_j$ in the external $r$-boson configuration and let $s_{ij}=x_i+x_j$.  The two annihilated momenta are replaced by a new pair $u,s_{ij}-u$ with the same total momentum, while all other momenta are spectators.  The square-root factors from Eq.~\eqref{SM-eq:continuum-oscillator-H} combine with the envelope in Eq.~\eqref{SM-eq:P-convention} so that
\begin{equation}
    \sqrt{x_ix_j\,u(s_{ij}-u)}
    \frac{\sqrt{u(s_{ij}-u)\prod_{\widehat{ij}}x}}
         {\sqrt{x_ix_j\prod_{\widehat{ij}}x}}
    =
    u(s_{ij}-u).
    \label{SM-eq:two-two-envelope-cancel}
\end{equation}
Including the identical-boson normalization factors from Eq.~\eqref{SM-eq:many-boson-state}, the raw coefficient $3/2$ in $H_{22}$ becomes $3$.  Thus
\begin{align}
    (H_{rr}P_r)&(x_1,\ldots,x_r)
    =
    \left(\sum_{i=1}^r x_i^3\right)P_r(x_1,\ldots,x_r)
    \nonumber\\
    &\quad
    +3\sum_{i<j}
    \int_0^{s_{ij}}\!\!du\,u(s_{ij}-u)\,
    P_r(u,s_{ij}-u,\widehat{x}_{ij}),
    \label{SM-eq:Hrr-continuum}
\end{align}
where $s_{ij}=x_i+x_j$ and $\widehat{x}_{ij}$ denotes the remaining $r-2$ spectator momenta.  The argument of $P_r$ in the second line is understood symmetrically, so its order is immaterial.

\subsubsection{Number-changing blocks}

The number-changing blocks come from $H_{13}$.  First consider the splitting process, which maps an $r$-boson wavefunction to an $(r+2)$-boson wavefunction.  In a final configuration $x_1,\ldots,x_{r+2}$, choose the triple $T$ of daughter momenta which came from one parent boson.  Momentum conservation fixes the parent to have fraction $s_T=\sum_{i\in T}x_i$.  The square-root factors from Eq.~\eqref{SM-eq:continuum-oscillator-H} combine with the envelope factors as
\begin{equation}
    \sqrt{s_T\prod_{i\in T}x_i}\,
    \frac{\sqrt{s_T\prod_{\overline{T}}x}}
         {\sqrt{\prod_{i\in T}x_i\prod_{\overline{T}}x}}
    =
    s_T .
    \label{SM-eq:one-three-envelope-cancel}
\end{equation}
After including the identical-boson normalization factors, the coefficient of this block is $\sqrt{6}$.  Hence
\begin{align}
    (H_{r+2,r}P_r)(x_1,\ldots,x_{r+2})
    &=
    \sqrt{6}
    \sum_{|T|=3}
    s_T\,
    P_r(s_T,x_{\overline{T}}),
    \label{SM-eq:Hrplus2r}
\end{align}
with $s_T=\sum_{i\in T}x_i$ and $x_{\overline{T}}$ the $r-1$ spectator coordinates not in $T$.  The reverse block is the adjoint with respect to Eq.~\eqref{SM-eq:P-inner-product}.  Equivalently, for $y\in\Delta_r$,
\begin{align}
    (H_{r,r+2}P_{r+2})(y_1,\ldots,y_r)
    &=
    \sqrt{6}
    \sum_{i=1}^r
    \int_{\substack{z_1,z_2>0\\ z_1+z_2<y_i}}
    dz_1\,dz_2
    \nonumber\\
    &\quad\times
    z_1z_2z_3
    P_{r+2}(z_1,z_2,z_3,y_{\widehat{i}}),
    \label{SM-eq:Hrrplus2}
\end{align}
where $z_3=y_i-z_1-z_2$.  The factor $z_1z_2z_3$ is the product of the three daughter momenta left by the weighted inner product measure after taking the adjoint of Eq.~\eqref{SM-eq:Hrplus2r}.

\subsubsection{One-boson sector}

The one-boson sector is one-dimensional.  In the normalization used here
\begin{equation}
    H_{11}=1,
    \qquad
    (H_{31}P_1)(x_1,x_2,x_3)=\sqrt{6}\,P_1 .
    \label{SM-eq:H11-H31}
\end{equation}
Equations \eqref{SM-eq:block-eigen-equation}--\eqref{SM-eq:H11-H31} give the desired $1+3+5+\cdots$ continuum Hamiltonian.

\subsection{Truncation and discretization}

A finite boson-number truncation keeps only
\begin{equation}
    r=1,3,5,\ldots,R_{\rm max}.
    \label{SM-eq:boson-number-truncation}
\end{equation}
For example, the $1+3+5$ approximation keeps $P_1,P_3,P_5$ and drops the coupling from $P_5$ to $P_7$; the $1+3+5+7$ approximation keeps one more block.  This is not a low-energy projection in the finite-size Virasoro sense.  Rather, it is a Fock-space truncation of the continuum integral equations at fixed total momentum.  Increasing $R_{\rm max}$ tests the effect of allowing the strongly interacting chiral boson to split into more partons of the total momentum.

To solve the integral equations numerically, we replace each simplex by a finite set of grid points.  Choose a positive integer $M$ and write
\begin{equation}
    x_i=\frac{m_i}{M},
    \qquad
    m_i\in \mathbb{Z}_{>0},
    \qquad
    \sum_{i=1}^r m_i=M .
    \label{SM-eq:simplex-grid}
\end{equation}
Thus the continuum simplex integral is approximated by a finite weighted sum over rational points with denominator $M$.  This weighted sum is the quadrature rule.  Since the wavefunctions are symmetric, all permutations of a tuple $(m_1,\ldots,m_r)$ represent the same bosonic configuration.  We therefore store only one sorted representative of each orbit.  If $\alpha=(m_1,\ldots,m_r)$ is such a representative, its quadrature weight is
\begin{equation}
    \mu_\alpha^{(r)}
    =
    \nu_\alpha\,
    \frac{{\rm Vol}(\Delta_r)}{N_r(M)}
    \prod_{i=1}^r \frac{m_i}{M},
    \qquad
    N_r(M)=\binom{M-1}{r-1},
    \label{SM-eq:grid-weight}
\end{equation}
where $\nu_\alpha$ is the number of distinct permutations of the tuple and ${\rm Vol}(\Delta_r)=1/(r-1)!$.  In other words, $\mu_\alpha^{(r)}$ is the volume element assigned to the grid point $\alpha$, including both the multiplicity of its permutations and the measure factor $\prod_i x_i$ in Eq.~\eqref{SM-eq:P-inner-product}.  With this rule,
\begin{equation}
    \langle P_r,Q_r\rangle_r
    \approx
    \sum_{\alpha\in\Delta_r(M)}
    \mu_\alpha^{(r)}
    P_r(\alpha)^*Q_r(\alpha),
    \label{SM-eq:grid-inner-product}
\end{equation}
where $\Delta_r(M)$ denotes the set of sorted grid representatives.

The same-sector integral in Eq.~\eqref{SM-eq:Hrr-continuum} is evaluated on the same denominator-$M$ grid.  If $s_{ij}=\sigma_{ij}/M$, then
\begin{align}
    &\int_0^{s_{ij}}du\,u(s_{ij}-u)\,
    \nonumber\\
    &\quad\times
    P_r(u,s_{ij}-u,x_{\widehat{ij}})
    \;\longrightarrow\;
    \frac{1}{M}
    \sum_{\ell=1}^{\sigma_{ij}-1}
    \frac{\ell}{M}\frac{\sigma_{ij}-\ell}{M}\,
    \nonumber\\
    &\quad\times
    P_r\!\left(\ell/M,(\sigma_{ij}-\ell)/M,x_{\widehat{ij}}\right).
    \label{SM-eq:grid-pair-rule}
\end{align}
The number-changing block Eq.~\eqref{SM-eq:Hrplus2r} needs no additional integration: every triple of daughter grid momenta has a parent momentum $s_T$ which is again on the denominator-$M$ grid.

After these replacements, each operator block is just a finite matrix acting on the values of the wavefunctions at the grid points.  Let $L^{(rs)}$ denote the raw matrix which maps values in the $s$-boson sector to values in the $r$-boson sector.  Because the grid inner product contains the weights $\mu_\alpha^{(r)}$, it is convenient to pass to an orthonormal weighted basis by defining
\begin{equation}
    A_{\alpha\beta}^{(rs)}
    =
    \sqrt{\mu_\alpha^{(r)}}\,
    L_{\alpha\beta}^{(rs)}
    \frac{1}{\sqrt{\mu_\beta^{(s)}}}.
    \label{SM-eq:orthonormal-grid-matrix}
\end{equation}
This is simply the matrix representation of the same operator in variables for which the discrete inner product is ordinary Euclidean dot product.  For the diagonal blocks one uses the weighted-symmetric part of this matrix, while the off-diagonal blocks are inserted as adjoint pairs.  The result is a real symmetric finite matrix in the truncated $1+3+5+\cdots+R_{\rm max}$ space.

\subsection{Spectral function}

We now connect the finite matrix problem to the right-moving spin-current spectral function.  For $q>0$, define the broadened real-frequency correlator
\begin{align}
    S_\eta(q,\omega)
    &=
    \frac{1}{\pi}\,
    {\rm Re}
    \int_{-\infty}^{\infty}\!\!dx
    \int_0^\infty\!\!dt\,
    e^{i\omega t-\eta t}e^{-iqx}
    \nonumber\\
    &\quad\times
    \langle 0|J_R(x,t)J_R(0,0)|0\rangle ,
    \label{SM-eq:broadened-current-definition}
\end{align}
with $\eta>0$.  With the continuum oscillator convention used above,
\begin{equation}
    J_R(x)
    =
    \int_0^\infty\frac{dp}{2\pi}
    \sqrt{\frac{p}{2}}\,
    \left(
        a_p^{\vphantom\dagger} e^{ipx}
        +
        a_p^\dagger e^{-ipx}
    \right),
    \label{SM-eq:current-continuum-final}
\end{equation}
so that the Fourier component which creates a positive-momentum excitation is
\begin{equation}
    J_q\equiv \int dx\,e^{iqx}J_R(x)
    =
    \sqrt{\frac{q}{2}}\,a_q^\dagger ,
    \qquad q>0 .
    \label{SM-eq:Jq-source}
\end{equation}
while $J_{-q}=\sqrt{q/2}\,a_q^{\vphantom\dagger}$ annihilates this excitation.  The opposite sign convention for the spatial Fourier transform simply interchanges $q$ and $-q$.
Thus the current creates a one-boson state.  In the first-quantized language of Eqs.~\eqref{SM-eq:many-boson-state}~and~\eqref{SM-eq:P-convention}, this source is simply the vector with support only in the $r=1$ sector, with $P_1=1$, up to the overall factor $\sqrt{q/2}$.

Using time evolution with the critical Hamiltonian and taking the ground-state energy to be zero, Eq.~\eqref{SM-eq:broadened-current-definition} becomes the resolvent matrix element
\begin{equation}
    S_\eta(q,\omega)
    =
    \frac{1}{\pi}\,
    {\rm Re}\,
    \langle 0|
    J_{-q}
    \frac{1}{\eta-i(\omega-H_3)}
    J_q
    |0\rangle .
    \label{SM-eq:current-resolvent}
\end{equation}
Since $J_q|0\rangle=\sqrt{q/2}\,|e_q\rangle$ and $H_3=\lambda q^3 A$ in the fixed-$q$ first-quantized space, this can be written as
\begin{equation}
    S_\eta(q,\omega)
    =
    \frac{1}{2\lambda q^2}\,
    G_0\!\left(
        \frac{\omega}{\lambda q^3},
        \frac{\eta}{\lambda q^3}
    \right),
    \label{SM-eq:S-to-G0}
\end{equation}
where the dimensionless line shape is
\begin{equation}
    G_0(\Omega,\bar{\eta})
    =
    \frac{1}{\pi}\,
    {\rm Re}\,
    \langle e|
    \frac{1}{\bar{\eta}-i(\Omega-A)}
    |e\rangle ,
    \label{SM-eq:grid-resolvent-spectral-function}
\end{equation}
and $|e\rangle$ is the normalized weighted-grid vector corresponding to the one-boson source $P_1=1$.  The prefactor in Eq.~\eqref{SM-eq:S-to-G0} is specific to the current normalization in Eq.~\eqref{SM-eq:current-continuum-final}; its $q^{-2}$ dependence is the scaling behavior quoted in Eq.~\eqref{eq:30}.

Equivalently, if $\epsilon_\alpha$ are the dimensionless eigenvalues of $A$ and $w_\alpha=|\langle \alpha|e\rangle|^2$ are the squared overlaps with the one-boson source, then
\begin{equation}
    G_0(\Omega,\bar{\eta})
    =
    \frac{1}{\pi}
    \sum_\alpha
    w_\alpha
    \frac{\bar{\eta}}{(\Omega-\epsilon_\alpha)^2+\bar{\eta}^2}.
    \label{SM-eq:grid-spectral-function}
\end{equation}
The continuum limit is approached by increasing $M$ at fixed $R_{\rm max}$, followed by increasing the maximum retained boson number.

\subsection{Lanczos evaluation}

For small grids one may diagonalize the finite matrix $A$ directly.  For the larger grids used in the continuum extrapolation, however, this is unnecessary.  The spectral function only requires the scalar resolvent in Eq.~\eqref{SM-eq:grid-resolvent-spectral-function}, i.e. the response of the normalized one-boson source $|e\rangle$.  We therefore apply the Lanczos algorithm with starting vector
\begin{equation}
    |q_1\rangle = |e\rangle .
    \label{SM-eq:lanczos-start}
\end{equation}
After $K$ Lanczos steps one obtains an orthonormal Krylov basis
\begin{equation}
    {\cal K}_K(A,e)
    =
    {\rm span}\left\{
        |e\rangle,
        A|e\rangle,
        A^2|e\rangle,
        \ldots,
        A^{K-1}|e\rangle
    \right\},
    \label{SM-eq:krylov-space}
\end{equation}
in which $A$ is represented by a real symmetric tridiagonal matrix
\begin{equation}
    T_K=
    \begin{pmatrix}
        \alpha_1 & \beta_1 \\
        \beta_1 & \alpha_2 & \beta_2 \\
        & \beta_2 & \alpha_3 & \ddots \\
        && \ddots & \ddots
    \end{pmatrix}.
    \label{SM-eq:lanczos-tridiagonal}
\end{equation}
The projected resolvent is then
\begin{equation}
    G_0^{(K)}(\Omega,\bar{\eta})
    =
    \frac{1}{\pi}\,
    {\rm Re}\,
    \langle e_1|
    \frac{1}{\bar{\eta}-i(\Omega-T_K)}
    |e_1\rangle ,
    \label{SM-eq:lanczos-resolvent}
\end{equation}
where $|e_1\rangle=(1,0,\ldots,0)^T$ is the first basis vector in Krylov space.  If $\theta_\alpha$ are the eigenvalues of $T_K$ and $u_{\alpha 1}$ is the first component of the corresponding normalized eigenvector, this becomes the pole expansion
\begin{equation}
    G_0^{(K)}(\Omega,\bar{\eta})
    =
    \frac{1}{\pi}
    \sum_{\alpha=1}^{K}
    |u_{\alpha 1}|^2
    \frac{\bar{\eta}}{(\Omega-\theta_\alpha)^2+\bar{\eta}^2}.
    \label{SM-eq:lanczos-pole-expansion}
\end{equation}
Thus the calculation stores the Lanczos pole positions $\theta_\alpha$ and weights $|u_{\alpha 1}|^2$.  Once these are known, curves for different broadenings $\bar{\eta}$ can be generated without repeating the matrix construction or the Lanczos recursion.

\begin{figure}[htbp]
  \centering
  \includegraphics[width=\columnwidth]{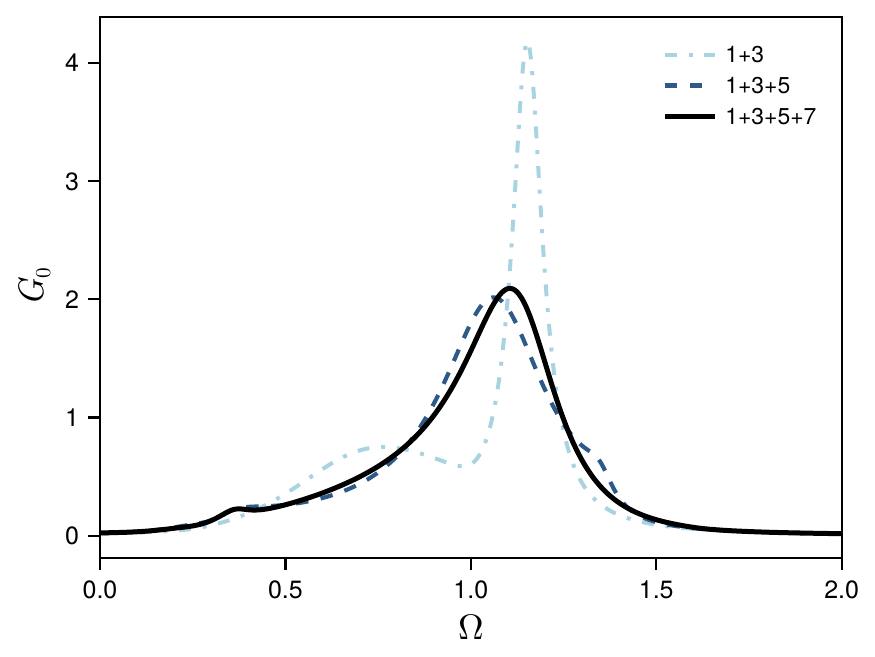}
  \caption{Convergence of the continuum many-boson approximation to the spin-current spectral function.  The curves compare calculations in which successively higher odd boson-number sectors are retained, using the same simplex grid with $M=140$ and broadening $\bar{\eta}=0.05$.  Including more sectors allows the current excitation to split into more right-moving bosons and tests convergence of the Fock-space truncation.}
  \label{SM-fig:many-boson-sector-compare}
\end{figure}

\section{Computational details and further results}
\label{SM-sec:comp-deta-furth}

\subsection{Method}
\label{SM-sec:method}

All matrix-product-state calculations were performed with the Julia
ITensor and ITensorMPS libraries, using codes written with AI assistance.
Spectral functions were obtained using a system of $N=300$ sites with open boundary conditions, and using quantum-number-conserving MPS
tensors that conserve total $S^z$.

The initial (ground) state was prepared by
DMRG as follows.  First, a probe DMRG calculation was performed in
the $S^z_{\rm tot}=0$ sector.  The resulting state was used to
evaluate $\langle {\bf S}_{\rm tot}^2\rangle$, from which the total
spin $S$ was inferred by solving
$S(S+1)=\langle {\bf S}_{\rm tot}^2\rangle$.  For $g \leq 2/\pi$, the total spin is zero, and the obtained state serves as the initial state.  For $g>2/\pi$, $S \neq 0$ and the calculation proceeded to a second stage.  We target the highest-weight sector $S^z_{\rm tot}=S$, which depends on
$g$.  A second DMRG calculation was then performed directly in this
highest-weight sector, yielding the state used for time evolution.
The DMRG truncation cutoff was $10^{-10}$.  We used an adaptive
sweep schedule with maximum bond dimensions
\begin{equation}
  \chi_{\rm max}=10,20,40,80,160,300,500,800,1200,
\end{equation}
minimum/maximum sweep counts $6/24$ at non-final bond dimensions and
$8/40$ at the final bond dimension.  The adaptive stopping criteria
used an absolute energy tolerance $10^{-8}$, relative energy tolerance
$10^{-10}$, final truncation-error target $10^{-8}$, and patience of
two consecutive converged sweeps.  In the production runs presented here
the final highest-weight states had $S^z_{\rm tot}=34$ for $g=1$ and
$S^z_{\rm tot}=57$ for $g=4$.  The realized maximum bond dimension
of the converged state was smaller than the largest allowed value when
the truncation target was reached before saturating $\chi_{\rm max}$.

Real-time dynamics were computed from this highest-weight ground state
using one-site TDVP.  For each correlator, a local operator was applied
at the center site $c=N/2=150$, producing the initial time-evolved state
$O_c|\psi_0\rangle$.  We computed the raw space-time correlators
\begin{equation}
  C_j^{AB}(t) =
  e^{i E_0 t}
  \langle \psi_0 | A_j e^{-iHt} B_c |\psi_0\rangle ,
\end{equation}
for the three choices $(A,B)=(S^+,S^-)$, $(S^-,S^+)$, and $(S^z,S^z)$.
The equal-time row was evaluated directly from the ground-state
correlation matrix.  The important parameters for the time evolution
were the time step $\Delta t=0.1$, the final time $t_{\rm max}=100$,
TDVP truncation cutoff $10^{-9}$, and maximum TDVP bond dimension
$\chi_{\rm max}=800$.

The spectral function was obtained from the connected space-time data
using a one-sided damped transform.  Before transforming in space, the
real-space correlator was multiplied by a Gaussian window centered on the
source site, with width $\sigma_x=N/4$, and shifted so that the source
site defined the origin.  For each spatial Fourier component $C_q(t)$ we
then evaluated
\begin{equation}
  S^{AB}_\eta(q,\omega) =
  \frac{1}{\pi}
  {\rm Re}
  \sum_n w_n e^{(i\omega-\eta)t_n} C_q^{AB}(t_n),
\end{equation}
where $w_n$ are trapezoidal time-integration weights.  The spectra shown
use $\eta=0.10$, frequency spacing $\Delta\omega=0.02$, and the positive
part $\max(S^{AB}_\eta(q,\omega),0)$ on $0\leq\omega\leq15$.

\subsection{Results for the ground state}
\label{SM-sec:ground-state-results}

We measured the single-site expectation value, $\langle S_i^z\rangle$ in the highest-weight ground state in the ordered phase, with open boundary conditions.  A characteristic Friedel-like oscillation appears, as is well known in other studies of quantum spin chains.  As shown by Hikihara and Furusaki\cite{hikihara2004}, the massless free boson theory of a magnetized spin chain leads to the form
\begin{equation}
\begin{aligned}
  \langle S^z_x\rangle & = m \\
  &+ A(-1)^x \sin\!\left[2\pi m\left(x-\tfrac{N+1}{2}\right)\right]
  \left(\tfrac{2N+1}{\pi}\sin\tfrac{\pi x}{N+1}\right)^{-a},
\end{aligned}
\label{SM-eq:Hikihara}
\end{equation}
where $N$ is the length of the chain, $m$ is the average magnetization, $A$ is an amplitude, and $a$ is an exponent related to the Luttinger parameter.  Although our system is chiral, the same form is expected to hold in our model. In terms of the parameterization \eqref{SM-eq:n-amended}, $A=C_1$ and $a=1/(8K)$, where $K$ is given by \eqref{SM-eq:eta-satellites}. The excellent fit is a validation of the $z=1$ Luttinger liquid sector of the present theory.
\begin{figure}[htbp]
  \centering
  \includegraphics[width=\columnwidth]{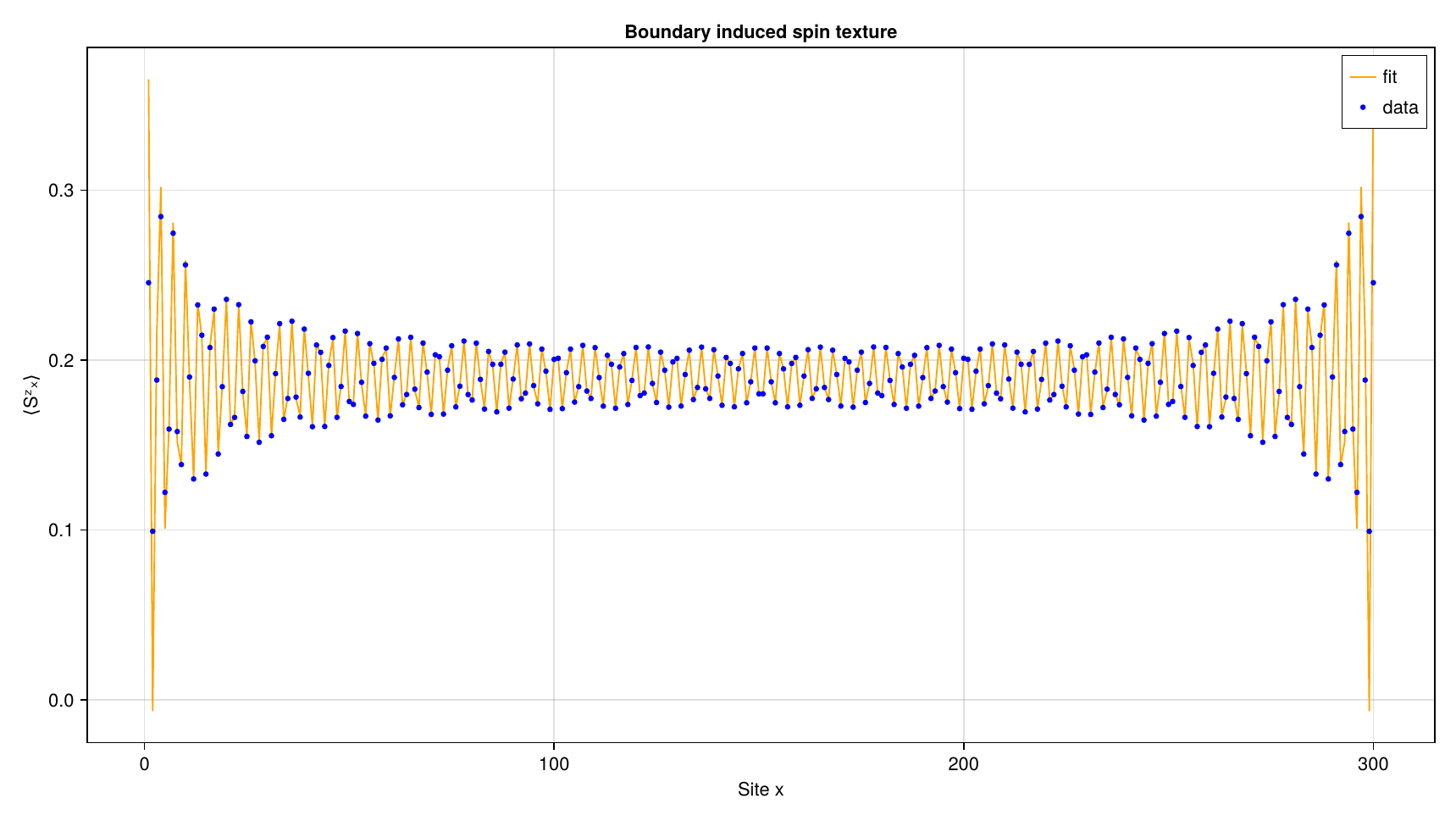}
  \caption{Boundary-induced spin texture in the open chain for $N=300$ and $g=4$.
The blue points show the measured ground-state profile $\langle S^z_x\rangle$.
The orange curve is a fit to Eq.~\eqref{SM-eq:Hikihara}
with $m$ fixed to the mean magnetization $m=0.19$.
The fit was performed using only sites $40 \le x \le 260$, giving
$A=0.482373$ and $a=0.630981$.}
  \label{SM-fig:friedel}
\end{figure}

\section{TBA calculations}
\label{SM-sec:tba-magnetization-magnon}

The TBA calculations were carried out with AI assistance.  Correctness was checked by comparison with known results for the Heisenberg model, agreement with DMRG calculations of the magnetization and of spectral functions, and other spot checks.  This appendix is not meant to be a derivation, but rather a summary of the equations necessary to reproduce the results.

We summarize the zero-temperature TBA formulae used to obtain the results shown elsewhere in this paper.  We keep an exchange constant $J$ in this appendix; the main text uses $J=1$.  Let
\begin{equation}
  s(x)=\frac{1}{4\cosh(\pi x/2)},\qquad
  \widehat R(k)=\frac{e^{-|k|}}{2\cosh k},
\end{equation}
where $\widehat f(k)=\int dx\,e^{-ikx}f(x)$ and $f*g(x)=\int dy\,f(x-y)g(y)$.  After the higher strings are eliminated, the dressed one-string pseudoenergy at field $h$ obeys
\begin{equation}
  \epsilon(x)=-2\pi J s(x)+2\pi g s'(x)+\frac{h}{2}
  +R*\epsilon^+(x),
  \label{SM-eq:tba-scalar-appendix}
\end{equation}
with $\epsilon^+(x)=\max[\epsilon(x),0]$.  The physical zero-field magnetization is obtained as the $h\to0^+$ response, not by counting the bare occupied interval.  Defining $\chi(x)=\Theta(\epsilon(x))$, where $\Theta$ is the step function, the dressed spin response $z(x)=\partial_h\epsilon(x)$ satisfies
\begin{equation}
  z(x)=\frac12+R*\left[\chi z\right](x),
  \label{SM-eq:tba-response-appendix}
\end{equation}
and the spontaneous magnetization per site in the highest-weight state is
\begin{equation}
  m=\int_{-\infty}^{\infty} dx\,s(x)\chi(x)z(x).
  \label{SM-eq:tba-mag-appendix}
\end{equation}

The mistake in the magnetization calculation of Ref.~\cite{sedrakyan2025magnetochiral} is to replace the full hierarchy by the undressed one-string equation once only the one-string pseudoenergy has a negative region.  Higher strings need not themselves be occupied to dress the one-string energy, and their elimination gives the convolution term in Eq.~\eqref{SM-eq:tba-scalar-appendix}, which changes the magnetization.

The spiral wavevector and the vector spin chirality require the dressed Bethe root density in the same background.  We denote the real-root sea by
\begin{equation}
  {\cal B}=\{x:\epsilon(x)<0\},
\end{equation}
and write $\chi_{\cal B}(x)=1$ for $x \in {\cal B}$ and $\chi_{\cal B}(x)=0$ otherwise.  Let $K_n(x)=n/[\pi(n^2+x^2)]$, so that $\widehat K_n(k)=e^{-n|k|}$.  The root density satisfies
\begin{equation}
  \rho(x)=\chi_{\cal B}(x)\left[K_1(x)-K_2*(\chi_{\cal B}\rho)(x)\right],
  \label{SM-eq:tba-rho-appendix}
\end{equation}
with $\int_{\cal B}dx\,\rho(x)=1/2-m$.  The counting function is
\begin{align}
  {\cal Z}(x_0)&=\frac{\theta_1(x_0)}{2\pi}
  -\int_{\cal B}dx\,\frac{\theta_2(x_0-x)}{2\pi}\rho(x),
  \nonumber\\
  \theta_n(x)&=2\tan^{-1}(x/n),
  \label{SM-eq:tba-counting-appendix}
\end{align}
determines the center of the occupied Bethe integers.  If $A$ and $B$ are the two edges of ${\cal B}$, we define ${\cal Z}_0=[{\cal Z}(A)+{\cal Z}(B)]/2$ and obtain the transverse spiral wavevector
\begin{equation}
  q_s = \left|\pi-2\pi{\cal Z}_0\right|\quad {\rm mod}\;2\pi.
  \label{SM-eq:tba-wavevectors-appendix}
\end{equation}
This is the TBA $q_s$ wavevector shown in Fig.~\ref{SM-fig:wavevectors}.   It can be understood as the displacement of the center of the Bethe sea.

The nearest-neighbor vector spin chirality is obtained by threading a uniform spin twist $\Phi$ through the Bethe equations,
\begin{equation}
  L\theta_1(x_j)+\Phi-\sum_k\theta_2(x_j-x_k)=2\pi I_j.
\end{equation}
Writing $\sigma(x)=\rho(x)L\,\partial_\Phi x$ in the thermodynamic limit gives the linear shift equation
\begin{equation}
  \sigma(x)=\chi_{\cal B}(x)\left[-\frac{1}{2\pi}
  -K_2*(\chi_{\cal B}\sigma)(x)\right].
  \label{SM-eq:tba-twist-shift-appendix}
\end{equation}
With the charge convention $q_2(x)=-2/(1+x^2)$, the chirality is the twist derivative of $Q_2$,
\begin{equation}
  \kappa^z=\frac{\partial Q_2}{\partial\Phi}
  =\int_{\cal B}dx\,q_2'(x)\sigma(x)
  =\int_{\cal B}dx\,\frac{4x}{(1+x^2)^2}\sigma(x).
  \label{SM-eq:tba-kappa-appendix}
\end{equation}
The sign is fixed by the twist convention in the preceding equation.  With this convention $\kappa^z$ is positive in the chosen highest-weight state for $g>g_c$, vanishes below $g_c$, and its scaled value $(2/\pi)\kappa^z$ approaches the magnetization near the transition, as shown in Fig.~\ref{fig:mag}.

The ferromagnetic magnon is the finite-rapidity continuation of the infinite-rapidity spin-lowering descendant root, including its backflow on the TBA sea.  If this root has large rapidity $y$, its momentum and energy have the tails
\begin{equation}
  q(y)\sim \frac{4m}{y},\qquad
  E(y)=\epsilon(-y)\sim \frac{1}{\pi y^2}
  \int_{-\infty}^{\infty} dx\,\epsilon^+(x).
\end{equation}
Thus, with the convention $E(q)=Dq^2=q^2/(2m^*)$,
\begin{equation}
  D=\frac{1}{2m^*}
  =\frac{1}{16\pi m^2}
  \int_{-\infty}^{\infty} dx\,\epsilon^+(x).
  \label{SM-eq:tba-stiffness-appendix}
\end{equation}

Numerically, Eq.~\eqref{SM-eq:tba-scalar-appendix} is solved on a large uniform rapidity grid using FFT convolution and Anderson mixing, and Eq.~\eqref{SM-eq:tba-response-appendix}, Eq.~\eqref{SM-eq:tba-rho-appendix}, and Eq.~\eqref{SM-eq:tba-twist-shift-appendix} are then solved on the same grid with the converged indicator functions.  The rapidity window and grid spacing are increased until $m$, $q$, $\kappa^z$, and $D$ are unchanged within the quoted precision; the $1/y^2$ tail entering Eq.~\eqref{SM-eq:tba-stiffness-appendix} is evaluated through the integral of $\epsilon^+$ rather than by fitting finite-rapidity data.

\begin{figure}[htbp]
  \centering
  \includegraphics[width=\columnwidth]{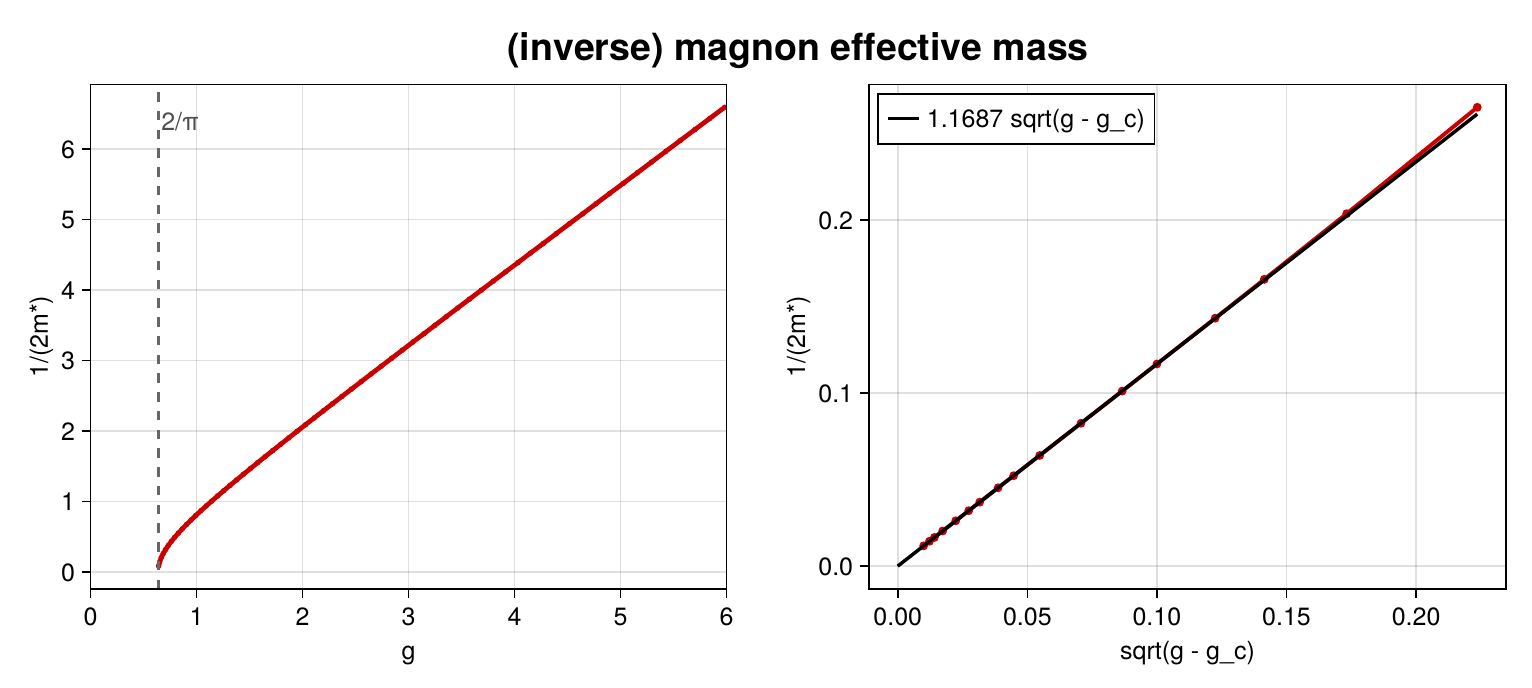}
  \caption{Magnon effective mass, plotted as $1/(2m^*)$, versus coupling $g$ determined from the TBA, evaluating Eq.~\eqref{SM-eq:tba-stiffness-appendix} numerically.  The right panel shows the square-root critical behavior expected from scaling for $g$ close to $g_c=2/\pi$.}
  \label{SM-fig:magnon_mass}
\end{figure}

\end{document}